\documentclass{ws-ijmpa}
\usepackage[super,compress]{cite}
\usepackage{graphicx}
\begin{document}
\markboth{Wilk and W\l odarczyk}{Intriguing aspects of  multiparticle production processes}

%
\catchline{}{}{}{}{}
%

\title{Some intriguing aspects of multiparticle production processes}

\author{Grzegorz Wilk}

\address{National Centre for Nuclear Research, Department of Fundamental Research\\
Warsaw 00-681, Poland\\
grzegorz.wilk@ncbj.gov.pl}

\author{Zbigniew W\l odarczyk}
\address{Institute of Physics, Jan Kochanowski University\\
Kielce 25-406, Poland\\
zbigniew.wlodarczyk@ujk.kielce.pl}

\maketitle

\begin{history}
\received{Day Month Year}
\revised{Day Month Year}
\end{history}

\begin{abstract}

Multiparticle production processes provide valuable information about the mechanism of the conversion of the initial energy of projectiles into a number of secondaries by measuring their multiplicity distributions and their distributions in phase space. They therefore serve as a reference point for more involved measurements. Distributions in phase space are usually investigated using the statistical approach, very successful in general but failing in cases of small colliding systems, small multiplicities, and at the edges of the allowed phase space, in which cases the underlying dynamical effects competing with the statistical distributions take over. We discuss an alternative approach, which applies to the whole phase space without detailed knowledge of dynamics. It is based on a modification of the usual statistics by generalizing it to a superstatistical form. We stress particularly the scaling and self-similar properties of such an approach manifesting themselves as the phenomena of the log-periodic oscillations and oscillations of temperature caused by sound waves in hadronic matter. Concerning the multiplicity distributions we discuss in detail the phenomenon of the oscillatory behaviour of the modified combinants apparently observed in experimental data.

\keywords{multiparticle production; statistical models; superstatistics, scaling, self-similarity}
\end{abstract}

\ccode{PACS numbers: 05.90.+m, 13.85.Hd, 24.10.Pa, 24.60.Ky, 24.60.-k }


\section{Introduction}	
\label{sec:I}

Multiparticle production, one of the first subjects studied in collision experiments, provides valuable information about the mechanism of the conversion of the initial energy of projectiles, $\sqrt{s}$, into a number $N$ of secondaries by measuring their multiplicity distributions, $P(N)$, and their distributions in the allowed phase space. It therefore serves as a reference point for more involved and detailed measurements. Conventionally, the produced particles are assigned to two categories: the majority, occupying the densely populated central region of small transverse momenta, $p_T \leq p_{T0}$, are supposed to be produced in {\it soft interactions}, the rest, with $p_T > p_{T0}$, are  associated with  quantum-chromodynamical (QCD-)based {\it hard collisions} and occupy the remaining part of the phase space. They differ significantly in the form of their distributions in $p_T$ which is exponential, $F (p_T )\sim \exp (-p_T/T)$, in the former case and power-like, $F \left(p_T\right) \sim p_T^{-n}$, in the latter. This, in turn, causes them usually to be investigated by, respectively, statistical methods with a scale parameter $T$ playing the role of "temperature"), or to be associated with QCD based hard scattering processes between partons (quarks and gluons) which are governed by the power index $n$.

In this review we present an approach combining both approaches in one purely phenomenological quasi-power law formula:
\begin{eqnarray}
  H(E) = C\cdot \left( 1 + \frac{E}{nT}\right)^{-n}
  \longrightarrow
  \left\{
 \begin{array}{l}
  \exp\left(-\frac{E}{T}\right)\quad \, \, \, \, {\rm for}\ E \to 0, \smallskip\\
  E^{-n}\qquad \qquad\quad {\rm for}\ E \to \infty .
 \end{array}
 \right .
 \label{eq:H}
\end{eqnarray}
It interpolates smoothly between the pure power-like behavior and the exponential formulation and was first proposed in \cite{CM-1,CM-2} , rediscovered in \cite{H} , and used for the first time to fit experimental data in \cite{UA1} (note that the parameter  $p_{T0}$, artificially dividing  the phase space into {\it soft} and {\it hard} parts, is no longer present). Widely known as the {\it QCD-based Hagedorn formula} it became one of the standard phenomenological formulas for $p_T$ data analysis. Eq. (\ref{eq:H}) coincides with the Tsallis distribution which is based on nonextensive statistical mechanics \cite{Tsallis-1,Tsallis-2} ,
\begin{equation}
h_q(E) = C_q \left[ 1- (1-q)\frac{E}{T}
\right]^{\frac{1}{1-q}}\quad \stackrel{def}{=}
C_q \exp_q\left(-\frac{E}{T}\right)
 \stackrel{q \rightarrow 1}{\Longrightarrow} C_1 \exp\left(-\frac{E}{T}\right),\label{eq:T}
\end{equation}
where $q=1+1/n$ is known as the nonextensivity parameter because in statistical mechanics it describes the nonextensivity of the  system. For $q \rightarrow 1$ (or $n \rightarrow \infty$) both Eqs. (\ref{eq:H}) and (\ref{eq:T}) become the usual Boltzmann-Gibbs (BG) exponential distributions with $T$ becoming the temperature and our system becomes extensive.  Both formulas are equivalent, therefore in what follows we shall use them interchangeably. Whereas Eq. (\ref{eq:T}) has been widely used in many other branches of physics \cite{Contemporary} , both Eqs. (\ref{eq:H}) and (\ref{eq:T}) have been used in data analysis of multiparticle production processes (cf., for example, \cite{UA1had,CMS-1,CMS-2,CMS-3,ATLAS-1,ATLAS-2,ATLAS-3,ALICE-1,ALICE-2,ALICE-3,UA1,PHENIX-1,PHENIX-2,STAR} and references therein) and in their phenomenological description (cf., for example, \cite{RWW-1,RWW-2,RWW-3,RWW-4,RWW-5,WWrev-1,WWrev-2,RWW-6,Wibig-1,Wibig-2,Betall-1,Betall-2,Betall-3,Betall-4,JCl-1,JCl-2,JCl-3,AD-1,AD-2,AD-3,AD-4,Others-1,Others-2,Others-3,WalRaf} and references therein).

The nonextensivity parameter $q$ plays an important role because the departure from unity of $|q-1|$ is a measure of how far we are from the situation described by BG statistics (which is recovered for $q=1$). It carries information on all factors, dynamical or otherwise, making our system {\it nonextensive}. Therefore, by adding new dynamical information concerning our system we are diminishing the value of $|q-1|$ (as was demonstrated in  \cite{BMNSW}). In this review we present some selected aspects of this problem with special emphasis on the sources of the quasi-power law distributions (Sections \ref{sec:Qple} and \ref{sec:Other}) and, among these, on the so called {\it superstatistical} approach  \cite{SS-BC,SS-S} (Section \ref{sec:SuperS}). We concentrate mainly on a couple of examples from our most recent work (see, for example, \cite{WWrev-1,WWrev-2,WW-APPB46,WW-Entropy14,WW-Entropy17,WW-CSF81} and references therein). They cover the generalities of fluctuation, correlation and nonextensivity (Section \ref{sec:Fluct}) and some specific examples of the quasi-power law in hadronic and nuclear collisions (Section \ref{sec:Power}). In Section \ref{sec:Surprises} we present some surprising features of quasi-power law distributions which distinguished themselves by their scaling (Section \ref{sec:complexN}) and self-similar properties (Section \ref{sec:oscillations}) resulting in, respectively, the occurrence of log-periodic oscillations of observed distributions as a function of $p_T$, or oscillations of the scale parameter $T$ which indicate the possible formation of sound waves in hadronic matter formed in the collision process. In Section \ref{sec:Combinants} we present a novel feature of multiplicity distributions, namely we show apparent oscillations of their modified combinants. Section \ref{sec:Summary} summarises our review.  We do not cover results based on nonextensive thermodynamics which can be found, for example, in \cite{JCl-1,JCl-2,AD-1,AD-2} and their theoretical justification is provided by \cite{Just-1,Just-2,Just-3,Bbook} . Also, a description of nonextensive dense hadronic or partonic matter is not addressed here, but may be found, for example, in \cite{JRGW-1,JRGW-2,AD-DM} (and references therein).

\section{Quasi-power law ensembles}
\label{sec:Qple}

\subsection{Statistical properties of small or constrained systems}
\label{sec:Stat}

We start with small or constrained systems which lead to Tsallis distributions with $q<1$. The problem of the applicability of statistical physics to small systems was first addressed a long time ago in \cite{F} (on the example of the nucleation reaction observed experimentally). For systems composed of a small number $N$ of particles a number of new effects occur caused, for example, by the surface terms or by rotational movements, which prevent the applicability of ordinary thermodynamics applicable only in the limit of $N\rightarrow \infty$. Therefore, this problem was initially approached by simply supplementing the ordinary thermodynamic relations by some correction terms (called {\it subdivision energies}) \cite{Hill-book} . Later, such {\it nanothermodynamics} (a synonym for the thermodynamics of small systems also called {\it nanosystems} and introduced in \cite{TG}) was formulated by focusing on the inherent characteristics of any nanosystem, namely on the fluctuations of its thermodynamic functions. It turns out that when averaged over they result in the Tsallis statistics description \cite{AKR-CSP-SA,HT,G-MCP} . A similar situation is encountered when one considers large but constrained systems \cite{Kodama-1,Kodama-2} . In both cases one encounters deviations from the usual Boltzmann-Gibbs formula. In what follows we address a number of such cases.

\subsubsection{All variables fixed}
\label{sec:Fixed}

The simplest situation is when out of the three main variables characterizing the system: the total energy $U$, the temperature $T$ and the multiplicity $N$, one or two are fixed and the rest fluctuate (in the case of $U$ or $T$ according to a gamma distribution, in the case of $N$ according to a Poisson distribution) \cite{WWcov} ; note that only in the thermodynamic limit of $N \rightarrow \infty$ do these fluctuations take the form of the usually considered Gaussian distributions. Fluctuations of all three variables always induce some correlations in the system  \cite{WWcov} . If all variables are fixed the energy $E$ of a constituent fluctuates according to a Tsallis distribution with $q < 1$:
\begin{eqnarray}
f(E) &=& \left(1 - \frac{E}{U}\right)^{N-2} = \left[ 1 - (1-q)\frac{E}{T}\right]^{\frac{1}{1-q}} \label{eq:ConstN}\\
&&{\rm where}\quad T = \frac{U}{N-2}\quad{\rm and}\quad q = \frac{N-3}{N-2} < 1 .\nonumber
\end{eqnarray}
Note that the same distribution emerges for the so called {\it induced partition} \cite{IndPart} in which $N-1$ randomly chosen independent points, $\left\{ U_{1},\dots, U_{N-1}\right\}$, break the segment $(0,U)$ into $N$ parts whose lengths are distributed according to Eq. (\ref{eq:ConstN}). For ordered $U_k$ the length of the $k^{th}$ part corresponds to
the value of the energy $E_k = U_{k+1}-U_k$. In our case it could correspond to the case of random breaks of the string in $N-1$ points in the energy space. Note that the induced partition differs from {\it successive sampling} from the uniform distribution, $E_k\in \left[ 0, U - \sum_{i=1}^{k-1} E_i\right]$, which results in $f(E) =1/E$~~\cite{1overE-1,1overE-2} .

\subsubsection{Conditional probability}
\label{sec:Conditional}

Let us consider now a system of $n$ independent particles with energies $\left\{E_{i=1,\dots,N}\right\}$, each of them
distributed according to a Boltzmann distribution $g_i\left( E_i\right)$ (their sum, $U = \sum_{i=1}^{N}E_i$, is then
distributed according to a gamma distribution $g_N(U)$):
\begin{equation}
g_i\left(E_i\right) = \frac{1}{T}\exp\left( -\frac{E_i}{T} \right)\quad{\rm and}\quad g_N(U) =
\frac{1}{T(N-1)!}\left(\frac{U}{T}\right)^{N-1} \exp \left( - \frac{U}{T}\right). \label{eq:BolGam}
\end{equation}
Suppose that the available energy is limited, $U = const$. In such a case the distribution of energy of a single particle follows some {\it conditional probability}, which again has the form of a Tsallis distribution:F
\begin{eqnarray}
f\left( E_i|U \right) &=& \frac{g_1\left( E_i \right)
g_{N-1}\left( U - E_i \right )}{g_N (U)} = \frac{(N-1)}{U}\left(1 -
\frac{E_i}{U}\right)^{N-2} = \nonumber\\
&=& \frac{2-q}{T}\left[ 1 - (1 - q) \frac{E_i}{T}\right]^{\frac{1}{1 - q}}, \label{eq:constraints}\\
&&{\rm where}\quad q = \frac{N-3} {N-2} < 1\qquad {\rm and}\qquad T = \frac{U}{N-2}, \label{eq:c1}
\end{eqnarray}
the same as in Eq. (\ref{eq:ConstN}) above.

\subsubsection{Statistical physics considerations}
\label{sec:Statistical}

Both Eqs. (\ref{eq:ConstN}) and (\ref{eq:constraints}) follow from  the statistical physics considerations of an isolated system with energy $U = const$ and with $\nu$ degrees of freedom (particles). Choose a single degree of freedom (a single, well-defined, state) with energy $E << U$ and denote the remaining (or reservoir) energy as $E_r = U - E$. Let $\Omega(U-E)$ be the number of states of the whole system and $P(E) \propto \Omega(U-E)$ (which is slowly varying) the probability that the energy of the chosen degree of freedom is $E$. Expanding $\ln \Omega(U-E)$ around $U$, and keeping only the first two terms one finds that
\begin{equation}
\ln P(E) \propto \ln \Omega(E) \propto - \beta E \quad {\rm with}\quad \beta = \frac{1}{k_B T} \stackrel{def}{=}
\frac{\partial \ln \Omega\left(E_r\right)}{\partial
E_r}.\label{eq:deriv}
\end{equation}
 This means that $P(E) \propto \exp( - \beta E)$ is an extensive ($q=1$) Boltzmann distribution. On the other hand, one expects that
\begin{equation}
\Omega\left(E_r\right) \propto \left(\frac{E_r}{\nu}\right)^{\alpha_1 \nu - \alpha_2}
\label{eq:Omega}
\end{equation}
(where $\alpha_{1,2}$ are of the order of unity; we put $\alpha_1 = 1$ and, to account for diminishing the number of states in the reservoir by one state, $\alpha_2 = 2$) \cite{Reif,Feynm} . This means that
\begin{equation}
\frac{\partial^k \beta}{\partial E_r^k}\, \propto\, (-1)^k k! \frac{\nu - 2}{E^{k+1}_r}\, =\, (-1)^k k! \frac{\beta^{k-1}}{(\nu - 2)^k} \label{eq:FR}
\end{equation}
and the probability of choosing energy $E$ can be written as:
\begin{eqnarray}
P(E) &\propto& \frac{\Omega(U-E)}{\Omega(U)} =
\exp\left[ \sum_{k=0}^{\infty}\frac{(-1)^k}{k+1}\frac{1}{(\nu - 2)^k}(- \beta E)^{k+1}\right] = \nonumber\\
&=&C\left(1 - \frac{1}{\nu - 2}\beta E\right)^{(\nu - 2)} = \beta(2-q)[1 - (1-q)\beta E]^{\frac{1}{1-q}}
\label{eq:statres}
\end{eqnarray}
(were we  used the equality $ \ln(1+x) = \sum_{k=0}^{\infty}(-1)^k \frac{x^{k+1}}{(k+1)}$ ). For
\begin{equation}
q= 1 - \frac{1}{\nu -2} \leq 1 \label{q-from-stat}
\end{equation}
Eq. (\ref{eq:statres}) coincides with the previous results obtained from the induced partition and conditional
probability.

\subsection{Superstatistics. Beyond B-G statistics}
\label{sec:SuperS}

Note that all the previous distributions, Eqs. (\ref{eq:ConstN}), (\ref{eq:constraints}) and (\ref{eq:statres}) are Tsallis distributions with $q<1$ for which
\begin{equation}
\langle E\rangle = \frac{T (N-2)}{N} = \frac{T}{3-2q}. \label{eq:MeanEq<1}
\end{equation}
For $N \rightarrow \infty$   (or $q\rightarrow 1$) we get $\langle E\rangle = T$, as in the BG distribution, $f(E) = (1/T)\exp(-E/T)$. We shall proceed now to the case of $q>1$. It is closely  connected with the presence of some intrinsic fluctuations in the system under consideration. In our case this idea was conceived when investigating some anomalies observed in cosmic ray experiments where the observed spectra visibly departed from the expected exponential behavior. We proposed at that time that this is just a manifestation of cross section fluctuations (which were, in fact, observed in nuclear collisions and in diffraction dissociation experiments at accelerators) \cite{CR} . Some time later we realized that the same data can be fitted by a Tsallis distribution with $q > 1$  which indicates that values of $q > 1$ are connected with the strength of such fluctuations \cite{CRLF} . In fact, it turned out that convoluting the usual Boltzmann-Gibbs exponential factor, $\exp (-E/T)$, with a gamma distribution,
\begin{equation}
g(1/T) = \frac{1}{\Gamma\left(\frac{1}{q -
1}\right)}\frac{T_0}{q - 1}\left(\frac{1}{q -
1}\frac{T_0}{T}\right)^{\frac{2 - q}{q - 1}}\cdot\exp\left( - \frac{1}{q - 1}\frac{T_0}{T}\right), \label{eq:Gamma}
\end{equation}
results in a Tsallis distribution, $h_q(E)$, with a new parameter $q$, which for $q \rightarrow 1$ becomes the usual BG distribution:
\begin{equation}
h_q(E) = \frac{2-q}{T}\exp_q \left(-\frac{E}{T}\right)= \frac{2-q}{T}\left[1 - (1-q)\frac{E}{T}\right]^{\frac{1}{1-q}}
 \stackrel{q \rightarrow 1}{\Longrightarrow}\, \frac{1}{T}\exp\left(-\frac{E}{T}\right).  \label{eq:Tsallis}
\end{equation}
The parameter $q$ is directly connected to the variance of $T$ (the prefactor $(2-q)/T$ occurring here reflects our definition of the distribution as a probability density function with standard normalization, $\int dE h_q(E) = 1$) \cite{SS-WW-1,SS-WW-2} :
\begin{equation}
q = 1 + \omega_T^2\qquad {\rm where}\qquad \omega_T^2 = \frac{Var(T)}{<T>^2}. \label{eq:q}
\end{equation}
The gamma function distribution of $1/T$ follows from the situation when the heat bath is not homogeneous and must be described by a local temperature, $T$, fluctuating from point to point around some equilibrium value, $T_0$. Assuming some simple diffusion picture as being responsible for equalization of this temperature \cite{WWrev-1,SS-WW-1,SS-WW-2} we get the evolution of $T$ in the form of a Langevin stochastic equation with the distribution $g(1/T)$ in the form of a gamma distribution (\ref{eq:Gamma}) emerging as a solution of the corresponding Fokker-Planck equation. Such an approach  allows for further generalization to cases where diffusion also accounts for some energy transfer to/from the heat bath \cite{WWrev-1} . In such cases $T$ becomes a $q$-dependent quantity:
\begin{equation}
T = T_{eff} = T_0 + (q-1)T_V . \label{eq:Teff}
\end{equation}
Depending on the type of energy transfer $T_V$ is either negative (the system under consideration loses energy like, for example, in nuclear collisions when energy is transferred to the spectators \cite{WWTout} ) or positive (energy is transferred into the system, for example to cosmic rays during their propagation through outer space \cite{WWTin}).

This idea was further developed in \cite{BJ1,BJ2,Bbook} (where problems connected with the notion of temperature in such cases were addressed). This forms the basis for the so-called {\it superstatistics} \cite{SS-B,SS-BS,SS-BC,SS-S}  understood as a superposition of two different statistics relevant to driven nonequilibrium systems with a stationary state and intensive parameter fluctuations. For BG statistics and others defined by $g(1/T)$ the resultant distribution is then given by the convolution of both,
\begin{equation}
h(E) = \int \exp\left( - \frac{E}{T}\right) g(1/T)) d(1/T). \label{eq:SSdef}
\end{equation}
Depending now on the statistical properties of the fluctuations, one obtains different effective statistical mechanical descriptions. Tsallis statistics follow from the above gamma distribution of an intensive variable, but other classes of generalized statistics can be obtained as well and, for small variance of the fluctuations, they all behave in a universal way but only the gamma distribution has such a simple physical interpretation as presented here.
We shall now discuss this situation in more detail (cf. \cite{PLA}). Note that the distribution $g\left(T'\right)$ in Eq. (\ref{eq:Gamma}) is in fact the product of two distributions: a scale free power law $g_1\left(T'\right)$ and  an exponential, $g_2\left(T'\right)$, which cuts off the $\left(T'\right)$ for small values of $T'$ with the scale parameter $T$
determining how effective this cut-off is:
\begin{equation}
g\left(T'\right) \sim g_1\left(T'\right)\cdot g_2\left(T'\right)~~~{\rm with}~~~g_1\left(T'\right) = \left( \frac{1}{T'}\right)^{\kappa},~~g_2\left( T' \right) = \exp \left( - \frac{nT}{T'}\right). \label{eq:g12}
\end{equation}
Fluctuating the BG distribution part of (\ref{eq:Tsallis}) using only the scale free power law $g_1\left(T'\right)$ results again in a scale free distribution:
\begin{equation}
h_1(E) = \int_0^{\infty} dT'\, g_1\left(T'\right) \exp\left( - \frac{E}{T'}\right) \propto E^{-\kappa + 1}. \label{eq:h1}
\end{equation}
The scale appears only when one cuts-off somehow small values of $T'$. For example, a sharp cut-off, i.e., limiting $T'$ to $T' > T$ only, results in the following distribution:
\begin{equation}
h_2(E) = \int_T^{\infty} dT'\, g_1\left(T'\right) \exp\left( -
\frac{E}{T'}\right) \propto  E^{-\kappa + 1} \left[ \Gamma(\kappa - 1) - \Gamma
\left(\kappa - 1, \frac{E}{T}\right)\right], \label{eq:h2}
\end{equation}
where $\Gamma(x,y)$ is an incomplete gamma function. The factor $\Gamma(\kappa - 1) - \Gamma(\kappa - 1, E/T)$ now suppresses the power distribution (\ref{eq:h1}) for small values of $E$. However, the form of this suppression is not the same as in the Tsallis distribution. This can be seen by comparing the expansion
\begin{equation}
\left( \frac{E}{T}\right)^{-\kappa}\left[ \Gamma(\kappa - 1) - \Gamma \left(\kappa - 1, \frac{E}{T}\right)\right] =
\frac{1}{\kappa} +  \sum_{i=1}^{\infty}\frac{\Gamma(i + \kappa)}{\Gamma(i + \kappa + 1)\Gamma(i + 1)}\left( -
\frac{E}{T}\right)^i \label{eq:comp1}
\end{equation}
with the corresponding expansion of the Tsallis factor:
\begin{equation}
\left( 1 + \frac{E}{\kappa T}\right)^{-\kappa} = \frac{1}{\kappa} +  \sum_{i=1}^{\infty}\frac{\Gamma(i + \kappa)}{\Gamma(1 + \kappa)\Gamma(i+1)}\left( - \frac{E}{\kappa T}\right)^i.
\label{eq:comp2}
\end{equation}
However, if we smooth out this suppression by replacing the previous sharp limitation of the integrand by some smooth
suppression factor, provided, for example, by the exponential function $g_2\left( T'\right)$ from Eq. (\ref{eq:g12}), we get, as a result, the Tsallis distribution defined in  Eq. (\ref{eq:Tsallis}).

\section{Other possible mechanisms leading to the Tsallis distribution}
\label{sec:Other}

\subsection{Preferential attachment}
\label{sec:Pa}

The effect of $q>1$ can also be obtained if the system under consideration exhibits correlations of the preferential attachment type, corresponding to the "rich-get-richer" phenomenon in networks \cite{WWnets-1,WWnets-2,WWrev-2,nets-1,nets-2} , and if the scale parameter $T$ depends on the variable under consideration, for example, if
\begin{equation}
T\rightarrow T_0'(E) = T_0 + (q-1)E. \label{eq:pat}
\end{equation}
In such cases on obtains from the evolution equation for the network \cite{WWnets-1,WWnets-2,WWrev-2}, as its solution, a probability distribution function which has the form of a Tsallis distribution (with $q > 1$). In our case it corresponds to
\begin{equation}
\frac{df(E)}{dE} = - \frac{1}{T_0'(E)}f(E)\quad \Longrightarrow\quad  f(E) = \frac{2-q}{T_0}\left[ 1 -
(1-q)\frac{E}{T_0}\right]^{\frac{1}{1-q}}. \label{eq:nets}
\end{equation}
For $T_0'(E) = T_0$ (i.e., for $q=1$) one again gets the usual exponential distribution. In \cite{WWnets-3} we applied this approach to the analysis of multiparticle production processes (demonstrating that the "power law", assumed {\it ad hoc} in \cite{MGMG} as a kind of opposition to Tsallis statistics, has its explanation using a stochastic networks approach).

Note that "preferential attachment" is also present in superstatistics where the convolution
\begin{equation}
f(E) = \int g(T) \exp\left( - \frac{E}{T}\right) dT \label{eq:SSt}
\end{equation}
becomes a Tsallis distribution, Eq. (\ref{eq:T}), for
\begin{equation}
g(T) = \frac{1}{\Gamma(n)T}\left(\frac{nT_0}{T}\right)^n
\exp\left( - \frac{nT_0}{T}\right). \label{eq:gamma}
\end{equation}
Differentiating Eq. (\ref{eq:SSt}) one obtains Eq. (\ref{eq:nets}).
\begin{equation}
\frac{df(E)}{dE} = - \frac{1}{T(E)} f(E)\qquad {\rm where}\quad
T(E) = T_0 + \frac{E}{n}. \label{eq:diffSSt}
\end{equation}
We close this part by mentioning that this is not the only place where such a form of $T(E)$ appears. For example, in \cite{WalRaf} it occurs in a description of the thermalization of quarks in a quark-gluon plasma by a collision process treated within Fokker-Planck dynamics.

\subsection{Multiplicative noise}
\label{sec:Mn}

Suppose that we treat our system as being formed in a stochastic process defined by the following Langevin equation:
\begin{equation}
\frac{dp}{dt} + \gamma(t)p = \xi(t), \label{eq:Le}
\end{equation}
where $\gamma(t)$ and $\xi(t)$ denote multiplicative and additive noise \cite{SS-WW-1,BJ1,Bbook,WWAIP-2013} . The resulting Fokker-Planck equation for the distribution function $f$,
\begin{eqnarray}
\frac{\partial f}{\partial t} &=& - \frac{\partial \left( K_1 f\right)}{\partial p} + \frac{\partial^2 \left( K_2
f\right)}{\partial p^2}, \label{eq:FPe}\\
K_1 &=& \langle \xi\rangle - \langle \gamma\rangle p\quad {\rm
and}\quad K_2 = Var(\xi) - 2 Cov(\xi, \gamma) p + Var(\gamma )
p^2, \label{eq:K1K2}
\end{eqnarray}
has a stationary solution $f$, which satisfies
\begin{equation}
\frac{d \left( K_2 f\right)}{dp} = K_1 f. \label{eq:K1vK2}
\end{equation}
If there is no correlation between the different types of noise and no drift term due to the additive noise, i.e., for $Cov(\xi,\gamma) = \langle
\xi\rangle = 0$ ~\cite{AT} , the solution of this equation is a Tsallis distribution for $p^2$,
\begin{equation}
f(p) = \left[1 + (q - 1)\frac{p^2}{T}\right]^{\frac{q}{1-q}}~~{\rm with}~~ T = \frac{2Var(\xi)}{\langle \gamma \rangle};~~q = 1 + \frac{2Var(\gamma)}{\langle \gamma\rangle}. \label{eq:solutionpp}
\end{equation}
If we insist on a solution in the form of Eq. (\ref{eq:H}),
\begin{equation}
f(p) = \left[1 + \frac{p}{nT}\right]^n\qquad {\rm with}\qquad n =
\frac{1}{q-1}, \label{eq:singlep}
\end{equation}
then the condition to be satisfied has the form:
\begin{equation}
K_2(p) = \frac{nT+p}{n}\left[ K_1(p) - \frac{dK_1(p)}{dp}\right]
\label{eq:K1vK2p}
\end{equation}
resulting in
\begin{eqnarray}
&&n = 2 + \frac{\langle \gamma\rangle}{Var(\gamma)}\quad{\rm or}\quad q = 1 + \frac{Var(\gamma)}{\langle \gamma\rangle + 2
Var(\gamma)}, \label{eq:n-q}\\
&&T(q) = T_0 + (q-1)T_1 - \frac{(q - 1)^2}{2-q}T_2, \label{eq:Tq}\\
&& {\rm where}\quad T_2 = \frac{\langle \xi\rangle}{2\langle \gamma\rangle};\quad T_1 = T_2 - 2 T_0,\quad T_0= - \frac{Cov(\xi,\gamma)}{\langle \gamma\rangle}. \label{eq:T2T1}
\label{eq:TeffN}
\end{eqnarray}
Note that $T$ depends now on $q$, we face therefore a similar situation as for the $T_{eff}$ encountered before in  Section \ref{sec:SuperS}. Some comments on the effective temperature in action are therefore in order here (see also Section \ref{sec:effT} for experimental examples). If $\langle \xi\rangle  =0$ (no drift due to the additive noise) positivity of temperature $T$ requires that $Cov(\xi,\gamma) < 0$, i.e., that one has an anticorrelation between the different types of noise. For $\langle \xi\rangle \neq 0$ one can have $T > 0$ even for $Cov(\xi,\gamma) = 0$. For the linear dependence of $T$ on $(q-1)$ apparently observed experimentally \cite{WWrev-1,WWrev-2,RWW-6} one needs $Cov(\xi,\gamma) < 0$ and $\langle \xi\rangle = 0$. For a quadratic dependence one can have $\langle \xi\rangle \neq 0$.

\subsection{Tsallis distribution from Shannon entropy}
\label{sec:TsShannon}

The common practice in selecting the most suitable distribution law to describe the available data is to use the principle of maximum entropy as a robust theoretical basis. To make the resulting distribution unique one has to decide on the entropic form to be used and specify the constraints to be imposed. The usual choice for entropy is Shannon entropy but for our purpose here Tsallis entropy seems to be the better choice \cite{WWrev-1,WWrev-2,Tsallis-2} . However, as demonstrated in \cite{GR} , it seems that the effects of changing the form of entropy can be accommodated by the right choice of constraints. In particular, in \cite{TfromS} it was shown that a Tsallis distribution emerges in a natural way from the usual Shannon entropy, $S$ (for some probability density $f(x)$), by means of the usual MaxEnt approach, if only one imposes the right constraint provided by some function of $x$, $h(x)$:
\begin{equation}
S = - \int dx f(x)\ln[f(x)]\quad{\rm where}\quad \int dx f(x)h(x) = \langle h(x)\rangle = const. \label{eq:StT}
\end{equation}
Because this approach contains the same information as that based on Tsallis entropy one can either use Tsallis entropy with relatively simple constraints, or Shannon entropy with rather complicated ones (an example of the list of possible
distributions one can obtain in this way is provided in \cite{GR}). One therefore finds that
\begin{equation}
f(x) = \exp\left[ \lambda_0 + \lambda h(x)\right],
\label{eq:TfromS}
\end{equation}
with constants $\lambda_0$ and $\lambda$ calculated from the normalization of $f(x)$ and from the constraint equation. The
constraint leading to the Tsallis distribution has the form of
\begin{equation}
<z> = z_0 = \frac{q-1}{2-q}\qquad {\rm where}\qquad  z =
\ln\left[1 - (1 - q)\frac{E}{T_0}\right], \label{eq:constrT}
\end{equation}
(remember that $f(z)dz=f(E)dE$) and the corresponding distribution is
\begin{eqnarray}
f(z) = \frac{1}{z_0}\exp\left( -\frac{z}{z_0}\right)~~ \Longrightarrow~~ f(E) &=& \frac{1}{\left(1 +
z_0\right)T_0}\left( 1 + \frac{z_0}{1+z_0}\frac{E}{T_0}\right)^{ - \frac{1 + z_0}{z_0}} = \nonumber \\
 &=& \frac{2 - q}{T_0}\left[ 1 - (1 - q)\frac{E}{T_0}\right]^{\frac{1}{1 - q}}. \label{eq:TfS}
\end{eqnarray}
To obtain the scale parameter $T_0$ ("temperature") one has additionally to assume knowledge of $\langle E\rangle$. Note that in the case of the BG distribution it would be the only constraint, but here it is an additional condition to be accounted for.

To be fully acceptable and useful this approach must be further endowed with a proper understanding of the meaning of the necessary constraint. So far its physical significance is not fully understood and we can offer only some hints as a basis for future discussion. For example, note that the form of constraint (\ref{eq:constrT}) can be deduced from the idea of varying the scale parameter in the form of the preferential attachment, Eq. (\ref{eq:pat}), which in the present notation means $T \rightarrow T(E)=T_0+(q-1)E$. As shown in Section \ref{sec:Pa} this results in a Tsallis distribution (\ref{eq:TfS}). This suggests the use of $z =\ln\left[ T(E)/T_0\right]$ constrained as in Eq. (\ref{eq:constrT}). In such an approach $\ln f(E) = - [1/(q-1)]\ln [T(E)] + [(2-q)/(q-1)]\ln \left( T_0\right)$ and, because $S = - \langle \ln f(E)\rangle$, therefore $S = 1/(2-q) + \ln \left( T_0\right)$ for for Tsallis distribution becomes $S = 1 + \ln \left( T_0\right)$ for the Boltzmann-Gibbs (BG) distribution ($q=1$). The other possible explanation is to note that the constraint (\ref{eq:constrT}) seems to be natural for a multiplicative noise described by the Langevine equation: $dp/dt + \gamma(t)p=\xi(t)$, with traditional multiplicative noise $\gamma (t)$ and additive noise (stochastic processes) $\xi(t)$) (see \cite{WWAIP-2013} for details). This is because there is a connection between the kind of noise in this process and the condition imposed in the MaxEnt approach. For processes described by an  additive noise, $dx/dt = \xi(t)$, the natural condition is that imposed on the arithmetic mean,   $<x>= c+\langle \xi\rangle t$, which results in an exponential distribution. For the multiplicative noise, $dx/dt = x\gamma(t)$, the natural condition is that imposed on the geometric mean, $ <\ln x> = c+\langle \gamma\rangle t$, which results in a power law distribution \cite{R} . Apparently, condition (\ref{eq:constrT}) combines both possibilities and leads to a quasi-power law Tsallis distribution combining both types of behavior.

\section{Fluctuation, correlation and nonextensivity}
\label{sec:Fluct}

\subsection{Tsallis distribution in energy from fluctuations of multipicity}
\label{sec:TtoP}

We shall now continue the discussion initiated in Section \ref{sec:Fixed} (all variables fixed) and continued in Section   \ref{sec:SuperS} (fluctuations of $T$ or superstatistics) by allowing in our system, described by the total energy $U$, temperature $T$ and multiplicity $N$, for fluctuations of multiplicity $N$ according to some distribution $P(N)$. In this case, the resulting distribution is
\begin{eqnarray}
f(E) &=& \sum f_N(E) P(N), \label{eq:resdistr}\\
{\rm where}&& f_N(E) = \left(1 - \frac{E}{U}\right)^N\qquad {\rm and}\qquad U =
\sum E = const. \label{eq:fixedN}
\end{eqnarray}
$f_N(E)$ is a distribution for fixed $N$ (to simplify the notation we changed $N-2$ in Eq. (\ref{eq:ConstN}) to $N$). The most characteristic distributions for our discussion are the following: Binomial Distribution, $P_{BD}$, Poissonian Distribution, $P_{PD}$ and Negative Binomial Distribution, $P_{NBD}$ (cf. \cite{WWjets}):
\begin{eqnarray}
P_{BD}(N)  &=& \frac{M!}{N!(M-N)!} \left(\frac{<N>}{M}\right)^N\left(1 - \frac{<N>}{M}\right)^{M-N}; \label{eq:BD}\\
P_{PD}(N)  &=& \frac{<N>^N}{N!} e^{-\langle N\rangle}; \label{eq:PD}\\
P_{NBD}(N) &=& \frac{\Gamma(N+k)}{\Gamma(N+1)\Gamma(k)}\left( \frac{<N>}{k}\right)^N \left( 1 + \frac{<N>}{k}\right)^{-k-N}.
\label{eq:NBD}
\end{eqnarray}
When used in Eq. (\ref{eq:resdistr}) they lead to Tsallis distributions with $q$ ranging from $q<1$ to $q>1$ (in all cases $\beta = \langle N\rangle/U$):
\begin{eqnarray}
f_{BD}(E) &=& \left(1 - \frac{\beta
E}{M}\right)^M,\qquad q = 1 - \frac{1}{M}  < 1,\quad M = \frac{1}{1-q};  \label{eq:BDf}\\
f_{PD}(E) &=& \exp( - \beta E),\qquad\quad \, \, \, q = 1; \label{eq:PDf}\\
f_{NBD}(E) &=& \left( 1 + \frac{\beta E}{k}\right)^{-k}, \qquad q = 1 + \frac{1}{k} > 1\qquad k = \frac{1}{q-1}. \label{eq:NBDf}
\end{eqnarray}
The physical meaning of $q$ remains the same in all three cases, it measures the strength of the corresponding multiplicity fluctuation,
\begin{equation}
q - 1 = \frac{Var(N)}{<N>^2} - \frac{1}{<N>}. \label{eq:FluctN}
\end{equation}
This means therefore that in the BD $Var(N)/<N> < 1$, in the PD $Var(N)/<N> = 1$, and in the NBD $Var(N)/<N>~ > 1$.

Let us concentrate now on the NBD $q > 1$ case where fluctuations of multiplicity $N$ can be translated into fluctuations of temperature $T$. This is possible because, as shown in \cite{WWrev-1,NBDder-1,NBDder-2,NBDder-3} ,
\begin{eqnarray}
P_{NBD}(N) &=& \int f\left( \bar{N}\right) P_{P}\left(N;\bar{N}\right) = \frac{\Gamma(N+k)}{[\Gamma(N+1)\Gamma(k)]}\cdot\gamma^k(1+\gamma)^{-k-N},\label{eq:NBDq}\\
{\rm where}~~P_{P}(N) &=& \frac{\bar{N}^N}{N!} e^{-{\bar{N}}},\quad f(\bar{N}) = \frac{\gamma^k{\bar{N}}^{k-1}}{\Gamma(k)}\cdot e^{-\gamma \bar{N}}\quad
{\rm  with}\quad \gamma = \frac{k}{<\bar{N}>}, \label{eq:Pfg}
\end{eqnarray}
i.e., the NBD appears as a result of fluctuations in the mean multiplicity in the PD using the gamma distribution (note that we have two types of average here: $\bar{X}$ means the average value for a given event whereas $<X>$ denotes averages over all events (or ensembles)). Identifying now fluctuations of the mean multiplicity $\bar{N}$ with fluctuations of the temperature $T$, namely, noting that $\bar{\beta} = \bar{N}/U,~\langle \bar{N}\rangle = U \langle \bar{\beta}\rangle $ and $\gamma = k/( U<\bar{\beta}>)$, one can rewrite the gamma distribution for mean multiplicity, $f(\bar{N})$, as a gamma distribution of mean inverse temperature $\bar{\beta}$,
\begin{eqnarray}
f\left( \bar{\beta}\right) &=& \frac{k}{<\bar{\beta}>\Gamma(k)}\left( \frac{k\bar{\beta}}{<\bar{\beta}>}\right)^{k-1}
\exp\left(-\frac{k\bar{\beta}}{<\bar{\beta}>}\right) =\nonumber\\
&=& \frac{ \left(\frac{1}{q-1}\frac{\bar{\beta}}{{<\bar{\beta}>}} \right)^{\frac{1}{q - 1} - 1}}{(q-1)<\bar{\beta}>\Gamma\left(\frac{1}{q-1}\right)}
\exp\left(-\frac{1}{q-1}\frac{\bar{\beta}}{<\bar{\beta}>}\right), \label{eq:NtoT}
\end{eqnarray}
where $q = 1 + Var(\bar{\beta})/\langle \bar{\beta}\rangle$ now denotes the strength of the temperature fluctuations. This is precisely the gamma distribution describing temperature fluctuations investigated in {\it superstatistics}  (see Section \ref{sec:SuperS}). It presents yet another possible derivation of the gamma distribution form of the temperature fluctuations, in addition to that presented in \cite{SS-WW-1,SS-WW-2} . Note that for large values of $N$ and $\langle N\rangle$ one obtains the scaling form of Eq. (\ref{eq:NBD}),
\begin{equation}
\langle N\rangle P(N) \cong  \psi\left( z=\frac{N}{\langle N\rangle} \right) = \frac{k^k}{\Gamma(k)} z^{k-1}\exp( - kz),
\label{eq:scalingform}
\end{equation}
which is a particular expression of Koba-Nielsen-Olesen (KNO) scaling \cite{KNO-1,KNO-2} .

\subsection{Multiplicity distribution from Tsallis distribution in energy}
\label{sec:P(N)toT}

Whereas the previous Section demonstrates that the NBD form of fluctuations of the number of particles produced results in a Tsallis distribution form of the distribution of their energies, we demonstrate now that the opposite is also true, i.e., that a Tsallis distribution of the observed $f(E)$ goes together with $P(N)$ being of the NBD form \cite{WWrev-1} .

\subsubsection{Poisson multiplicity distribution}
\label{sec:P}

We start with the  situation where in some process one has $N$ independently produced secondaries with energies $\{ E_{1,\dots,N}\}$, each distributed according to the simple Boltzmann  distribution:
\begin{equation}
f\left(E_i\right) = \frac{1}{T}\cdot \exp\left( -
\frac{E_i}{T}\right)\quad {\rm where}\qquad  T =\langle E\rangle. \label{eq:f(E)}
\end{equation}
The corresponding joint probability distribution for such system is then given by:
\begin{equation}
f\left( \{ E_{1,\dots,N}\} \right) = \frac{1}{T^N}\cdot
\exp\left( - \frac{1}{T}\sum_{i=1}^N E_i\right).\label{eq:jointP}
\end{equation}
For independent energies $\{ E_{i=1,\dots,N}\}$ the sum $U = \sum_{i=1}^N E_i$ is then distributed according to the following gamma distribution,
\begin{equation}
g_N(U) = \frac{1}{T (N-1)!} \cdot \left(\frac{U}{T}\right)^{N-1}\cdot \exp\left( - \frac{U}{T}\right), \label{eq:Egamma}
\end{equation}
the distribuant of which is equal to
\begin{equation}
G_N(U) = 1 - \sum_{i=1}^{N-1}\frac{1}{(i-1)!}\cdot \left(\frac{U}{T}\right)^{i-1}\cdot \exp \left( - \frac{U}{T}\right) . \label{eq:distribuant}
\end{equation}
Eq. (\ref{eq:Egamma}) follows immediately either by using characteristic functions or by sequentially performing integration of the joint distribution (\ref{eq:jointP}) and noting that:
\begin{equation}
g_N(U) = g_{N-1}(U)\frac{U}{N-1} . \label{eq:E123ijkn}
\end{equation}
For energies such that
\begin{equation}
\sum_{i=0}^N E_i \leq U \le \sum_{i=0}^{N+1} E_i \label{eq:conditsumE}
\end{equation}
the corresponding multiplicity distribution has a Poissonian form (note that $U/T = \langle N\rangle$):
\begin{equation}
P(N) = G_{N+1}(U) - G_N(U) = \frac{1}{N!} \left(\frac{U}{T}\right)^N\cdot \exp( - \alpha U) = \frac{\langle N\rangle ^N}{N!}\cdot \exp( - \langle N\rangle ). \label{eq:Poisson}
\end{equation}
In other words, whenever we have variables $E_{1,\dots,N,N+1,\dots }$ taken from the exponential distribution $f\left(E_i\right)$ and whenever these variables satisfy the condition $\sum_{i=0}^N E_i \leq U \le \sum_{i=0}^{N+1} E_i$, then the corresponding multiplicity $N$ has a Poissonian distribution (note that this is the method for generating a Poisson distribution in numerical Monte-Carlo codes).

\subsubsection{Negative Binomial multiplicity distribution}
\label{sec:NB}

We switch now to the situation when in some process we have $N$ independent particles with energies $\{ E_{1,\dots,N} \}$
distributed according to the Tsallis distribution,
\begin{equation}
h\left( \{E_{1,\dots,N}\}\right) = C_N\left[1-(1-q)\frac{\sum_{i=1}^N E_i}{T}\right]^{\frac{1}{1-q}+1 -N}, \label{eq:Enq}
\end{equation}
with normalization constant $ C_N$ given by
 \begin{equation}
C_N = \frac{1}{T^N}\prod_{i=1}^N[(i-2)q - (i-3)] = \frac{(q-1)^N}{T^N}\cdot \frac{\Gamma\left( N +
\frac{2-q}{q-1}\right)}{\Gamma \left( \frac{2-q}{q-1}\right)}. \label{eq:CNORMP}
\end{equation}
According to our philosophy this means that there are some intrinsic (so far unspecified but summarily characterized by the parameter $q$) fluctuations present in the system under consideration. In this case we do not know the characteristic function for the Tsallis distribution. However, because we are dealing here only with variables $\{E_{i=1,\dots,N} \}$ occurring in the form of the sum, $U = \sum_{i=1}^N E_i$, one can still sequentially perform integrations of the joint probability distribution (\ref{eq:Enq}) and, noting that (as before, cf. eq. (\ref{eq:E123ijkn}))
\begin{equation}
h_N(U) = h_{N-1}(U)\frac{U}{N-1} = \frac{U^{N-1}}{(N-1)!}h\left(\{E_{1,\dots,N}\}\right), \label{eq:qE123ijkn}
\end{equation}
we arrive at a formula corresponding to eq. (\ref{eq:Egamma}), namely that
\begin{equation}
h_N(U) = \frac{U^{(N-1)}}{(N-1)!T^N}\prod^N_{i=1}[(i-1)q - (i-3)]\left[ 1 - (1-q)\frac{U}{T}\right]^{\frac{1}{1-q}+1-N} \label{eq:Hq}
\end{equation}
the distribuant of which is given by
\begin{eqnarray}
H_N(U) &=& 1 - \sum_{j=1}^{N-1}  \tilde{H}_i(U)\qquad {\rm where} \nonumber\\
\tilde{H}_i(U) &=& \frac{U^{i-1}}{(j-1)!T^j}\prod_{i=1}^j \left[ (i-1)q - (i-3)\right]\left[ 1 -
(1-q)\frac{U}{T} \right]^{\frac{1}{1-q}+1-j} . \label{eq:qdistribuant}
\end{eqnarray}
For energies $U$ satisfying the condition given by eq.(\ref{eq:conditsumE}), the corresponding multiplicity distribution
is now given by the Negative Binomial distribution :
\begin{eqnarray}
P(N) &=& H_{N+1}(U) - H_N(U) = \label{eq:qP(N)}\\
&=& \frac{(q-1)^N}{N!}\cdot\frac{q-1}{2-q}\cdot \frac{\Gamma\left(N+1+\frac{2-q}{q-1}\right)}{\Gamma \left(
\frac{2-q}{q-1}\right)}\times \left(\frac {U}{T}\right)^N \left[1-(1-q)\frac{U}{T} \right]^{-N+\frac{1}{1-q}} =
\nonumber\\
&=& \frac{\Gamma(N+k)}{\Gamma(N+1)\Gamma(k)}\cdot \frac{\left( \frac{\langle N\rangle}{k}\right)^N }{\left( 1 + \frac{\langle N\rangle}{k}\right)^{N+k} }\quad {\rm with}\quad k = \frac{1}{q-1}. \label{eq:NBDfinal}
\end{eqnarray}
where the mean multiplicity and variance are, respectively,
\begin{equation}
\langle N\rangle = \frac{U}{T}\quad{\rm and}\quad Var(N) = \frac{U}{T}\left[ 1 - (1-q)\frac{U}{T}\right] = \langle N\rangle + \langle
N\rangle^2 \cdot (q-1) .\label{eq:meanNVarq}
\end{equation}
Note that for $q \rightarrow 1$ one has $k\rightarrow \infty$ and $P(N)$ becomes a Poisson distribution, whereas for
$q\rightarrow 2$ one has $k\rightarrow 1$ and we obtain a geometrical distribution. The parameter $k$ in the NB distribution can also be simply expressed by the correlation coefficient $\rho$ for the two-particle energy correlations, $ k = (\rho + 1)/\rho$, see \cite{Q} for details.

\section{Some specific examples of a quasi-power law in hadronic and nuclear collisions}
\label{sec:Power}

\subsection{Quasi-power law in transverse momentum distributions obseved at LHC experiments}
\label{sec:QCD}

High energy experiments at the Large Hadron Collider at CERN (CMS \cite{CMS-1,CMS-2,CMS-3}, ATLAS \cite{ATLAS-1,ATLAS-2,ATLAS-3} and ALICE \cite{ALICE-1,ALICE-2}) enlarged unprecendently the range of the measured transverse momenta in comparison with, for example, previous UA1 data \cite{UA1had} . It now reaches $p_T \leq 180$ GeV. As a result the measured cross section spans a range of $\sim 14$ orders of magnitude. It is amazing therefore that, as shown in \cite{WWCT} , all these data can be successfully fitted by a single Tsallis distribution (\ref{eq:H}) or (\ref{eq:T}),
\begin{equation}
\frac{dN}{2\pi dy p_Tdp_t}\big|_{y=0} = A \exp_q\left( - \frac{E_T}{T}\right),\qquad E_T = \sqrt{m^2 + p_T^2}, \label{eq:TsallisFit}
\end{equation}
\begin{figure}[h]
\centering
\includegraphics[scale=0.5]{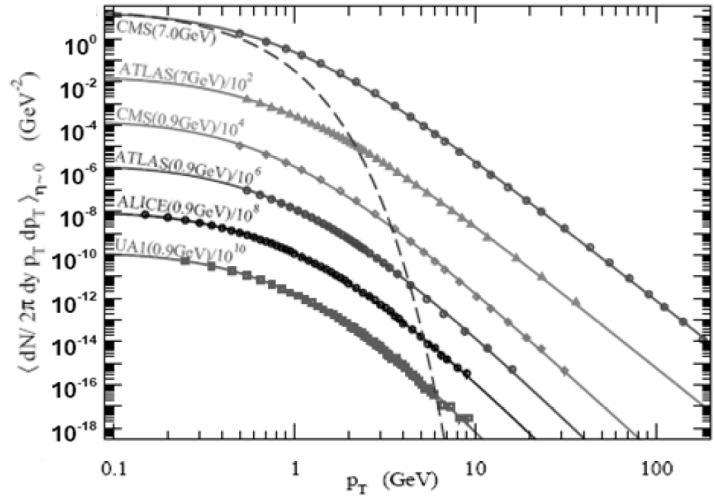}
\caption{Comparison of Eq.\ (\ref{eq:T}) with the experimental transverse momentum distribution of hadrons in~$pp$ collisions at central pseudorapidity $\eta$. The corresponding Boltzmann-Gibbs (purely exponential) distribution is illustrated by the dashed curve. For a better  visualization, both the data and the analytical curves have been divided by a constant factor as indicated. Data come from the UA1 \cite{UA1had}, CMS \cite{CMS-1,CMS-2,CMS-3}, ATLAS \cite{ATLAS-1} and ALICE \cite{ALICE-1,ALICE-2}Collaborations.  \label{TsFit}}
\end{figure}
\begin{table}[h]
\tbl{Parameters used to obtain the fits presented in  Fig. \ref{TsFit}.  The values of $A$ are in units of {GeV}$^{-2}$ and $T$ in GeV. }
{\begin{tabular}{@{}llcccc@{}} \toprule
Collaboration               & \hspace{5mm}$\sqrt{s}$  & $A$   & $q$   & $n = \frac{1}{q-1}$ & T     \\ \colrule
CMS\cite{CMS-1,CMS-2,CMS-3} & $pp$ at $7$ TeV         & 16.2  & 1.151 & 6.60                & 0.147 \\
ATLAS\cite{ATLAS-1}         & $pp$ at $7$ TeV         & 17.3  & 1.148 & 6.73                & 0.150 \\
CMS\cite{CMS-1,CMS-2,CMS-3} & $pp$ at $0.9$ TeV       & 15.8  & 1.130 & 7.65                & 0.128 \\
ATLAS\cite{ATLAS-1}         & $pp$ at $0.9$ TeV       & 13.6  & 1.124 & 8.09                & 0.140 \\
ALICE\cite{ALICE-1,ALICE-2} & $pp$ at $0.9$ TeV       & ~9.95 & 1.119 & 8.37                & 0.150  \\
UA1\cite{UA1had}            & $\bar{p}p$ at $0.9$ TeV & 13.1  & 1.109 &  9.21               & 0.154  \\ \botrule
\end{tabular} \label{TsFit-T}}
\end{table}
where $m$ is taken as the pion mass,  see Fig. \ref{TsFit}. The values of the parameters $A$, $q$ (and the corresponding $n$) and $T$ are given in Table \ref{TsFit-T}.  The exponential Boltzmann-Gibbs distribution (corresponding to $q=1$) is shown by the dashed curve for comparison.  These results show that Eq. (\ref{eq:TsallisFit}) adequately describes the hadron $p_T$ spectra at central rapidity in high-energy $pp$ collisions.  We verify that~$q$ increases slightly with the beam energy, but, for the present energies, remains $q\simeq 1.1$, corresponding to a power index $n$ in the range of $6-10$ that decreases as a function of $\sqrt{s}$. Note that the good agreement of the present phenomenological fit extends to the whole $p_T$ region (or at least for $p_T$ greater than $0.2\,\textrm{GeV}$, where reliable experimental data are available) \cite{WW12,WW13} .  This is achieved with a single nonextensive statistical mechanical distribution with only three degrees of freedom.

\subsection{Self-similarity in jet events}
\label{sec:SelfSim}

\begin{figure}[h]
\includegraphics[width=4.2cm]{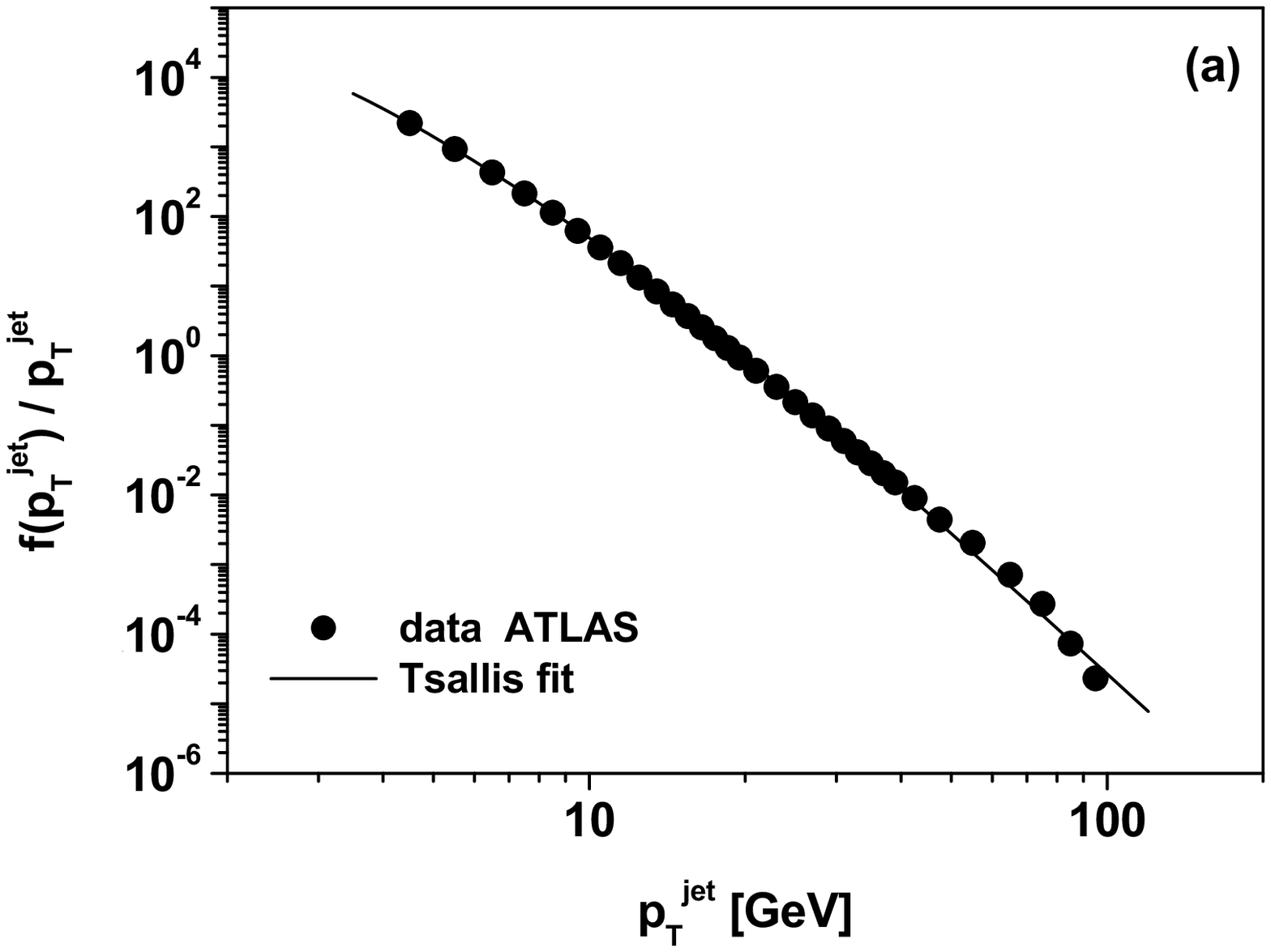}
\includegraphics[width=4.0cm]{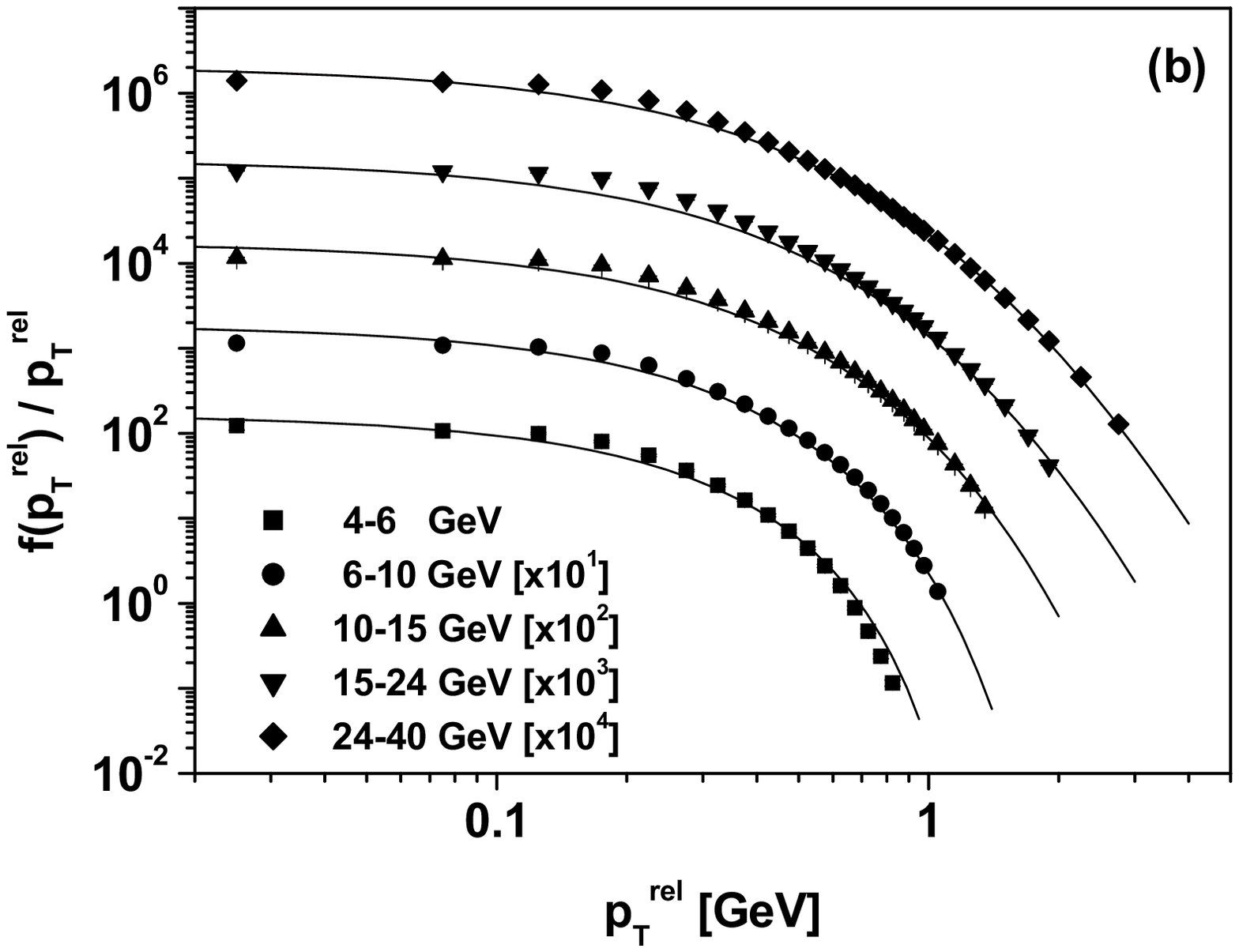}
\includegraphics[width=4.2cm]{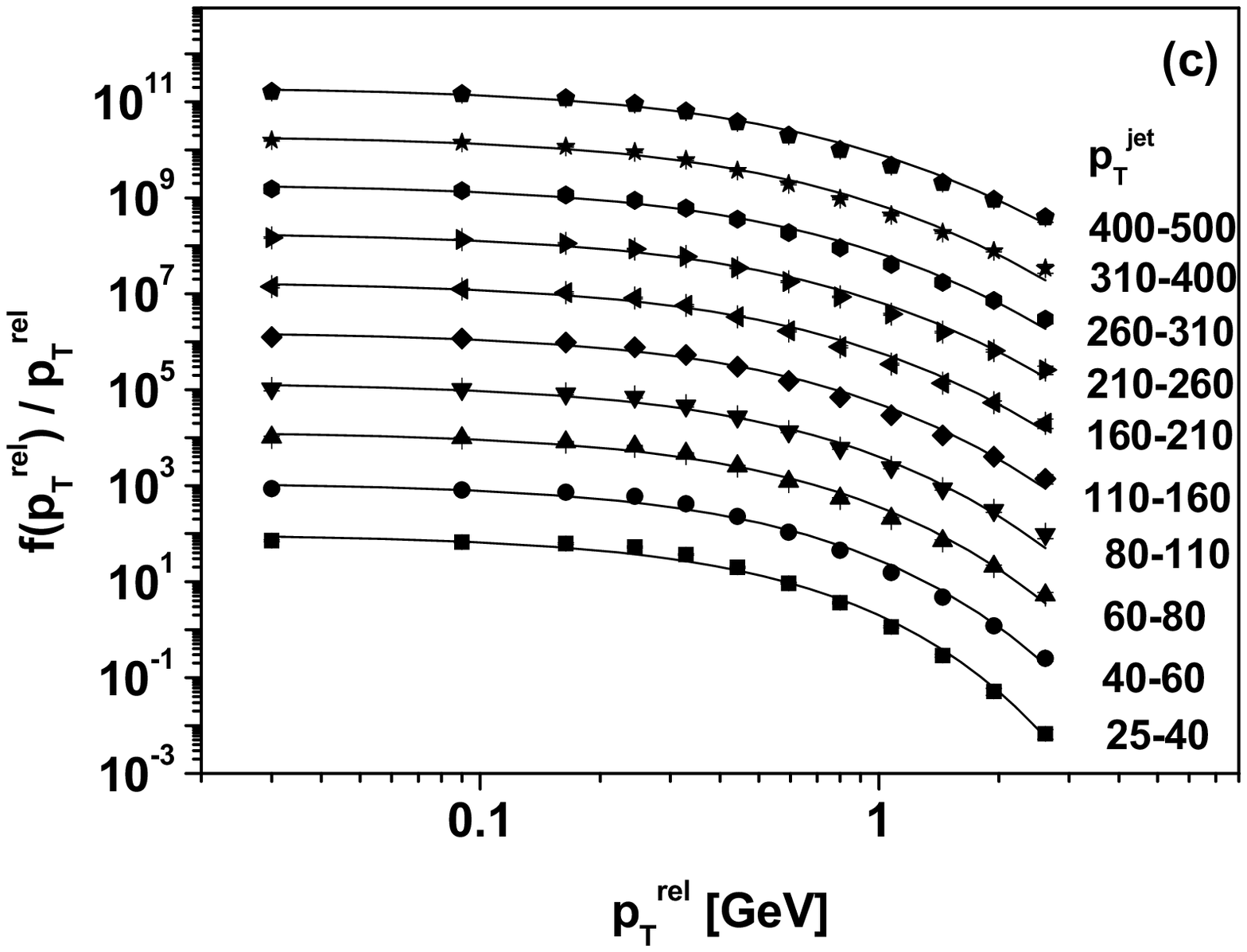}
\vspace{-3mm}
\caption{$(a)$ Distribution of $p_T^{jet}$ for jets at $\sqrt{s} = 7$ TeV fitted using Eq. (\ref{eq:H}) with $T = 0.45$ GeV and $n = 7$ (which corresponds to $q = 1.14$). $(b)$ Distributions of $p^{rel}_T$ particles inside the jets with different values of $p_T^{jet}$ fitted using Eq. (\ref{eq:H}). To make the results readable, the consecutive curves $i = 0,1,2,\dots$ in panels $(a)$ and $b)$ were multiplied by $10^i$. For all curves in $(a)$ and $(b)$ $T = 0.18$ GeV. The corresponding values of the parameter $n$ (and $q = 1+1/n$) are listed in Table \ref{Table1}. $(c)$ Distributions of $p^{rel}_T$ particles inside the jets with different values of $p_T^{jet}$ fitted using Eq. (\ref{eq:H}). The corresponding values of the parameter $n$ (and $q = 1+1/n$) are listed in Table \ref{Table2}. All data are taken from \cite{ATLAS-1,ATLAS-2,ATLAS-3} .}
\label{Fig-jets}
\end{figure}

The distribution of the longitudinal component of the momenta of particles within jets produced in $pp$ collisions \cite{Betall-2} turns out to be similar to what was found in $e^+e^-$ collisions \cite{Betall-1} . Recent ATLAS data \cite{ATLAS-1,ATLAS-2,ATLAS-3} offers the possibility to look for such similarities also in the transverse characteristics of jets and in the multiplicities of charged particles, $P(N)$, within them \cite{WWjets} . The corresponding nonextensive parameters can be deduced from the distributions of the transverse momenta of jets, $f\left(p_T^{jet}; q_{jet}\right) =\left(1/N_{jet}\right) dN_{jet}/dp^{jet}_T$, from distributions of transverse momenta of (only charged) particles within jets, $f\left(p_T^{rel}; q_{rel}\right) =\left(1/N\right) dN/dp^{rel}_T$ where $p_T^{rel} = \left| \vec{p}\times \vec{p}_{jet}\right|/\left| \vec{p}_{jet}\right|$, and from the multiplicity distributions of particles within observed jets, $P(N,q)$. The other two LHC experiments, ALICE and CMS, do not provide such results for the same experimental conditions and using the same criteria for data selection. Because the pure power law distribution, $f\left( p_T\right) \sim p_T^{-\gamma}$, is not experimentally observed for jets and the observed slope parameter $\gamma$ depends on $p_T$, $\gamma = \gamma\left( p_T\right)$, we account for this dependence by using the following two-parameter ($n$ and $T$) form,
\begin{equation}
\gamma\left( p_T \right) =
          n \frac{\ln\left( nT + p_T\right)}{\ln \left(
          p_T\right)}  +  \frac{ (n - 1)\ln(nT) + \ln(n-1)}{ \ln \left(
          p_T\right)}. \label{eq:qnT}
\end{equation}
In this case, the transverse momentum distribution for jets can be fitted by Eq. (\ref{eq:H}) with $n \simeq 7$ and $T = 0.45$ GeV, cf. Fig. \ref{Fig-jets}$a$.
\begin{table}[b]
\tbl{Fit parameters for Fig. \ref{Fig-jets}$b$; $T=0.18$ GeV.}
{\begin{tabular}{@{}l|ccccc@{}} \toprule

~$p_T^{jet}$[GeV]       & ~$4 - 6$~ &~ $6 - 10$~ & ~$10- 15$~  & ~$15 - 24$~  &~ $24 - 40$~ \\
~ $n$                   & ~$-8.5$  & $-17$       & $55$        & $16$         & $11.5$    \\
~ $q = 1 + \frac{1}{n}$ & ~$0.88$  & $094$       & $1.02$      & $1.06$       & $1.09$     \\
\botrule
\end{tabular} \label{Table1}}
\end{table}
\begin{table}[b]
\tbl{Fit parameters for Fig. \ref{Fig-jets} $c$; $T=0.25$ GeV.}
{\begin{tabular}{@{}l|ccccc@{}} \toprule

~$p_T^{jet}$[GeV]       & ~$25 - 40$~ &~ $40 - 60$~ & ~$60- 80$~  & ~$80 - 110$~  &~ $110 - 160$~ \\
~ $n$                   & ~$70$       & $25$        & $18$        & $15$          & $12$          \\
~ $q = 1 + \frac{1}{n}$ & ~$1.014$  & $1.04$       & $1.056$      & $1.067$       & $1.083$     \\\colrule
~$p_T^{jet}$[GeV]       & ~$160 - 210$~ &~ $210 - 260$~ & ~$260- 310$~  & ~$310 - 2400$~  &~ $400 - 500$~ \\
~ $n$                   & ~$10$  & $9$       & $9$        & $9$         & $7.5$    \\
~ $q = 1 + \frac{1}{n}$ & ~$1.100$  & $1.111$       & $1.111$      & $1.111$       & $1.133$     \\
\botrule
\end{tabular} \label{Table2}}
\end{table}

Data on distributions of transverse momenta of particles produced within the jet, $p_T^{rel}$, are presented in \cite{ATLAS-1,ATLAS-2} for  $p^{jet}_T \leq 40$ GeV and in  \cite{ATLAS-3} for $p_T^{jet} > 40$ GeV. They can all be fitted by Eq. (\ref{eq:H}) and the results are shown in Fig. \ref{Fig-jets}$b$ and Table \ref{Table1}, for the first set, and in Fig. \ref{Fig-jets}$c$ and Table \ref{Table2}, for the second one. Referring to \cite{WWjets} for further details we end this part by noting the negative values of the parameter $n$ (or, correspondingly, $q < 1$ values of the nonextensivity parameter) for small values of $p_T^{jet}$, i.e., for small values of the energy of such jets seen in Fig. \ref{Fig-jets}$b$. This fact is connected with the limitation of the available phase space in this case (cf. \cite{MW-limited},).

\begin{figure}[h]
\includegraphics[width=6.2cm]{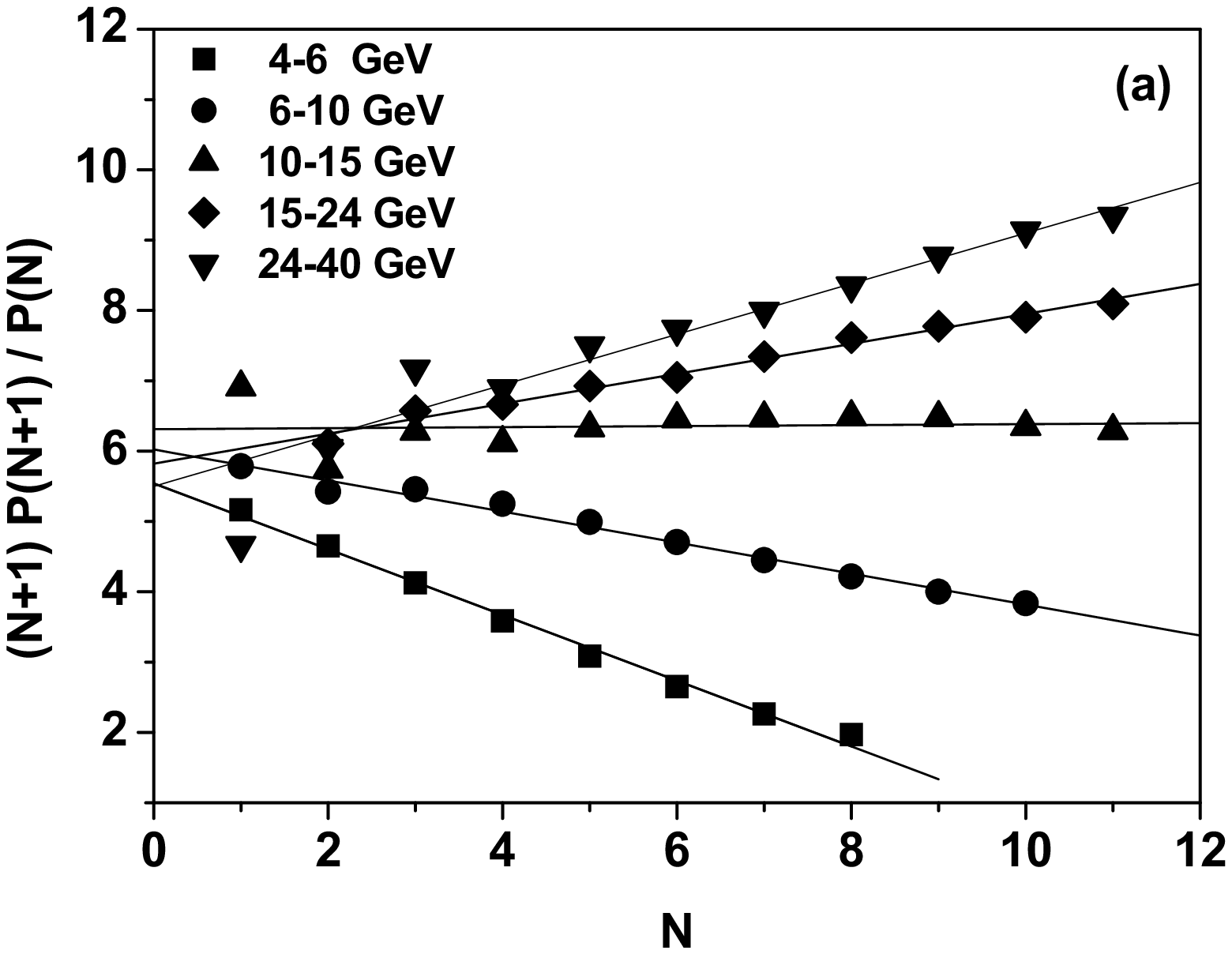}
\includegraphics[width=6.3cm]{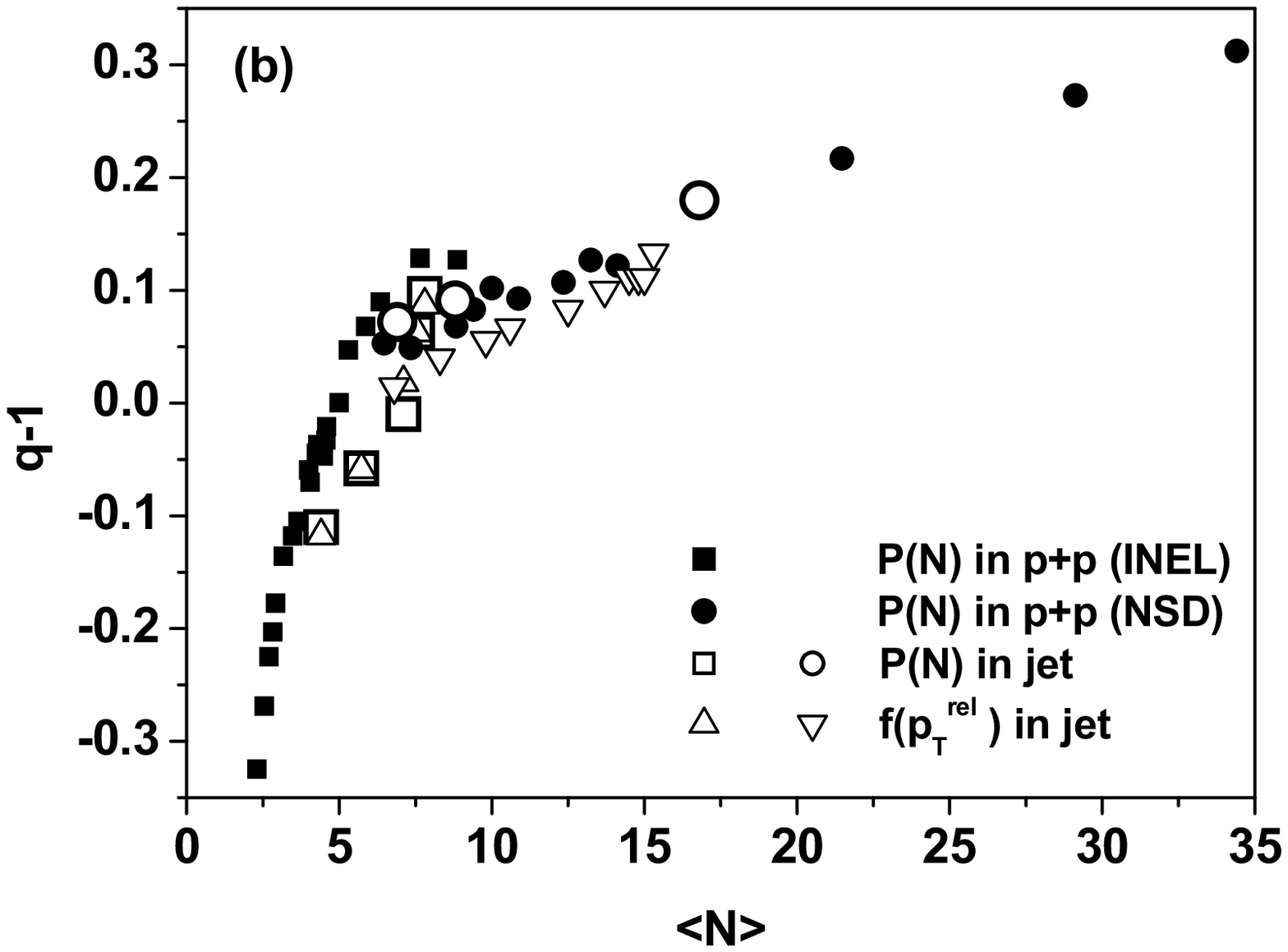}
\vspace{-3mm}
\caption{ $(a)$  $\frac{(N+1)P(N+1)}{P(N)}$ as a function of multiplicity $N$ in jets with different values of $p_T^{jet}$ as measured in \cite{ATLAS-1,ATLAS-2} and presented in Fig. \ref{Fig-jets}$b$. $(b)$ Compilation of values of $q$ as obtained from $p_T^{rel}$ distributions (triangles) and from multiplicity distributions (circles). Triangles at small $\langle N\rangle$ are obtained from data \cite{ATLAS-1,ATLAS-2}, those for larger $\langle N\rangle$ from \cite{ATLAS-3}. Full squares and circles are from data on multiparticle production in $p+p$ collisions and, correspondingly, squares (inelastic data) are from the compilation for LAB energy $3.7-303$ GeV presented in \cite{AKW} , whereas circles (non-single diffractive data) are from the compilation presented in \cite{G-G} . } \label{Fig-N}
\end{figure}

We shall proceed now to multiplicity distributions, $P(N)$, measured in jets \cite{ATLAS-1,ATLAS-2} . Although they are available only for $p_T^{jet} \leq 40$ GeV, they allow us to check which form of $P(N)$ discussed in Section \ref{sec:TtoP} is favored and what are the values of the corresponding $q=q_N$. They can all be described by the recurrence relation:
\begin{equation}
\frac{(N+1) P(N+1)}{P(N)} = a + b N \label{eq:dep}
\end{equation}
where for NBD $a=\frac{\langle N\rangle k}{k + \langle N\rangle}$ and $b = \frac{a}{k}$, for PD $a=\langle N\rangle$ and $b = 0$, for BD $a = \frac{\langle N\rangle \kappa}{\kappa - \langle N\rangle}$ with $b = \frac{a}{\kappa}$. The linear form of (\ref{eq:dep}) observed in Fig. \ref{Fig-N}$a$ tells us that the corresponding $P(N)$ are of the NBD or BD type (the deviation from linearity occurs only for $N=1$, for which one encounters experimental difficulties and we omitted it from our analysis). From the parameters $a$ and $b$ obtained in this way we can deduce values of $\langle N\rangle$, $Var(N)$ and $k$ or $\kappa$ (i.e., values of the corresponding nonextensivity parameter $q_N$) which are presented in Table \ref{Table3}. Note that their values correspond closely to those obtained from the distributions of $p_T$ in jets presented in Table \ref{Table1}.

We may now go further and compare the values of the nonextensivity parameters obtained from analysis of $P(N)$, $f\left(p_T^{jet}\right)$ and $f\left( p_T^{rel}\right)$ with the respective nonextensivity parameters obtained in measurements of $p_T$ distributions in other experiments on minimum bias $pp$ collisions in which the range of $p_T$ and multiplicities were similar and the energies of which were similar to the energies of the jets investigated. One must only remember that so far we have been estimating the parameter $q$ from distributions of $p_T$ or $N$ and discussing its energy dependence, $q(s)$, as obtained from different experiments \cite{WWrev-2,WWcov} .  Here we would like to compare distributions of particles in $p+p$ collisions to those in jets, for which, unfortunately, we do not know the corresponding energy $\sqrt{s}$. Nevertheless, we know $\langle N\rangle$ both for $p+p$ collisions and for particles produced in jets, so it is reasonable instead to show $q$ as a function of $\langle N\rangle$. This was done in Fig. \ref{Fig-N}$b$. The approximate similarity of these results is clearly visible.
\begin{table}[t]
\tbl{$P(N)$ characteristics for jets with different $p_T^{jet}$.}
{\begin{tabular}{@{}l|ccccc@{}} \toprule
~$p_T^{jet}$[GeV]       & ~$4 - 6$~  &~ $6 - 10$~  & ~$610 - 15$~  & ~$15 - 24$~     &~ $24 - 40$~ \\
~ $\langle N\rangle$    & ~$4.41$    &   $5.72$    & $7.11$        & $7.56$          & $7.80$          \\
~ $Var(N)$              & ~$2.31$    &  $3.83$     & $6.61$        & $11.2$          & $18.1$       \\
~$q_N - 1$              & ~$-0.11$~  &~ $-0.058$~  & ~$-0.0098$~   & ~$0.063$~       &~ $0.097$~ \\
\botrule
\end{tabular} \label{Table3}}
\end{table}
All this means that: $(i)$ A Tsallis distribution successfully describes inclusive $p_T$ distributions over a wide range of transverse momenta for all energies measured so far \cite{WWrev-2,WWrev-1,RWW-6,WW12} . $(ii)$ This is also true for the distribution of transverse momenta of jets as shown in Fig. \ref{Fig-jets} with $q = 1.14$, comparable to $q = 1.15$ describing transverse momenta distributions of particles at the same energy, $7$ TeV \cite{WW12} . $(iii)$ The Tsallis distribution also describes the transverse momenta distributions of particles in jets with $q$ roughly the same as those obtained from analyses of multiplicity distributions in these jets. This means that one observes a kind of similarity (in what concerns the values of the nonextensivity parameter $q$) of $P(N)$ and $f\left( p_T\right)$ of particles produced in minimum bias $pp$ collisions and particles in jets of comparable energies. This indicates that the mechanisms of particle production in both cases are either the same or are similar and contain some {\it common  part} \cite{Satz} which can be identified with the self-similarity character of the production process in both cases, resulting in a kind of cascade process, which always results in a Tsallis distribution \cite{Kas-1,Kas-2,BJ1} . Actually, the idea of a {\it self-similarity} was introduced a long time ago by Hagedorn who assumed that hadrons are produced through the formation of {\it fireballs which are a statistical equilibrium of an undetermined number of all kinds of fireballs, each of which in turn is considered to be a fireball} \cite{SsH-1,SsH-2} . In fact, one encounters the same idea in the pure dynamical QCD approach to hadronization  where partons fragment into final state hadrons through multiple sub-jet production. As a result one has a {\it self-similar behavior of the cascade of jets to sub-jets to sub-sub-jets \dots to final state hadrons} (see, for example, \cite{SsQCD-1,SsQCD-2,SsQCD-3} ; the above results were also an inspiration for the recent idea of the thermofractal nature of interactions \cite{TF-1,TF-2,TF-3} ).

\subsection{Effective temperature $T_{eff}$}
\label{sec:effT}

\begin{figure}
\centerline{
\includegraphics[width=6.2cm]{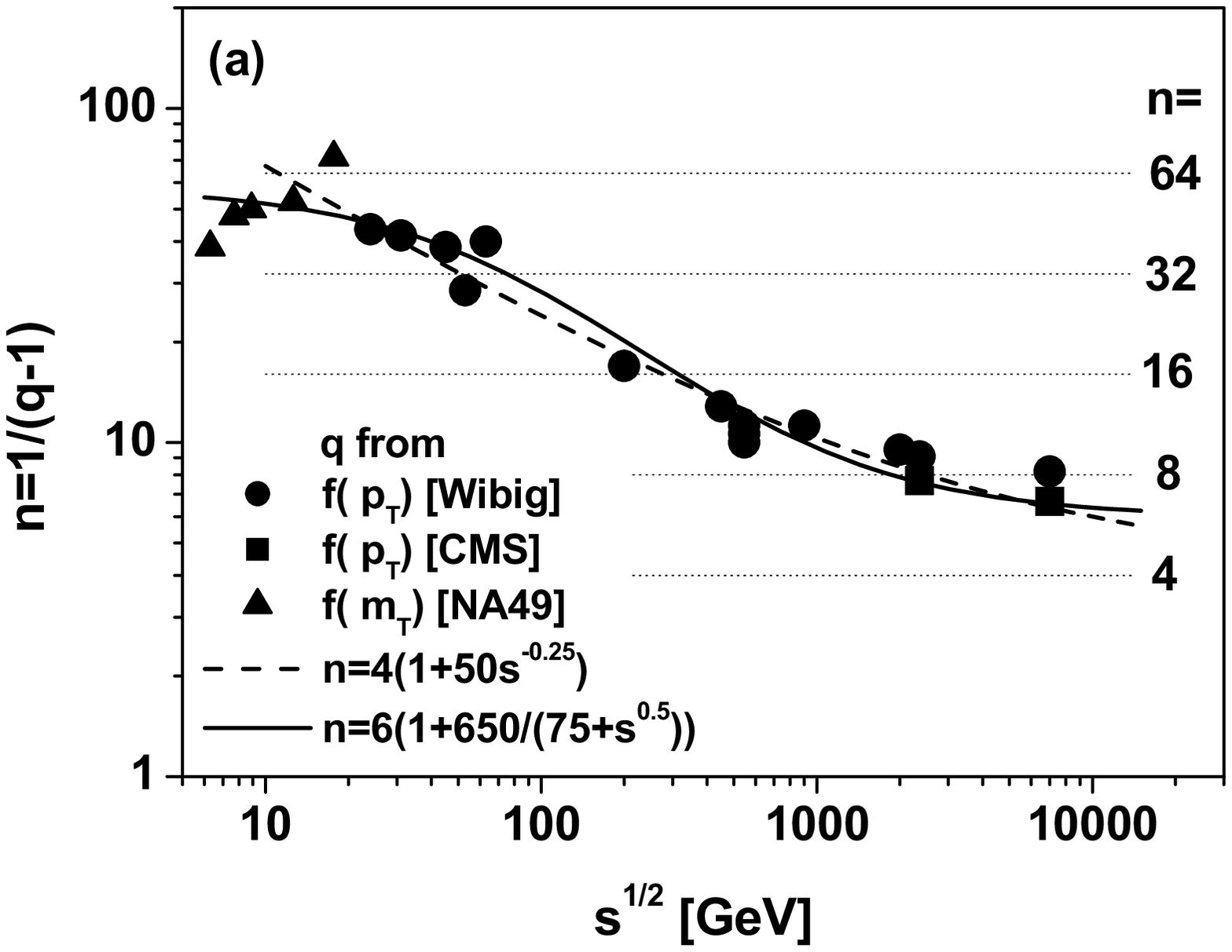}
\includegraphics[width=6.5cm]{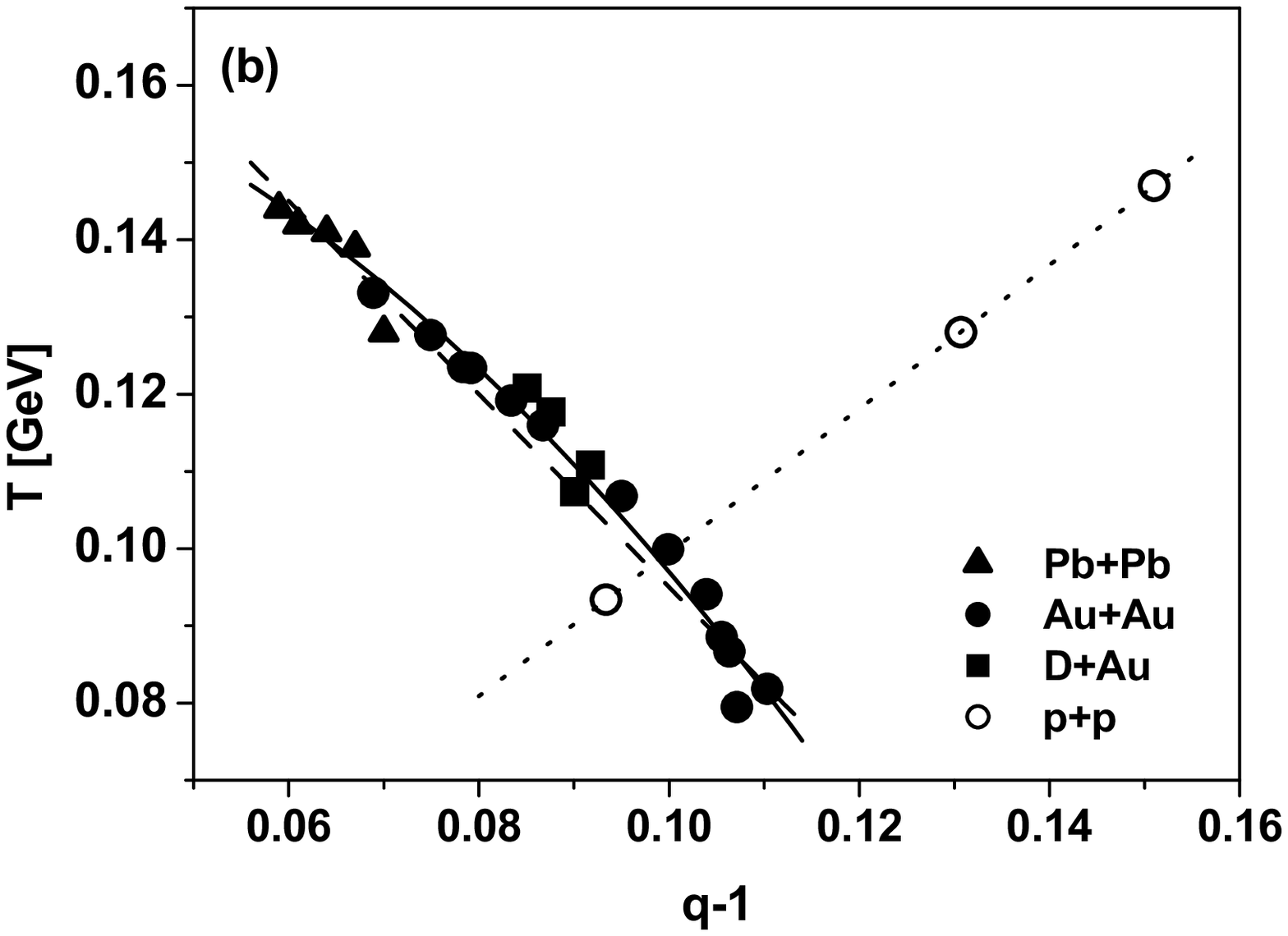}}
\vspace{-3mm}
\caption{ $(a)$ Two possible fits for the energy ($\sqrt{s}$) dependence of the   exponent $n=1/(q-1)$ deduced from NA49 \cite{NA49-1,NA49-2} and CMS \cite{CMS-1,CMS-2,CMS-3} experiments and from the compilation \cite{Wibig-1} . $(b)$ $T$ as a function of $(q-1)$  deduced from $Pb+Pb$ collisions at energies of  $6.3,~7.6,~8.8,~12.3,~17.3$ GeV \cite{NA49-1,NA49-1} , from $Au+Au$ and $D+Au$ collisions at an energy of $200$ GeV \cite{RHIC} , from $p+p$ collisions at energies of $200$ GeV \cite{RHIC} , $900$ and $7000$ GeV \cite{CMS-1,CMS-2,CMS-3,ATLAS-1} .  The nuclear results can be fitted either by a linear form,   $T(q) = 0.22-1.25(q-1)$ (dashed line), or by a quadratic, $T(q) = 0.17-7.3(q-1)^2$ (full line).   The best fit to $p+p$ results is linear: $T(q) = 0.0065 + 0.93(q-1)$.   \label{EffT}}
\end{figure}

We show now some experimental results substantiating the idea of effective temperature $T_{eff}$ from Eqs. (\ref{eq:Teff}) and (\ref{eq:Tq}). The examples of energy dependence of $n=1/(q-1)$ deduced from the available experimental data and $T$ as a function of $q-1$ for different energies are presented in Fig. \ref{EffT}. Notice the opposite behavior of $T(q)$ for $A+A$ and $p+p$ collisions. Experimental data show that in the presented range of $q$ for $p+p$ collisions $\langle \xi\rangle > 2|Cov(\xi,\gamma)|$ and for $A+A$ collisions $\langle \xi\rangle << 2|Cov(\xi,\gamma)|$ (and therefore it can become negative). Actually, already in \cite{WWTout} where we analyzed $T(q) = T_{eff}$ for the first time, the point for $p+p$ for $200$ GeV did not follow the $A+A$ results. In Fig. \ref{EffT}$a$ the new points from CMS $p+p$ data at higher energies were added. It turned out that $T(q)$ for $p+p$ collisions is just opposite to that for $A+A$ data. The only explanation we can offer at this time is to recall that behind $T_{eff} = T(q)$ used in \cite{WWTout} was the idea of energy transfer between the interaction region and its surroundings. For the $A+A$ collisions this means energy transfer to the spectator nucleons (not participating in the collision process). Therefore spectators, which have small transverse energy, effectively "cool" the interacting system. For $p+p$ collisions the region of interaction is immersed in the quark-gluon environment, which has transverse energy comparable with that of the colliding system and because of this it additionally "heats" it. As a result, the corresponding parameter describing this energy transfer is negative for $A+A$ collisions and positive for $p+p$ collisions. However, before reaching any final conclusions one has to stress the difficulties in the evaluation of the parameters $T$ and $q$ from the experimental data which must be clarified first \cite{WWAIP-2013} . We close by mentioning that the subject of $T_{eff}$ has also been recently addressed in \cite{BBB,SBW} with similar conclusions.

\subsection{Generalized thermodynamic uncertainty relations}
\label{sec:Generalized}

In the case of an ensemble in which the energy ($E$), temperature ($T$) and multiplicity ($N$) can all fluctuate nonextensive statistics allows us to connect all the fluctuating variables by means of a generalization of the thermodynamic uncertainty relations introduced in \cite{JL} . Referring for details to \cite{WWcov} we provide here a short outline of this problem. The idea that the thermodynamical quantities, temperature $T$ and energy $U$, could be regarded as being complementary (in the same way as are position and momentum in quantum mechanics) occurred already in  \cite{BH} . It is based on the observation that in order to attribute a definite temperature to a physical system one has to bring it into thermal contact and equilibrium with some heat bath. But in this case our system will freely exchange energy with the heat bath and one cannot control its energy. Conversely, the system has a definite energy only if it is isolated from its environment in which case one cannot determine its temperature. From dimensional analysis $\Delta U\, \Delta \beta \ge k $ ($\beta = 1/T$ and $k$ is Boltzmann's constant). Isolation ($U$ definite) and contact with a heat bath ($T$ definite) then represent two extreme cases of this complementarity. Because of the disputable meaning of the increment $\Delta\beta$ this idea  has so far not received much recognition (see \cite{UL,L}) . We propose therefore to use a nonextensive Tsallis statistics \cite{Tsallis-1,Tsallis-2} and identify  these increments, $\Delta x$, with  measures of fluctuation of the corresponding physical quantities in an ensemble, expressed by the variance $Var(x)$, in which the energy ($U$), temperature ($T$) and multiplicity ($N$) ($x = U,~T,~N$, respectively), can all fluctuate. This allows us to generalize the relation between fluctuations of $U$ and $T$ derived in thermodynamics \cite{JL} expressed by their relative variances, $Var(x)/\langle x\rangle^2 = \omega^2_x$ as:
\begin{equation}
\omega_U^2\, +\, \omega^2_T\, =\, \frac{1}{\langle N\rangle}.
\label{eq:JL}
\end{equation}
Eq. (\ref{eq:JL}) describes a small system remaining in thermal contact with a heat bath of varying size and represents the kind of uncertainty relation mentioned before (that the standard deviation of one variable can be made small only at the expense of increasing the corresponding standard deviation of the conjugate variable \cite{UL,L}). This relation is supposed to be valid all the way from the canonical ensemble, for which $Var(T) = 0$ and $Var(U) = 1/\langle N\rangle$, up to the
microcanonical ensemble for which $Var(T) = 1/\langle N\rangle$ and $Var(U) = 0$. It expresses the complementarity between temperature and energy and the canonical and microcanonical description of the system.

Following Sections \ref{sec:Fixed}, \ref{sec:SuperS} and \ref{sec:Fluct} note that when two variables (out of three: $T$, $N$ and $U$) are fixed, the third fluctuates according to a gamma distribution \cite{WWcov} :
\begin{equation}
U g_{T,N}(U)\, =\, N g_{T,U}(N)\, =\, \beta g_{U,N}(\beta) = \frac{(\beta U)^N}{\Gamma(N)}\, \exp( - \beta U), \label{eq:all} \end{equation}
and all respective relative fluctuations are identical:
\begin{equation}
\frac{Var(U)}{\langle U\rangle^2}\, =\, \frac{Var(\beta )}{\langle
\beta\rangle^2}\, =\, \frac{Var(N)}{\langle N\rangle^2}\, =\,
\frac{1}{\langle N\rangle} .\label{eq:allfluct}
\end{equation}
Assuming that $Var(\beta)/\langle \beta \rangle^2 \simeq Var(T)/\langle T\rangle^2$)  we get from Eq. (\ref{eq:q}) that
\begin{equation}
\frac{Var(T)}{\langle T\rangle^2} = \omega^2_T = q - 1.
\label{eq:defq}
\end{equation}
From the connection between NBD and $q > 1$ (cf. Section (\ref{sec:NB}) we also know that
\begin{equation}
\frac{1}{k} = \frac{Var(N)}{\langle N\rangle^2} - \frac{1}{\langle
N\rangle} = \omega_N^2 - \frac{1}{\langle N\rangle}. \label{eq:qN}
\end{equation}
Therefore, fluctuations of $N$ and $T$ are not independent, but related to each other:
\begin{equation}
\omega_N^2 - \frac{1}{\langle N\rangle} = \omega_T^2.
\label{eq:relNT}
\end{equation}
The NB multiplicity distribution can also be obtained by fluctuating (using a gamma function) the mean multiplicity, $\bar{N} = \langle N\rangle$, in the Poisson distribution (cf., Eq. (\ref{eq:NBDq})). This means that, in addition to Eq.
(\ref{eq:qN}) one also has that
\begin{equation}
\frac{1}{k} = \frac{Var(\bar{N})}{\langle \bar{N}\rangle^2}.
\label{eq:anotherNB}
\end{equation}
For $\bar{N} = const$ ($ k = \infty$) we have a Poisson distribution. Note that fluctuating $1/T$ according to a gamma distribution and keeping $U = const$ results in $Var(\bar{N})/\langle \bar{N}\rangle^2 = \omega^2_T$ and we recover Eq.
(\ref{eq:relNT}). Analogously, fluctuating $U$ while keeping $T=const$ gives us $Var(\bar{N})/\langle \bar{N}\rangle^2 =
\omega^2_U$. Fluctuating both $U$ and $T$ (and taking into account  that $\bar{N} = U/T$) one has that
\begin{equation}
\frac{Var(\bar{N})}{\langle \bar{N}\rangle^2} = Var\left(\frac{U}{T}\right)\cdot \left(\frac{\langle
T\rangle}{\langle U\rangle}\right)^2 \cong \left[ \frac{Var(U)}{\langle U \rangle^2} + \frac{Var(T)}{\langle
T\rangle^2} - 2\frac{Cov(U,T)}{\langle U\rangle\langle T\rangle} \right]\label{eq:covN}
\end{equation}
or, in terms of the scaled variances introduced before,
\begin{equation}
\omega^2_{\bar{N}} \cong \omega^2_U\, +\, \omega^2_T\, -\, 2\rho
\omega_U\omega_T, \label{eq:correl}
\end{equation}
where $\rho = \rho (U,T) \in [-1,1]$ is the correlation coefficient (this generalizes a century old relation: $\omega^2_P = \omega^2_V + \omega^2_T$  \cite{Q1} ). From  Eqs. (\ref{eq:qN}), (\ref{eq:anotherNB}) and (\ref{eq:correl}) one obtains the following general relation between all fluctuating variables:
\begin{equation}
\Big| \omega^2_N - \frac{1}{\langle N\rangle}\Big| = \omega^2_U + \omega^2_T - 2\rho \omega_U \omega_T =
\left(
\omega_U - \omega_T\right)^2 + 2 \omega_U \omega_T (1 - \rho). \label{eq:corq}
\end{equation}
This generalizes Linhard's thermodynamic uncertainty relation given by Eq. (\ref{eq:JL}) and allows us to express the
nonextensivity parameter $q$ in terms of the respective fluctuations and correlations (with $\quad \xi = \omega_U/\omega_T$:)
\begin{equation}
| q - 1 | = \left( \omega_U - \omega_T\right)^2 + 2 \omega_U \omega_T ( 1 - \rho ) = \omega^2_T \left[(1 - \xi)^2 + 2 \xi (1 - \rho)\right]. \label{eq:qNg}
\end{equation}
A few comments are needed here. We use $|\dots|$ to makes (\ref{eq:corq}) general, i.e., valid for both $\omega^2_N \ge 1/\langle N\rangle$ and for $\omega^2_N =0$ if $N=const$. Note that when all variables fluctuate the fluctuations of $N$ must be greater than Poissonian because sub-Poissonian fluctuations , corresponding to $q < 1$, signal the presence of some additional constraints (like conservation of some quantum numbers, cf., \cite{Gor}). We restrict ourselves to the case $q \ge 1$ and do not describe the region $0 < \omega^2_N < 1/\langle N\rangle$. So far there are no data which would necessitate the use of nonzero correlations, therefore, combining Eqs. (\ref{eq:qNg}) and (\ref{eq:qN}),  we use such a relation:
\begin{equation}
\frac{1}{k} = q - 1 = \omega^2_U + \omega^2_T. \label{eq:OMUOMT}
\end{equation}
Note that Eq. (\ref{eq:corq}) connects fluctuations of different observables, but defined in the same fragment of allowed phase space, whereas the available data usually refer to different parts of this phase space, therefore the corresponding $q$ parameters are difficult to compare. For example, $q = q_L$ obtained from rapidity ($y$) distributions, $dN/dy$, defined in the longitudinal phase space, are comparable with $q$ evaluated from the multiplicity distributions, $P(N)$, which are defined in the full phase space \cite{RWW-2,NA49-1,NA49-2,NA49-3,WWW} . On the other hand, transverse momentum ($p_T$) distributions, $dN/dp_T$, defined in the so-called transverse space, are described by much smaller values of $q = q_T$.
\begin{figure}[h]
  \begin{center}
   \includegraphics[width=8cm]{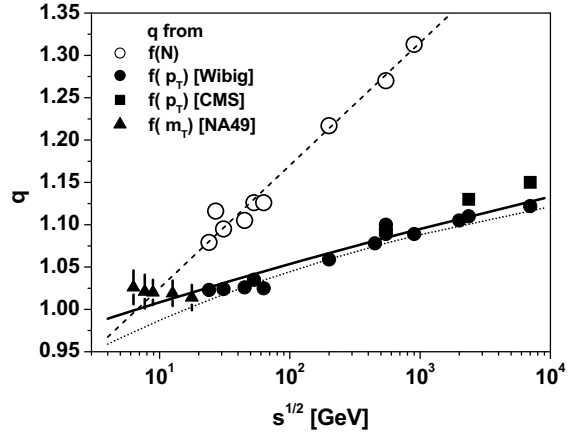}
   \caption{An example of the energy dependencies of the nonextensivity parameters $q$ obtained from
             different observables. See text for details.}
   \label{Figure2}
  \end{center}
\end{figure}

Referring to \cite{WWTout,WWcov} for details we present, as an example, in Fig. \ref{Figure2} the energy dependencies of the nonextensivity parameter $q$ obtained from different sources: from the multiplicity distributions, $f(N) = P(N)$, i.e., from the full phase space \cite{PN,G-G} and from a different analysis of transverse momenta distributions, $f\left( p_T\right)$ (i.e., from the transverse phase space \cite{Wibig-1,CMS-1,CMS-2,NA49-1,NA49-2,NA49-3}). The characteristic feature seen there is that whereas the former show substantial energy dependence (and essentially follow the results for $q = q_L$ obtained in \cite{RWW-2} from the analysis of $dN/dy$), the latter $q = q_T$ are only weakly dependent on the interaction energy (note also that $q_T(s)$ from different sources plotted in Fig. \ref{Figure2} are roughly the same). An attempt to compare $q$ and $q_T$ was presented in \cite{WWcov} . It turns out that, approximately (assuming that the $q_T$ are roughly independent of the energy fluctuations, that fluctuations of temperature contribute equally to each of the components of momenta, that fluctuations of energy $U$ are entirely given by its thermal part, and that one approximately accounts for some additional kinematical factors) one can write that $\omega^2_U \simeq \omega^2_T$ and both parameters, $q$ and $q_T$, are connected by the relation:
\begin{equation}
q_T = \frac{1 + 2q}{2 + q}. \label{eq:qqTn}
\end{equation}
Using $q = 1 + 1/k = 0.896 + 0.029\ln s$ obtained in \cite{PN} from $P(N)$ the respective $q_T$ is shown in Fig.
\ref{Figure2} (solid line) and compared to $q_T$ extracted from transverse momenta distributions $f(p_T)$  \cite{Wibig-1,CMS-1,CMS-2,NA49-1,NA49-2,NA49-3}. This can be compared with the dotted line representing $q_T = 1.25 - 0.33s^{-0.054}$  obtained in \cite{Wibig-1} . The noticeably good agreement of Eq. (\ref{eq:qqTn}) with data means that it really connects the fluctuations in different parts of phase space (modulo additional assumptions).

\subsection{The QCD origin of the quasi-power law at LHC experiments}
\label{sec:QCD-1}

We return now to the QCD origin of the quasi-power fits to large $p_T$ distributions observed at LHC experiments presented in Section \ref{sec:QCD}. The point to be stressed is that the simple Tsallis-like parametrization of data (spanning $\sim 14$ orders of magnitude), Eq. (\ref{eq:TsallisFit}) and Fig. \ref{TsFit} (from  \cite{WWCT}), are a first order approximation obtained from a very detailed approach based on QCD (initiated in \cite{WW12} and continued to its final form in \cite{WW13,WWCT}). It starts from the pure QCD partonic picture of elementary collisions, with {\it hard} scatterings between the quarks and gluons of incoming protons proceeding with high momentum transfer (all masses are neglected) and results in the power like distributions of partonic jets with index $n \simeq 4 - 4.5$. Jets are subsequently fragmented using a QCD based parton showering mechanism. It turns out that the results for $p_T$ distributions obtained in this way can be parameterized in a relatively simple way by modifying only the power index $n$ (which can easily reach values $n \simeq 7-8$ as observed in the experiment; note that at small energies one can observe much larger values of $n$ but also $n <0$, as shown in Figs. \ref{Fig-N} and \ref{EffT}). This is because the distributions of quarks and gluons and the QCD coupling constant, all of which depend on $p_T$, can be cast in simple power like forms, adding therefore to the modifications of $n$ and building some  prefactor (both mildly $p_T$-dependent), the final form of which will be our parameter $A$ (taken, in the first approximation, to be $p_T$-independent). However, in this way one reproduces properly only the power index $n$ (or $q$) and the resulting distribution is of {\it pure power-like} type, $\sim 1/p_T^n$, diverging for $p_T \rightarrow 0$ instead of being exponential there. The remedy (proposed in \cite{WW13}) is based on the realization that in the QCD approach large $p_T$ partons probe small distances (with small cross sections). With the diminishing of $p_T$, these distances become larger (and the cross sections are increasing) and, eventually, when $p_T$ approaches some value $p_{T0}$, they start to be of the order of the nucleon size. At this moment the cross section should stop rising, i.e., it should not depend anymore on the further decrease of the transverse momentum $p_T$. This can be modelled by introducing a scale parameter $p_{T0}$ and changing
\begin{equation}
\left( \frac{1}{p_T}\right)^{n} \rightarrow \left[ p_{T0}\cdot \left( 1 + \frac{p_T}{p_{T0}} \right)\right]^{-n}. \label{eq:change}
 \end{equation}
The scale parameter $p_{T0}$ can then be identified with $p_{T0}$ introduced in Section \ref{sec:I} separating "hard" and "soft" parts of phase space. A similar procedure was also used to regularize the QCD coupling, $\alpha \left( p_T \right)$ (with a scale term which is not necessarly the same as $p_{T0}$). Note that this corresponds, in a sense, to accounting for some (effective) masses which were so far neglected in the corresponding formulas for cross sections (as not important, but this is not true for $p_T \rightarrow 0$).

The hadron production can thus be viewed from two different and complementary perspectives. $(i)$  Microscopic description (based on perturbative QCD, parton-parton hard scattering, their structure functions, fragmentation and showering, the running coupling constant and other QCD processe). $(ii)$  Description based on statistical mechanics represented here by the single-particle distribution (\ref{eq:TsallisFit}). It exhibits all the essential features of the process with only three degrees of freedom, which, in the lowest-order approximation, are: a power index $n$ (or nonextensivity parameter $q$=$(n+1)/n$), the average transverse momentum $m_{T0}$ (or an effective temperature $T$=$m_{T0}/n$, (which turns out to be close to the mass of the pion), and a constant $A$ that is related to the multiplicity per unit rapidity when integrated over $p_T$. The fact that one can adequately describe  the experimental data indicates that the scenario proposed in \cite{CM-2,H} appears to be essentially correct.  However,  to deduce the underlying nonextensive parameters from the basic physical quantities of the collision process further, more rigorous, investigations are needed.

\subsection{Effects of the limitation of phase space}
\label{sec:Limitation}

Consider now the situation when the scale parameter $T$ in Eq. (\ref{eq:constraints}) can fluctuate. To this end, let us consider the joint probability distribution
\begin{equation}
g\left( \{E_{1,\dots,\nu}\}\right) = \prod^{\nu}_{i=1}g_1\left( E_i\right)
\label{eq:joint}
\end{equation}
and let the parameter $T$ in each $g_i\left( E_i\right)$ fluctuate according to a gamma distribution from Eq.~(\ref{eq:BolGam}) \cite{MW-limited} . In this case we have a single-particle Tsallis distribution
\begin{equation}
h_i\left(E_i\right) = \frac{n-1}{n T} \left( 1 + \frac{E_i}{n T}\right)^{-n},
\label{eq:hTsallis}
\end{equation}
whereas the distribution of  $U=\sum^{N}_{i=1} E_i$  is given by \cite{WWcov} :
\begin{equation}
h_{N}(U) = \frac{\Gamma\left( N + n-1\right)} {U \Gamma(N) \Gamma\left( n-1\right)}
\left( \frac{U}{T}\right)^{N}\left( 1 + \frac{U}{n T} \right)^{1 - N -n}.  \label{eq:hNE}
\end{equation}
If  the energy $U$ is limited we have the following conditional probability:
\begin{eqnarray}
h\left( E_i|U\right) &=& \frac{ h_1\left( E_i\right)h_{N -1}\left(U - E_i\right)}{h_{N}(U)} = \frac{(N - 1)(n-1)}{(n-2 + N)} \times\nonumber\\
\times && \frac{(nT+U)}{nTU}\left( \frac{U - E_i}{U}\right)^{N -1} \left( 1 +\frac{E_i}{n T}\right)^{-n}\left( 1 -\frac{E_i}{n T+U}\right)^{2 - N -n}.
\label{eq:fEiE}
\end{eqnarray}
For $n \rightarrow \infty$ Eq. (\ref{eq:fEiE}) reduces to Eq. (\ref{eq:constraints}) whereas for large energy ($U\rightarrow \infty$) and large number of degrees of freedom ($N \rightarrow\infty$), the conditional probability distribution (\ref{eq:fEiE}) reduces to the single particle distribution Eq.~(\ref{eq:hTsallis}). For $E_i \ll U$ the conditional probability (\ref{eq:fEiE}) can be rewritten as
\begin{eqnarray}
h\left( E_i|U\right) &\simeq& \frac{\left(N +2\right)(n-1)}
{n T(n-2 + N)} \left(1 +\frac{E_i}{n T}\right)^{-n} \label{eq:eismallaE}
\end{eqnarray}
which, when additionally $N \gg 1$,  reduces to Eq. (\ref{eq:hTsallis}).
\begin{figure}[h]
\centerline{
\includegraphics[width=6.4cm]{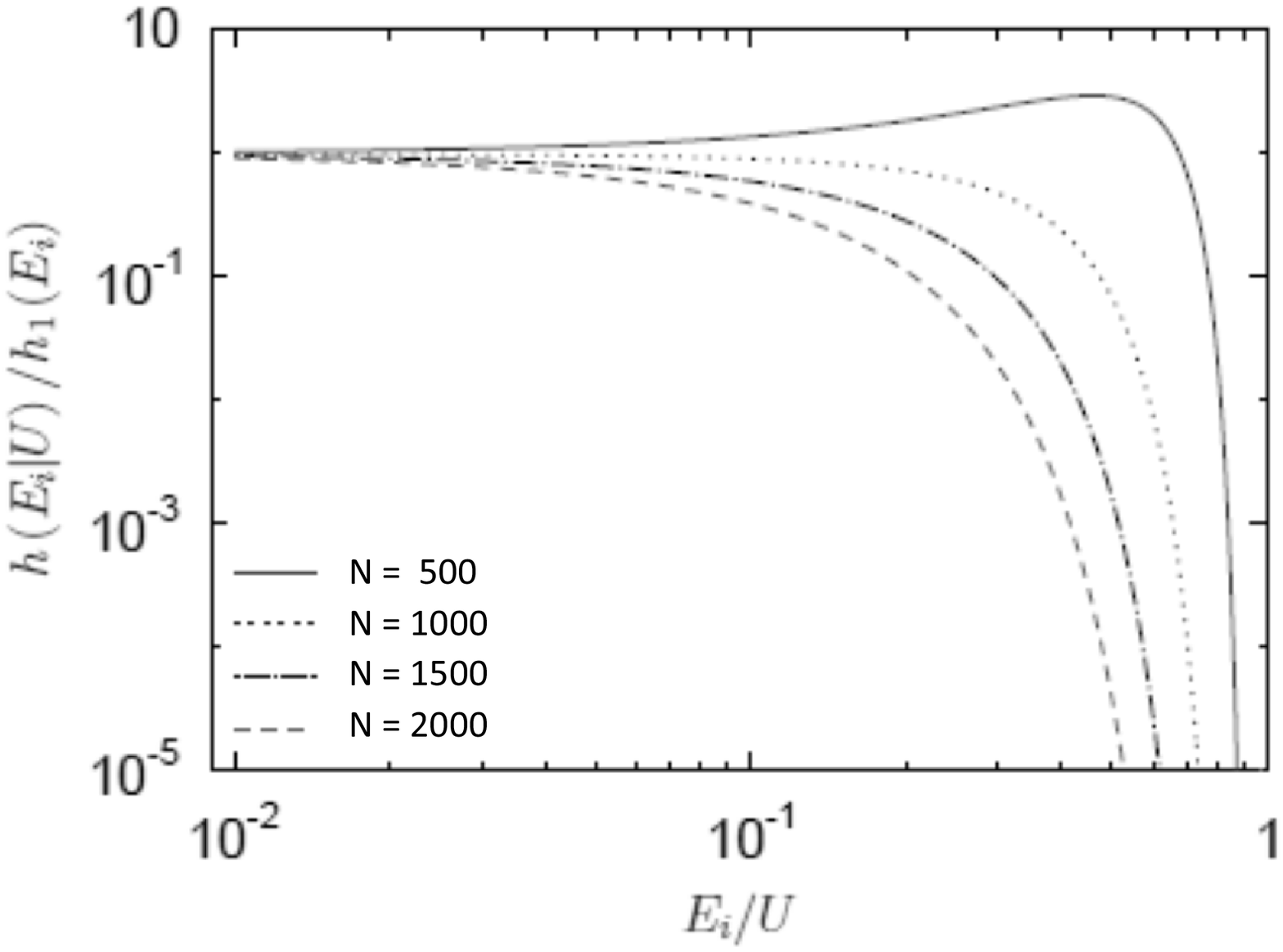}
\includegraphics[width=6.2cm]{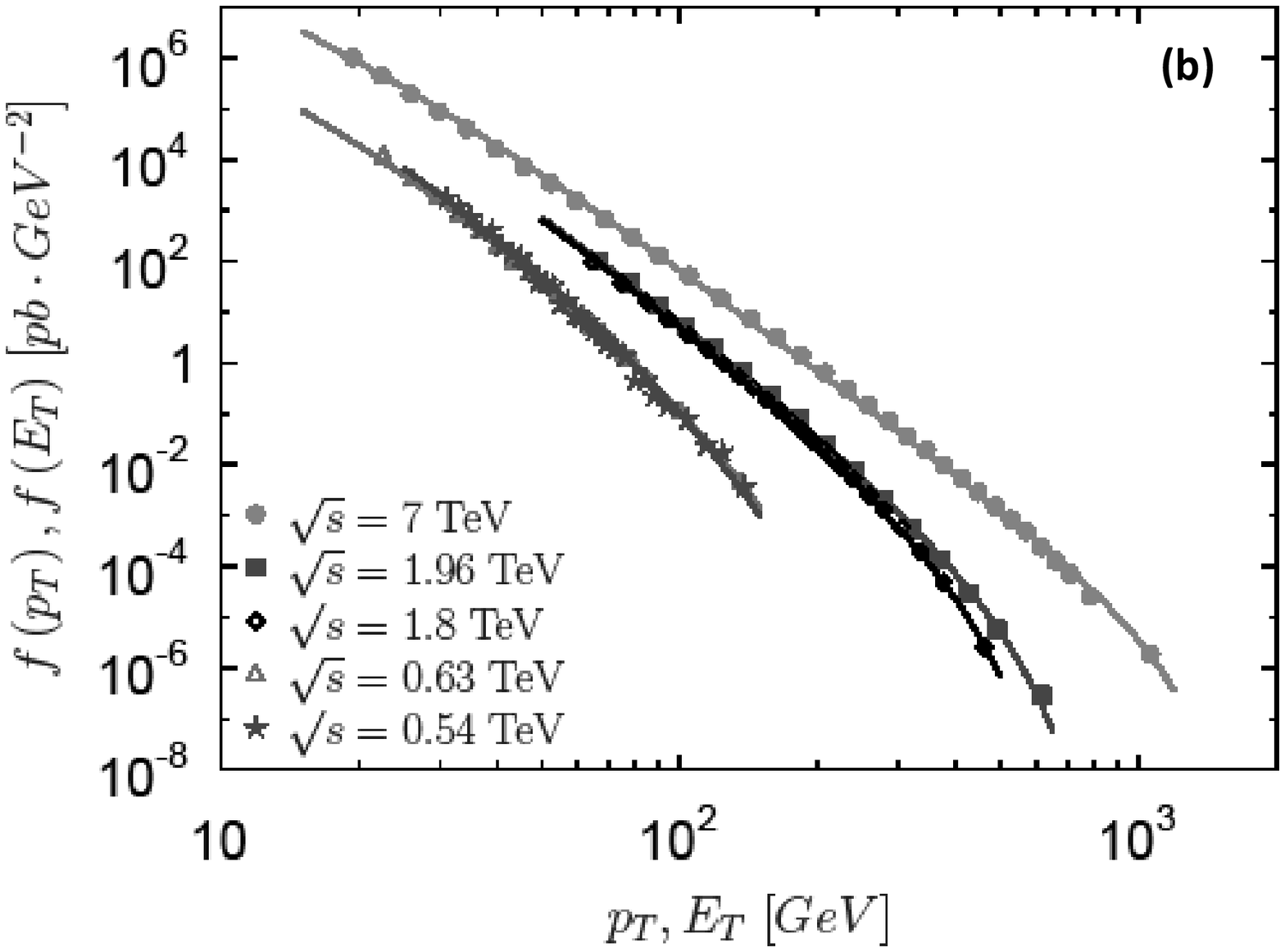}}
\vspace{-3mm}
\caption{ $(a)$  Ratio of the conditional distribution function $h(E_i|U)$ and the single particle distribution $h_{1}(E_i)$ as a function of $E_i/U$, for Tsallis statistics ($n=7$ and $T = 1$ GeV). $(b)$   Transverse spectra for jets ($p_T$ and $E_T$ distribution are shown by full and open symbols, respectively) fitted by the conditional Tsallis distribution given by Eq. (\ref{eq:fEiE}).
\label{Limit}}
\end{figure}

The above results of the limitation of the allowed phase space are illustrated in Fig. \ref{Limit}. Fig. \ref{Limit} $(a)$  shows how large are the differences between the {\it conditional} Tsallis distribution $h(E_i|U)$, Eq. (\ref{eq:fEiE}) and the {\it usual} $h_1(E_i)$, Eq. (\ref{eq:hTsallis}). In Fig. \ref{Limit} $(b)$ this is further exemplified by a number of fits to data for $p+\bar{p}$ and $p+p$ interactions, covering a wide energy range from $0.54$~TeV up to $7$ TeV \cite{CMS-5,CDF,D0,UA2} (cf., \cite{MW-limited} for details). These results can be compared with those discussed so far, especially those based on QCD in which jets were described without imposing the above limitations, these were hidden in the form of the parton distribution used. At the moment the only work addressing such a problem is \cite{MW-limited} but this subject deserves further investigation.

\section{Surprising effects: log-periodic oscillations in Tsallis distributions}
\label{sec:Surprises}

\begin{figure}[h]
\begin{center}
\resizebox{1.05\textwidth}{!}{%
  \includegraphics{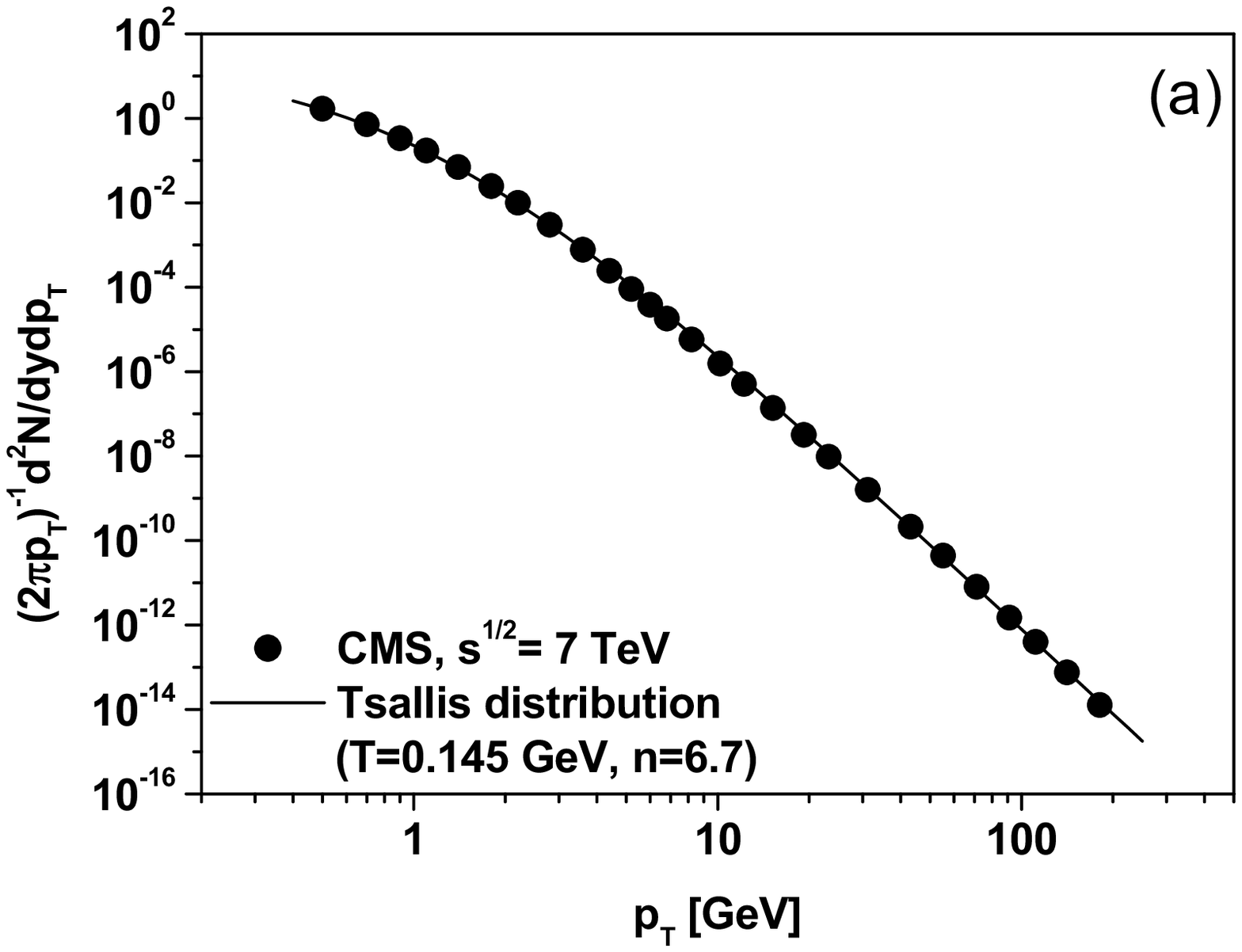}
  \includegraphics{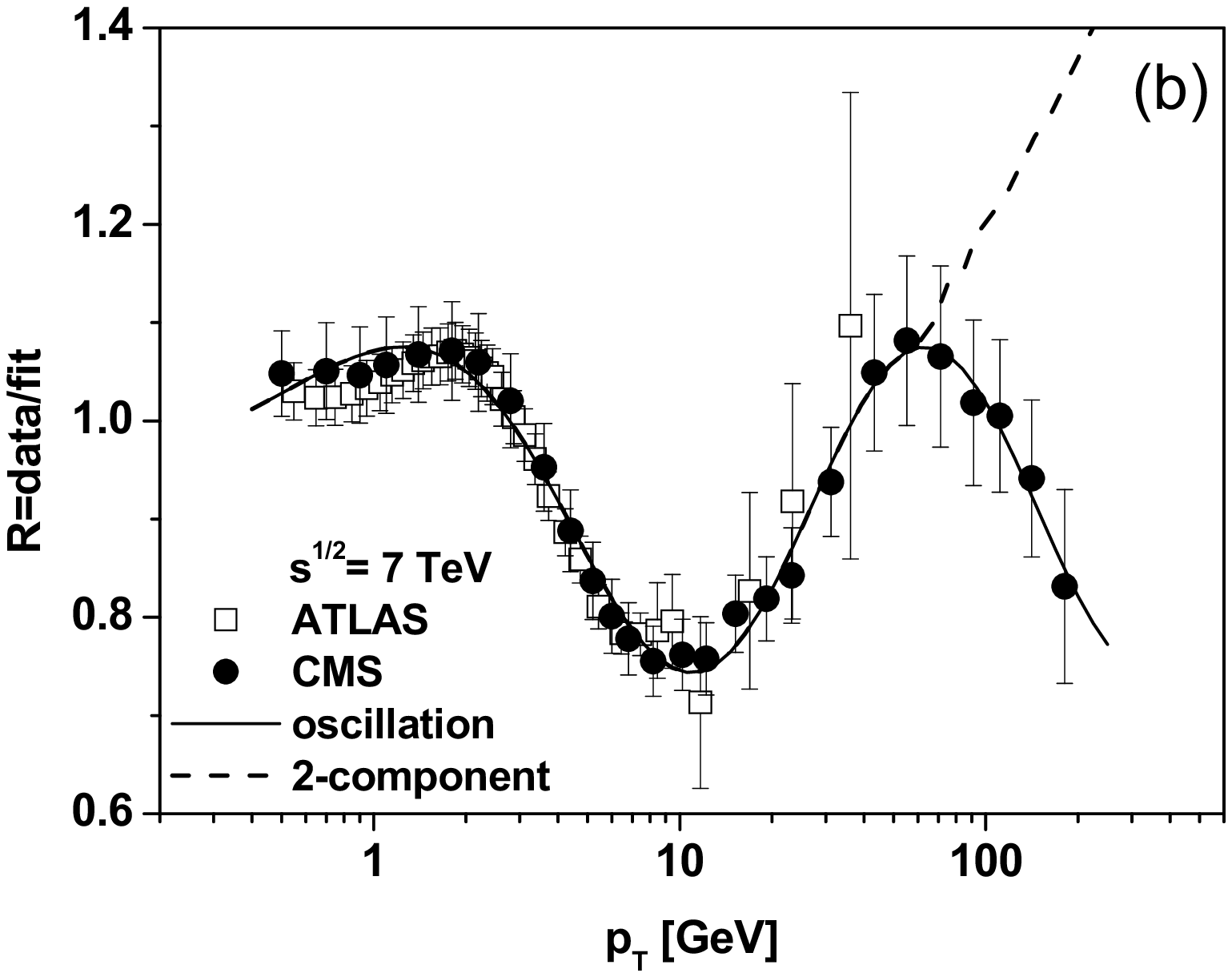}
  }
\vspace{-4mm}
\resizebox{1.05\textwidth}{!}{%
  \includegraphics{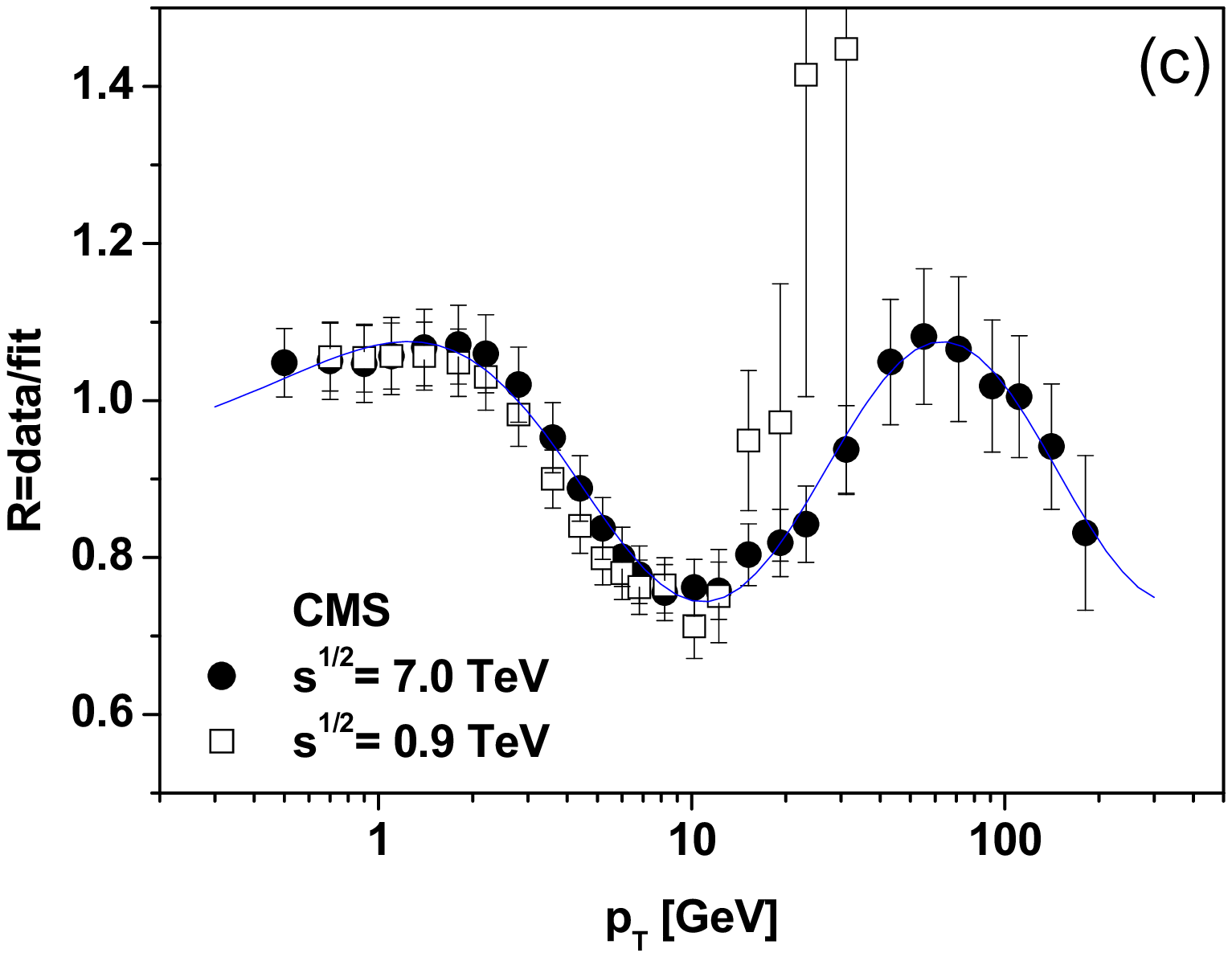}
  \includegraphics{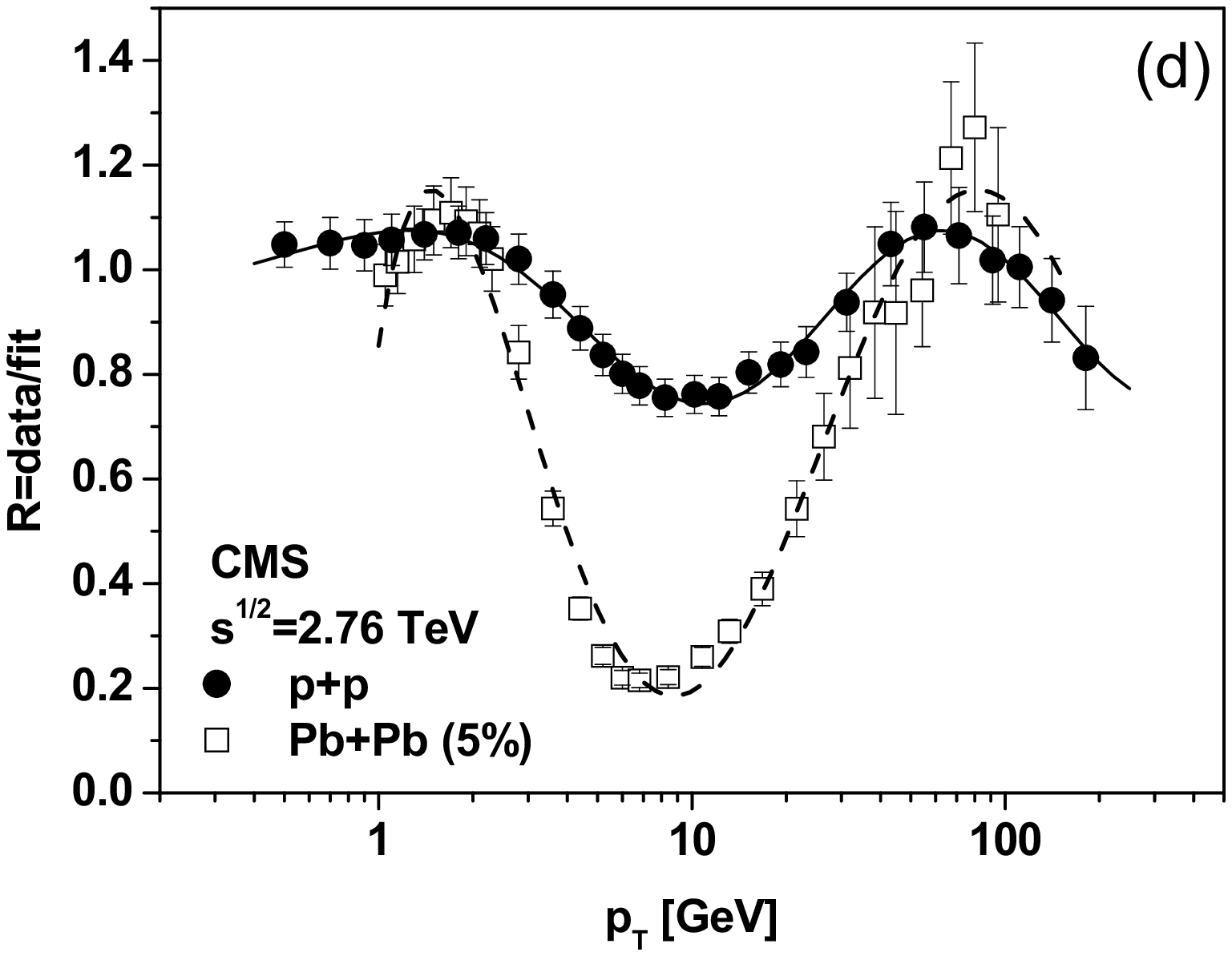}
}
\caption{$(a)$ Example of a typical $p+p$ large $p_T$ distribution observed in LHC experiments. $(b)$ Log periodically oscillating factor $R$ for different LHC experiments.  The dashed line denotes the prediction of the two-component approach proposed in \cite{SoftHard1,SoftHard2}. The other lines in all panels are obtained using Eq. (\ref{eq:approx} (see \cite{WWln,RWWlnA} for details). $(c)$ and $(d)$ - The same factor $R$ at two different energies. Also in $(d)$ we have an example of the $R$ factor for $Pb+Pb$ collisions. }
\label{FF1}
\end{center}
\end{figure}

We present now some surprising results, apparently hidden in the large transverse momenta distributions observed in CMS \cite{CMS-1,CMS-3} , ATLAS \cite{ATLAS-2} and ALICE \cite{ALICE-2,ALICE-3} experiments, which have so far remained unnoticed or not fully appreciated; they are displayed in Fig. \ref{FF1}. In panel $(a)$ we have a typical $pp$ large $p_T$ cross section which nicely follows the Tsallis distribution, Eq. (\ref{eq:T}). However, the ratios $R = \sigma_{data}\left( p_T\right)/\sigma_{fit}\left( p_T\right)$ plotted as a function of $p_T$ in panels $(b)-(d)$ exhibit distinct log-periodic oscillations. Albeit rather small they are persistent and show up in the results from different experiments (panel $(b)$), at different energies (panels $(c)-(d)$) and in different systems (panels $(d)$). Panel $(b)$ presents a comparison of $p_T$ distributions for $p+p$ and $Pb+Pb$ collisions \cite{WWln,RWWlnA} . The dashed line in panel $(b)$ shows the behavior of the alternative approach  proposed in \cite{SoftHard1,SoftHard2} and based on a two component, {\it soft+hard}, picture of the production process, each taken in the form of Eq. ({\ref{eq:T}). As can be seen it is not able to describe the observed effect in the whole region of $p_T$. Because the observed oscillations cannot be erased by any reasonable change of fitting parameters, we feel entitled to assume that this is a real effect which deserves to be investigated in detail. The Tsallis distribution (\ref{eq:T}) has two parameters, $n$ (or $q$) and $T$, therefore the observed log-periodic oscillations can be assigned to either of them (or to both, but such a possibility will not be discussed). We shall now discuss both possibilities referring for details to\cite{WWln,RWWlnA,WW-Entropy14,WW-APPB46,WW-CSF81} .

\subsection{Complex power index $n$ - scaling behavior of Tsallis distributions}
\label{sec:complexN}

The pure power-like distributions are known to be in many cases decorated by specific log-periodic oscillations (i.e., they are multiplied by some dressing factor $R$). This fact suggests that the system under consideration has some hierarchical structure and is usually regarded as indicating some kind of multifractality \cite{Sornette} . For simple power laws, if the function $O(x)$ is scale invariant, this means that $O(\lambda x) = \mu O(x)$ and $O(x) = Cx^{-m}$ with $m = - \ln \mu/\ln \lambda$ or $\mu \lambda^{m} = 1 = e^{i2\pi k}$, where $k$ is an arbitrary integer, resulting in a whole family of complex powers, $m_k$, with $m_k = - \ln \mu/\ln \lambda + i 2\pi k/\ln \lambda$. Their imaginary part signals a hierarchy of scales leading to log-periodic oscillations. This means that $O(x) = \sum_{k=0}w_k Re\left(x^{-m_k}\right) = x^{-Re \left( m_k \right)} \sum_{k=0} w_k\cos\left[Im\left(m_k\right)\ln(x)\right]$ (where the $w_k$ are coefficients of the expansion). This is the origin of the so called {\it dressing factor} $R$ appearing in \cite{Sornette} and used to describe data (with only $w_0$ and $w_1$ terms kept):
\begin{equation}
R(E)= a + b\cos\left[ c\ln(E + d) + f\right]. \label{eq:R}
\end{equation}
It turns out that a similar scaling solution can also be obtained for the Tsallis  distribution in Eq. (\ref{eq:T}) \cite{WWln} . Namely, using the preferential attachment rule from Section \ref{sec:Pa}, in which the Tsallis distribution is obtained by introducing a scale parameter depending on the variable considered, we have that $df(E)/dE = - f(E)/T(E)$ and $f(E) = \left[(n-1)/\left( nT_0\right)\right]\left[ 1 + E/\left(nT_0\right)\right]^{-n}$ with $T(E) = T_0 + E/n$. In finite difference form (with change in notation: $T_0 \rightarrow T$) we have that
\begin{equation}
\frac{df(E)}{dE} = - \frac{f(E)}{T(E)}~~ \Longrightarrow ~~f(E +
\delta E) = \frac{ - n\delta E + nT +E}{nT + E} f(E).
\label{eq:deltaE}
\end{equation}
Consider the situation in which $\delta E = \alpha nT(E) = \alpha (nT + E)$ with a new scale parameter $\alpha$ ($\alpha < 1/n$ in order to keep changes in $\delta E$ to be of the order of $T$; it can be very small but always remains finite).
In this case $f[E + \alpha(nT + E)] = (1 - \alpha n)f(E)$ \cite{WWln} , which formally corresponds to the following scale invariant relation:
\begin{equation}
g[(1 + \alpha) x] = (1 - \alpha n)g(x) \label{eq:gscaling}
\end{equation}
where $x = 1 + E/(nT)$. Following the procedure used previously for the function $O(x)$ and keeping only $k=1,2$ terms, one arrives at the following dressed Tsallis distribution
\begin{equation}
g(E) \simeq \left( 1 + \frac{E}{nT}\right)^{-m_0}\left\{ w_0 + w_1\cos\left[ \frac{2\pi}{\ln (1 + \alpha)} \ln \left( 1 +
\frac{E}{nT}\right)\right]\right\}. \label{eq:approx}
\end{equation}
with $m_0 = - \ln(1 - \alpha n)/\ln(1 + \alpha ) \stackrel{\alpha \rightarrow 0}{\longrightarrow} n$ and with the dressing factor in the form of (\ref{eq:R}) with parameters $\alpha$, $w_0$ and $w_1$. The remaining parameters can be expressed by them as follows: $a/b = w_0/w_1$, $c = 2\pi/\ln(1 + \alpha)$, $d=nT$ and $f = -2\pi\ln(nT)/\ln(1+\alpha) = -c\ln d$. In the case of a more involved evolution process, proceeding via $\kappa$ sequential cascades, a new additional parameter $\kappa$ appears and $c \rightarrow c'=c/\kappa$. This does not affect the slope parameter $m_0$ but changes the frequency of the oscillations which now decrease as $1/\kappa$ (cf. \cite{WWln} for details). Eq. (\ref{eq:approx}) was used to fit the results presented in Fig. \ref{FF1} (see \cite{WWln,RWWlnA} for details). To end this part let us mention that a complex $q$ has a number of interesting consequences discussed in \cite{WW-CSF81}  like a complex heat capacity of the system, or the possibilities of complex probability and complex multiplicative noise, all of them known already from other branches of physics.

\subsection{Temperature oscillation and sound waves - self-similarity of Tsallis distributions}
\label{sec:oscillations}

The other possibility is to keep the power index $n$ real but allow the scale parameter $T$ to vary with $p_T$ in such a way as to reproduce the observed log-periodic oscillations. This can be done using a log-periodically oscillating form of $T$,
\begin{equation}
T = T\left( p_T\right) = a + b \sin\left[ c \ln\left( p_T + d\right) + e \right]. \label{TpT}
\end{equation}
\begin{figure}[h]
  \begin{center}
   \includegraphics[width=8cm]{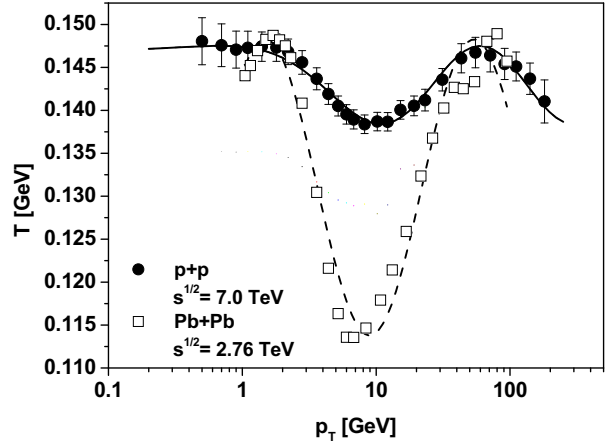}
   \caption{Example of the observed log-periodicity of the $p_T$-dependent scale parameter $T$ for $p+p$ and $Pb+Pb$  collisions deduced from ALICE data \cite{ALICE-1,ALICE-3} . The fits follow Eq. (\ref{TpT}) with parameters $a = 0.143$, $b=0.0045$, $c=2.0$, $d=2.0$, $e=-0.4$ for $p+p$ collisions and $a=0.131$, $b=0.019$, $c=1.7$, $d=0.05$, $e=0.98$ for $Pb+Pb$ collisions.}
   \label{FF1a}
  \end{center}
\end{figure}
which can fit data very nicely, see Fig. \ref{FF1a} \cite{WW-APPB46,WW-CSF81} . Such oscillations (and, respectively, the meaning of the parameters $a$, $b$, $c$, $d$ and $e$ in Eq. (\ref{TpT})) can originate from a suitable $p_T$ dependence of the noise $\xi$ defining the corresponding stochastic processes in the stochastic equation for the evolution of temperature $T$ written in the Langevin formulation \cite{Lang} . Assuming $\xi\left(t, p_T\right)$, and allowing for time-dependent $p_T = p_T(t)$  \cite{Lang} , one has that \cite{WW-APPB46,WW-CSF81}
\begin{equation}
\frac{dT}{dt} + \frac{1}{\tau} T + \xi\left( t,p_T\right) = \Phi, \label{LE}
\end{equation}
where $\tau$ is the relaxation time. Assuming now that $p_T=p_T(t)$ increases in time following the scenario of the preferential growth of networks \cite{WWnets-1} ,
\begin{equation}
\frac{dp_T}{dt} = \frac{1}{\tau_0}\left( \frac{p_T}{n} \pm T\right) \label{PrefG}
\end{equation}
(where $n$ coincides with the power index in Eq. (\ref{eq:H}) and $\tau_0$ is some characteristic time step), we get the stationary dependence of $T\left( p_T\right)$ represented by Eq. (\ref{TpT}) either when the noise term increases logarithmically with transverse momentum while the relaxation time $\tau$ remains constant,
\begin{equation}
\xi\left(t, p_T\right) = \xi_0(t) + \frac{\omega^2}{n}\ln \left( p_T\right), \label{p_T_increase}
\end{equation}
or when the white noise is kept constant, $\xi\left( t, p_T\right) = \xi_0(t)$, but the relaxation time becomes $p_T$-dependent, for example
\begin{equation}
\tau = \tau\left( p_T\right) = \frac{n\tau_0}{n + \omega^2 \ln\left( p_T\right)} \label{Tau_dep}
\end{equation}
(in both cases $\omega$ is some new parameter) \cite{WW-APPB46,WW-CSF81,WW-Entropy17}  . To fit data one needs only a rather small admixture of the stochastic processes with noise depending on the transverse momentum (defined by the ratio $b/a \sim 3\%$). The main contribution comes from the usual energy-independent Gaussian white noise.

Eq. (\ref{TpT}) can be obtained in yet another way in which oscillating $T$ is considered to be a signal of sound waves propagating in hadronic matter \cite{WWwaves,WW-Entropy17} . Such a phenomenon occurs in the hydrodynamical models used to describe strongly interacting systems. They have recently become increasingly sophisticated and popular in response to the continuous support received from experiment, especially from the observation of the so called  elliptic flow of secondaries produced in multiparticle production processes. This phenomenon is not easy to explain in other approaches, which are more successful in the description of the distributions of the measured transverse momenta but use a fixed scale parameter $T$. Also, a number of observations strongly indicate that hadronic matter produced in heavy ion collisions at RHIC and the LHC behaves as a kind of perfect fluid. The observation of sound waves would therefore provide additional support  for a hydrodynamical description of the multiparticle production process. Such waves could arise because in the initial phase of the collision process a number of highly energetic partons are created which subsequently lose their energy. This can proceed in two ways: either by exciting modes of the medium in a {\it collisional energy loss}, or by radiating gluons in a {\it radiative energy loss}. The further dissipation of the released energy resulting in thermalization depends on the character of the medium: in a {\it weakly coupled medium} it proceeds  through a cascade of collisions among quarks and gluons in the quark-gluon plasma, whereas in a {\it strongly coupled medium} the released energy is dissipated directly into thermal excitations and sound waves. Such phenomena were, in fact, already studied in \cite{HydroWaves} . Actually, in relativistic heavy ion collisions we may also have hard parton-parton collisions in which the outgoing partons have to traverse the surrounding fluid before escaping and forming jets which subsequently hadronize. Such partons may therefore form Mach shock waves during their passage and this, in turn, will affect the transverse momentum distribution of the observed final particles.

Another possible consequence of the hydrodynamical picture of the production process explored in \cite{WWwaves} is the possible formation and propagation of sound waves in hadronic matter. This idea is based on the form of  $T=T\left( p_T\right)$ in Eq. (\ref{TpT}) and on its Fourier transform,
 \begin{equation}
T(r) = \sqrt{\frac{2}{\pi}}\int^{\infty}_0 \, T\left( p_T\right) e^{ip_T r} dp_T = T_0 + T'(r),   \label{FT}
\end{equation}
which allows us to gain further insight into  the space-time structure of the collision process. Because the oscillations are in transverse momentum $p_T$, $r$ is defined in the plane perpendicular to the collision axis and located at the collision point and denotes the distance from the collision axis.
\begin{figure}[t]
\begin{center}
\includegraphics[scale=0.24]{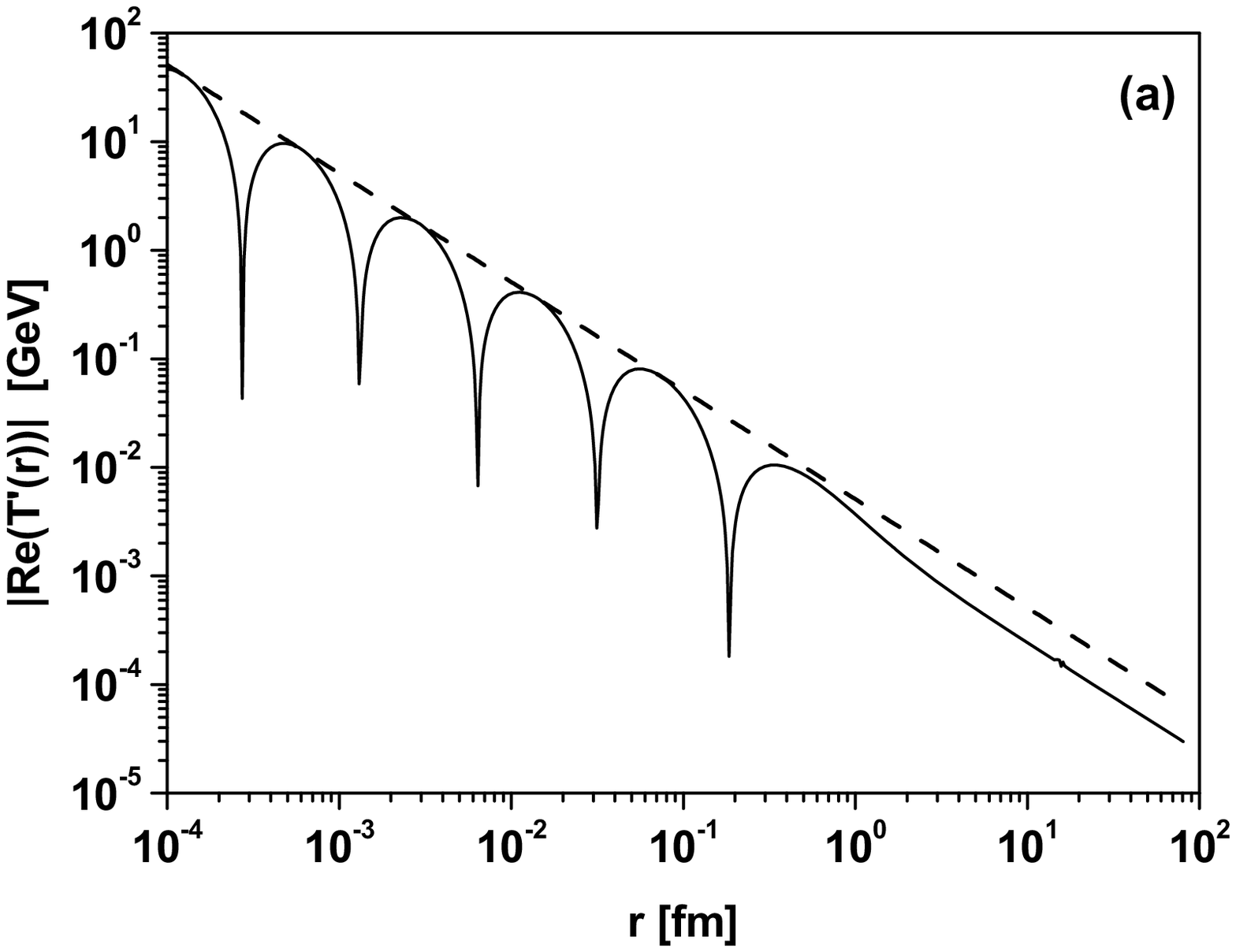}
\includegraphics[scale=0.255]{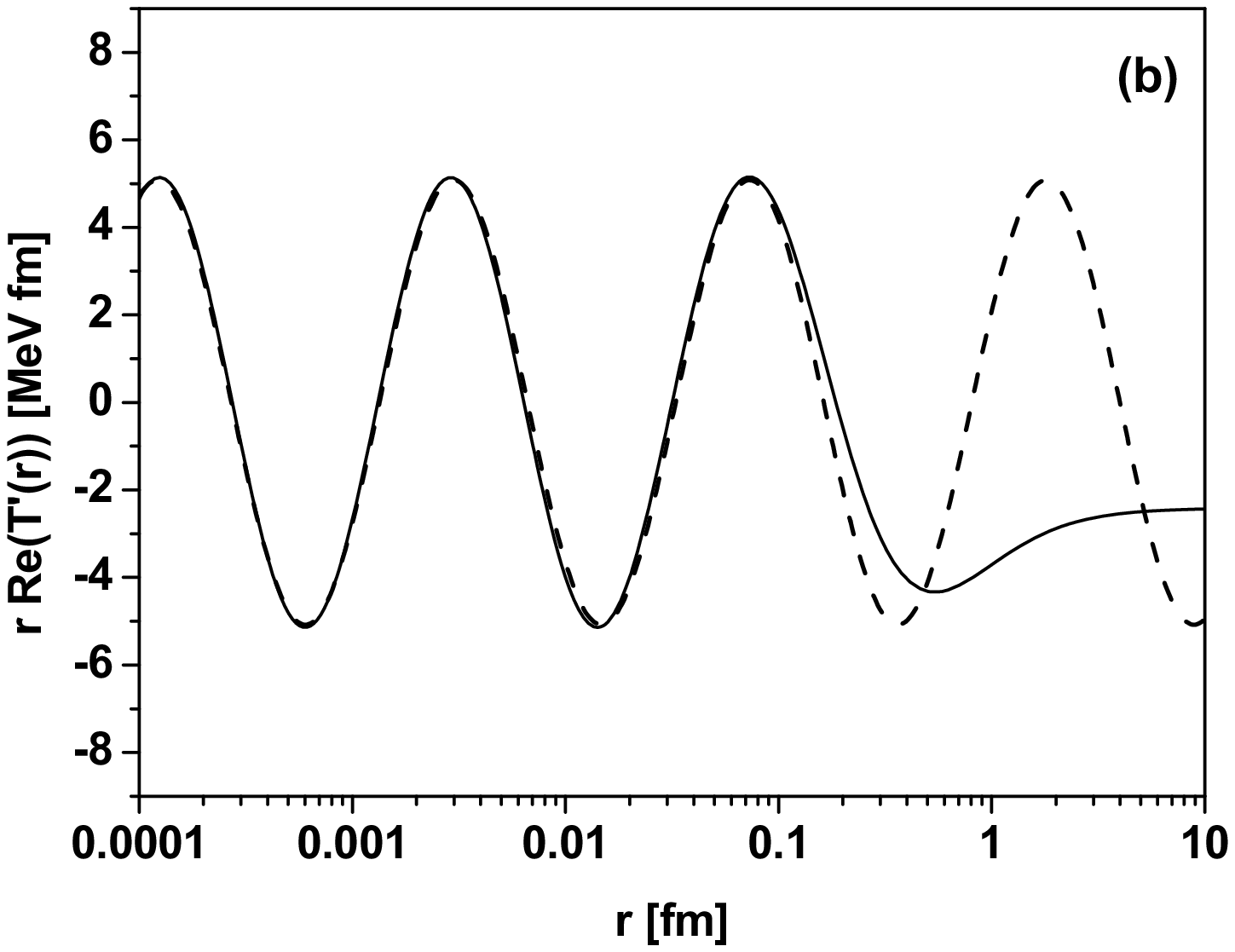}
\includegraphics[scale=0.24]{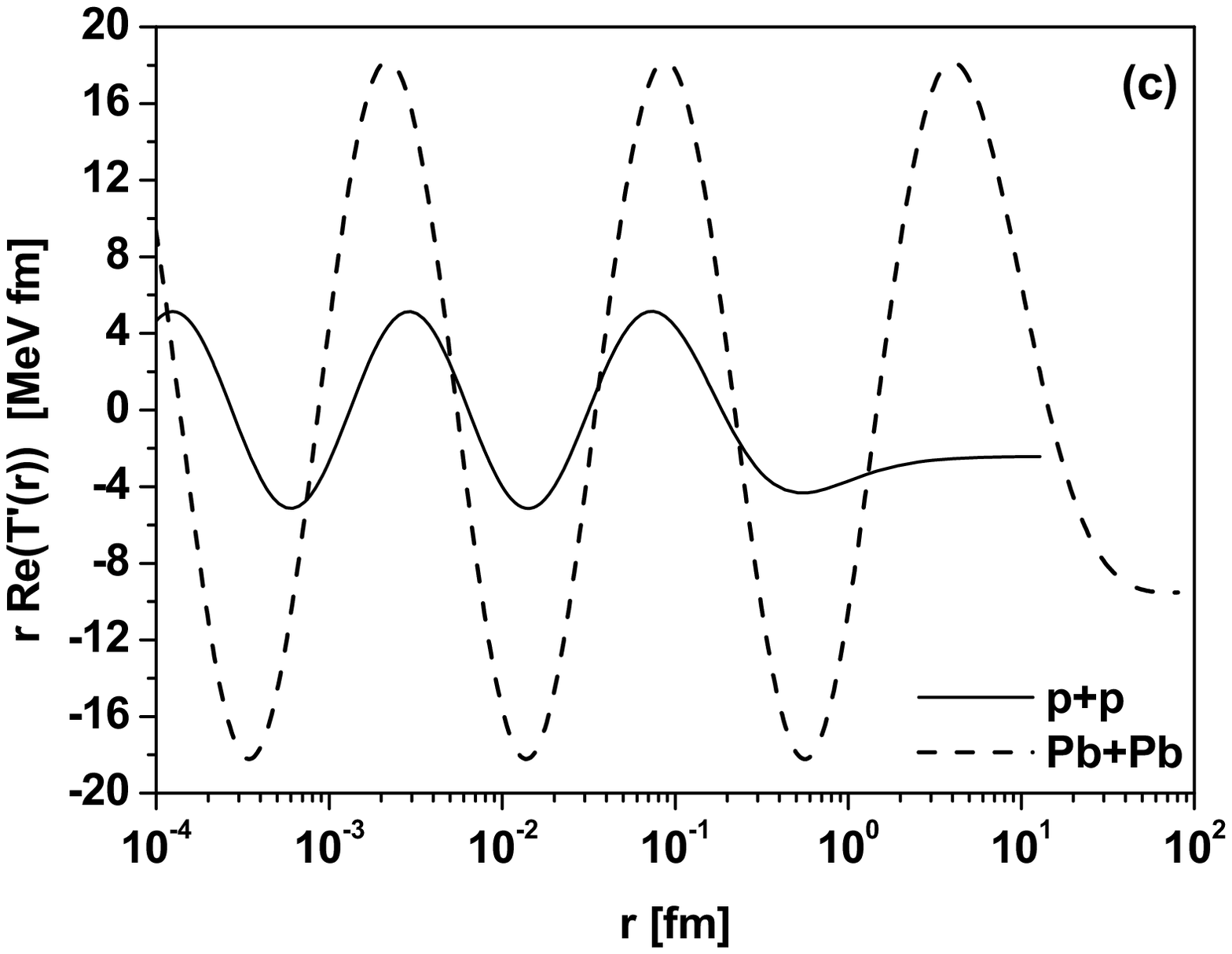}
\end{center}
\vspace{-5mm}
\caption{$(a)$ and $(b)$  The results of the Fourier transform of $T\left(p_T\right)$ from Eq. (\ref{TpT}) describing the results inferred from the CMS data \cite{CMS-1,CMS-3} . $(c)$ oscillating behavior of $r T'(r)$ for $p+p$ collisions at $7$ TeV presented in previous panels compared with similar results for the most central $Pb+Pb$ collisions at $2.76$ TeV (inferred from the ALICE data \cite{ALICE-2,ALICE-3}). See text for details.}
\label{FF1x}
\end{figure}
The results are shown in Fig. \ref{FF1x}. In panel $(a)$ we have $T'(r)$ (continuous line) as a function of $r$ confronted with $T'(r) = 0.0051/r$ dependence (dashed line).  In panel $(b)$ the continuous line represents $r T'(r)$ versus $r$ and  is confronted with the dashed line denoting the function $rT'(r) = 5.1 \sin [(2\pi)/3.2 \ln(1.24 r)]$ fitting it for small values of $r$. Panel $(c)$ presents a comparison of the oscillations of $T(r)$ for $p+p$ and the most central $Pb+Pb$ collisions deduced from the ALICE data \cite{ALICE-2,ALICE-3} at $2.76$ TeV \cite{RWWlnA} (the parameters of $T\left( p_T\right)$ used in both cases are the same as used in Fig. \ref{FF1a}). For the most central $Pb+Pb$ collisions the amplitude is $\sim 3.6$ times bigger and the period of the oscillations is $\sim 1.15$ times  longer than in $pp$ collisions.  With decreasing centrality the amplitude in central $Pb+Pb$ collisions decreases smoothly reaching practically the same value as for $p+p$ collisions \cite{RWWlnA} . In  the case considered here, the region of regular oscillations can be fitted by $ rT'(r) = 5.1 \sin \left[ \frac{2\pi}{3.2}\ln(r) + 0.42\right]$ for $p+p$ collisions and by $ rT'(r) = 18.53 \sin \left[ \frac{2\pi}{3.68}\ln(r) - 0.51\right]$ for $Pb+Pb$ collisions.

To see how such a log-periodically oscillating structure of $T(r)$ occurs we study the flow of a compressible fluid in a cylindrical source assuming small oscillations and oscillatory motion with small amplitude in a compressible fluid (i.e., a {\it sound wave}, since at each point in the fluid it causes alternate compression and rarefaction of the matter). Since the oscillations are small, their velocity $v$ is also small and the term $( {\bold v}\cdot {\bold {grad}}){\bold v}$ in Euler's equation may be neglected. For the same reason, the relative changes in the fluid density and pressure are small and we may write as \cite{Landau}
\begin{equation}
 P = P_0 + P',\quad\quad \rho = \rho_0 + \rho', \label{Prho}
\end{equation}
where $P_0$ and $\rho_0$ are the constant equilibrium pressure and density, and $P'$ and $\rho'$ are their variations in the sound wave $\left( \rho' << \rho,~~P' << P\right)$. Neglecting small quantities of the second order ($P'$, $\rho'$ and $v$ are regarded to be of the first order) the equation of continuity, $\partial \rho/\partial t + {\bold {div}}\cdot(\rho {\bold v}) =0$ becomes
\begin{equation}
\frac{\partial \rho'}{\partial t} + \rho_0 {\bold {div}}({\bold v}) = 0 \label{lindiv}
\end{equation}
and  Euler's equation reduces in this approximation to
\begin{equation}
\frac{\partial {\bold v}}{\partial t} +  \frac{1}{\rho_0}{\bold {grad}} P' = 0. \label{Euler}
\end{equation}
The linearized forms of (\ref{lindiv}) and (\ref{Euler}) are applicable to the propagation of sound waves if $v<<c$ (where $c$ is the velocity of sound), which  means that $P' << P_0$. Note that a sound wave in an ideal fluid is adiabatic, therefore a small change $P'$ in the pressure is related to a small change $\rho'$ in the density by ($c$ is the velocity of sound),
\begin{equation}
P' = \left(\frac{\partial P}{\partial \rho_0} \right)_S \rho'  = c^2 \rho'\qquad {\rm where}\qquad c = \sqrt{\left( \frac{\partial P}{\partial \rho}\right)_S} .\label{PpRr}
\end{equation}
Substituting $\rho'$ from Eq. (\ref{PpRr}) into Eq. (\ref{lindiv}) one gets
\begin{equation}
\frac{\partial P'}{\partial t} + \rho_0 \left( \frac{\partial P}{\partial \rho_0}\right)_S  {\bold {div}}\cdot{\bold v} = 0 \label{Eq1}
\end{equation}
which, together with Eq. (\ref{Euler}), using the unknowns ${\bold v}$ and $P'$, provides a complete description of the sound wave we are looking for. To express all the unknowns in terms of one of them, it is convenient to introduce the velocity potential $f$ by putting ${\bold v} = {\bold {grad}} f$.  From Eq. (\ref{Euler}) we have the relation between  $P'$ and $f$:
\begin{equation}
P' = - \rho \frac{\partial f}{\partial t}. \label{Ppf}
\end{equation}
which, used together with Eq. (\ref{Eq1}), results in the following cylindrical wave equation which the potential $f$ must satisfy,
\begin{equation}
\frac{1}{r} \frac{\partial}{\partial r}\left( r \frac{\partial f}{\partial r}\right) - \frac{1}{c^2}\frac{\partial^2 f}{\partial t^2} =0.    \label{partialwe}
\end{equation}
It can be shown that this represents a travelling longitudinal plane sound wave with velocity ${\bold v}$ in the direction of propagation. It is related to the pressure $P'$ and the density $\rho'$ in a simple manner, namely
\begin{equation}
v = \frac{P'}{\rho c} = c \frac{\rho'}{\rho}. \label{vel}
\end{equation}
To relate the above results to the temperature note that we may write $T(r)$ as consisting of a constant term, $T_0$, and some oscillating addition, $T'(r)$:
\begin{equation}
 T(r) = T_0 + T'(r)\qquad {\rm where}\qquad T' = \left( \frac{\partial T}{\partial P}\right)_S P'.  \label{oscillatingT}
\end{equation}
Using the well known thermodynamic formula $\left( \partial T/\partial P\right)_S = \left(T/c_P\right)\left(\partial V/\partial T\right)_P$ and Eq. (\ref{vel}) one obtains that
\begin{equation}
T' = \frac{c \kappa T}{c_P} v \qquad {\rm where}\qquad \kappa = \frac{1}{V}\left( \frac{\partial V}{\partial T}\right)_P \label{ResLandau}
\end{equation}
where $\kappa$ is the coefficient of thermal expansion and $c_P$ denotes the specific heat  at constant pressure  \cite{Landau} .
In the case of a monochromatic wave, when $f(r,t) = f(r)\exp(-i\omega t)$, we have that
\begin{equation}
\frac{\partial^2 f(r)}{\partial r^2} + \frac{1}{r} \frac{\partial f(r)}{\partial r} + K^2 f(r) = 0,\qquad \qquad
K=K(r) = \frac{\omega}{c(r)}  \label{eqn}
\end{equation}
where $K$ is the wave number which in inhomogeneous media depends on $r$. For the wave number given by
\begin{equation}
K(r) = \frac{\alpha}{r} \label{example}
\end{equation}
the solution of Eq. (\ref{eqn}) is given by a log-periodic oscillation in the form
\begin{equation}
f(r) \propto \sin [ \alpha \ln(r)]. \label{solution}
\end{equation}
Because ${\bold v} = {\bold {grad}} f$ we have $f(r) \propto v r$. Therefore, using Eq. (\ref{ResLandau}), we can write that
\begin{equation}
rT'(r) \propto \frac{c\kappa T_0}{c_P}f(r) = \frac{c \kappa T_0}{c_P} \sin[ \alpha \ln (r)]. \label{final}
\end{equation}
This is the solution we have used in describing the $T'(r)$ deduced from data and presented in Fig. \ref{FF1x}.

The above problem can be considered from yet another point of view by noting that Eq. (\ref{eqn}) with the wave number (\ref{example}) has a so-called {\it self similar solution of the second kind} \cite{BZ1,BZ2,B1,B2} . Such a solution is known from other branches of physics and is connected with the so called {\it intermediate asymptotic} encountered whenever dependence on the initial conditions disappears (because sufficient time has already passed from the beginning of the process considered), but our system has not yet reached the state of equilibrium \cite{BZ1,BZ2,B1,B2} . Introducing the variable
\begin{equation}
\xi = \ln r \label{ksi}
\end{equation}
we find that Eq. (\ref{eqn}) for the wave number (\ref{example}) represents a travelling wave equation,
\begin{equation}
\frac{\partial^2 F(\xi)}{\partial \xi^2} + \alpha^2 F(\xi) = 0, \label{TWE}
\end{equation}
the solution of which is
\begin{equation}
F(\xi) \propto \sin(\alpha \xi). \label{TWS}
\end{equation}
This self similarity of Eq. (\ref{eqn}), which can be written as $F(\xi + \ln \lambda) = F(\xi)$, can now  be confronted with a kind of scale invariance of this equation, namely with the fact that $f(\lambda\cdot r) = f(r)$.

We end this part by concluding that the space picture of the collision (in the plane perpendicular to the collision axis and located at the collision point) presented in Fig. \ref{FF1x} (panels $(b)$ and $(c)$) shows us the existence of some regular (on the logarithmic scale) structure for small distances. For $p+p$ collisions it starts to weaken quite early (at $ r \sim 0.1$ fm) and essentially disappears when $r$ reaches the dimension of the nucleon, i.e., for $r \sim 1$ fm. For $Pb+Pb$ collisions it seems to last longer, to around $r \sim 10$ fm (i.e., to a typical dimension of the nucleus).

A few practical remarks are in order here. The longer period of the oscillations in the $Pb+Pb$ collisions means that the values of the parameter $\alpha$ in Eq. (\ref{solution}) in nuclear collisions are smaller than those for $p+p$ collisions. Furthermore, considering the form of $K$ from Eq. (\ref{eqn}) or Eq. (\ref{example}), and remembering that $\omega/c(r)=\alpha/r$, one may deduce that the velocity of sound, $c(r) = (\omega/\alpha)r$, is greater in the nuclear environment  (for $Pb+Pb$) than in the $p+p$ case\footnote{This relationship is similar to the Hubble expansion of the Universe (with $\omega/\alpha$ corresponding to the Hubble constant). Also, in the blast-wave model the particles closer to the center of the fireball move slower than those on the edge \cite{BLASTMODEL} .}. This, in turn, means that the refractive index $n(r) = c_0/c(r)$ at position $r$ in nuclear collisions is smaller than in proton collisions. Therefore, in both cases we encounter an inhomogeneous medium with $r$-dependent  velocity of sound, $c(r)$, and refractive index, $n(r)$. These
findings agree with the fact that in nuclear collisions a higher speed of sound is actually observed, as demonstrated
by the NA61/SHINE collaboration at SPS energies \cite{NA61,GGS} (note, however, that what is measured is a parameter in the equation of state of hadronic matter described by a hydrodynamical model, $c^2_s$). Actually, this is not unexpected because, considering the connection of the isothermal compressibility  of nuclear matter, $\kappa_T = - (1/V)\left( \partial V/\partial P\right)_T$, and remembering that fluctuations of the multiplicity of the produced secondaries is represented by the relative variance, $\varpi$, of multiplicity fluctuations, one finds that \cite{Hill,Balescu}
\begin{equation}
T\kappa_T \rho_0 = \frac{\langle N^2 \rangle - \langle N\rangle^2}{\langle N\rangle} = \varpi \label{IC-NF}
\end{equation}
(where $\rho_0 = \langle N\rangle/V$ denotes the equilibrium density for $N$ particles with mass $m$ located in volume $V$). This allows the velocity of sound to be expressed by fluctuations of multiplicity:
\begin{equation}
c = \sqrt{\frac{\gamma}{\kappa_T \rho_0 m}} = \sqrt{\frac{\gamma T}{\varpi m}}\quad {\rm or}\quad \varpi = \frac{\gamma T}{m}\cdot \frac{1}{c^2}\quad{\rm where}\quad \gamma = \frac{c_P}{c_V}. \label{SV}
\end{equation}
Note that higher velocity of sound $c$ corresponds to lower fluctuations of multiplicity $\varpi$. From the experimental data \cite{NA61,GGS,RecentNA61} one can deduce that
\begin{equation}
\frac{c_{Pb+Pb}}{c_{p+p}} \simeq 1.04\quad{\rm and}\quad \frac{\varpi_{p+p}}{\varpi_{Pb+Pb}} \simeq 1.29 \pm 0.04 . \label{results}
\end{equation}
Therefore, using Eq. (\ref{SV}), the expected ratio of $\gamma T$ for $p+p$ and $Pb+Pb$ collisions at a beam energy of $158$ GeV is $\left(\gamma T\right)_{p+p}/\left( \gamma T\right)_{Pb+Pb} \sim 1.2 $. This could be the subject of further experimental investigations. Note that the above results may be connected with the pair correlation function, $g^{(2)}(r)$, because the scaled variance can be written as \cite{FS}
\begin{equation}
\rho_0 \kappa_T T = 1 + \rho_0 \int d \vec{r} \left[ g^{(2)}(r) - 1\right]. \label{Corrrel_g}
\end{equation}
As shown in \cite{RW} , for central nuclear collisions the number of binary collisions exceeds that of wounded nucleons (each nucleon participates in a number of collisions with other nucleons). This results in the correlation function becoming negative which, in turn, leads to a diminishing of fluctuations of multiplicity (because the variance of the total multiplicity from a number of particular collisions is smaller that the sum of the variances of independent nucleon-nucleon collisions).

\section{Oscillation of modified combinants for multiplicity distributions}
\label{sec:Combinants}

We shall concentrate now on the the multiplicity distribution, $P(N)$, which is another important characteristic of the  multiparticle  production process and is among  the first observables measured in any multiparticle production experiment \cite{Kittel} . The most popular, the Binomial Distribution (BD), Poisson Distribution (PD) and Negative Binomial Distribution (NBD), were already discussed from the point of view of their relations with nonextensivity in Section \ref{sec:Fluct} (cf. Eqs. (\ref{eq:BD}), (\ref{eq:PD}) and (\ref{eq:NBD}) for definitions). They all satisfy the recurrence relation (\ref{eq:dep}) introduced in Section \ref{sec:SelfSim}. For convenience we gather them together in Table \ref{TablePN} in slightly different notation. Suitable modifications of $g(N)$ result in more involved distributions $P(N)$ (cf. \cite{JPG} for references).

\begin{table}[h]
\tbl{The most popular forms of $P(N)$ emerging from the recurrence relation $(N+1)P(N+1) = g(N)P(N)$ where $g(N) = \alpha + \beta N$ }
{\begin{tabular}{@{}l|clcll@{}} \toprule
                                                                               &                          &  \\
BD      & ~~ & $P(N) = \frac{K!}{N!(K - N)!} p^N (1 - p)^{K-N}$                & ~~ & $\alpha = \frac{Kp}{1-p}$ & $\beta = -\frac{\alpha}{K}$  \\
        &    &                                                                 &    &                      &   \\
PD      & ~~ &$P(N) = \frac{\lambda^N}{N!} \exp( - \lambda)$                   & ~~ & $\alpha = \lambda$        & $\beta = 0$        \\
        &    &                                                                 &    &                      &    \\
NBD     & ~~ &$P(N) = \frac{\Gamma(N+k)}{\Gamma(N+1)\Gamma(k)} p^N (1 - p)^k$  & ~~ & $\alpha = kp$             & $\beta = \frac{\alpha}{k}$   \\
        &    &                                                                 &     &                      &     \\
\botrule
\end{tabular} \label{TablePN}}
\end{table}
\begin{figure}[h]
\begin{center}
\includegraphics[scale=0.25]{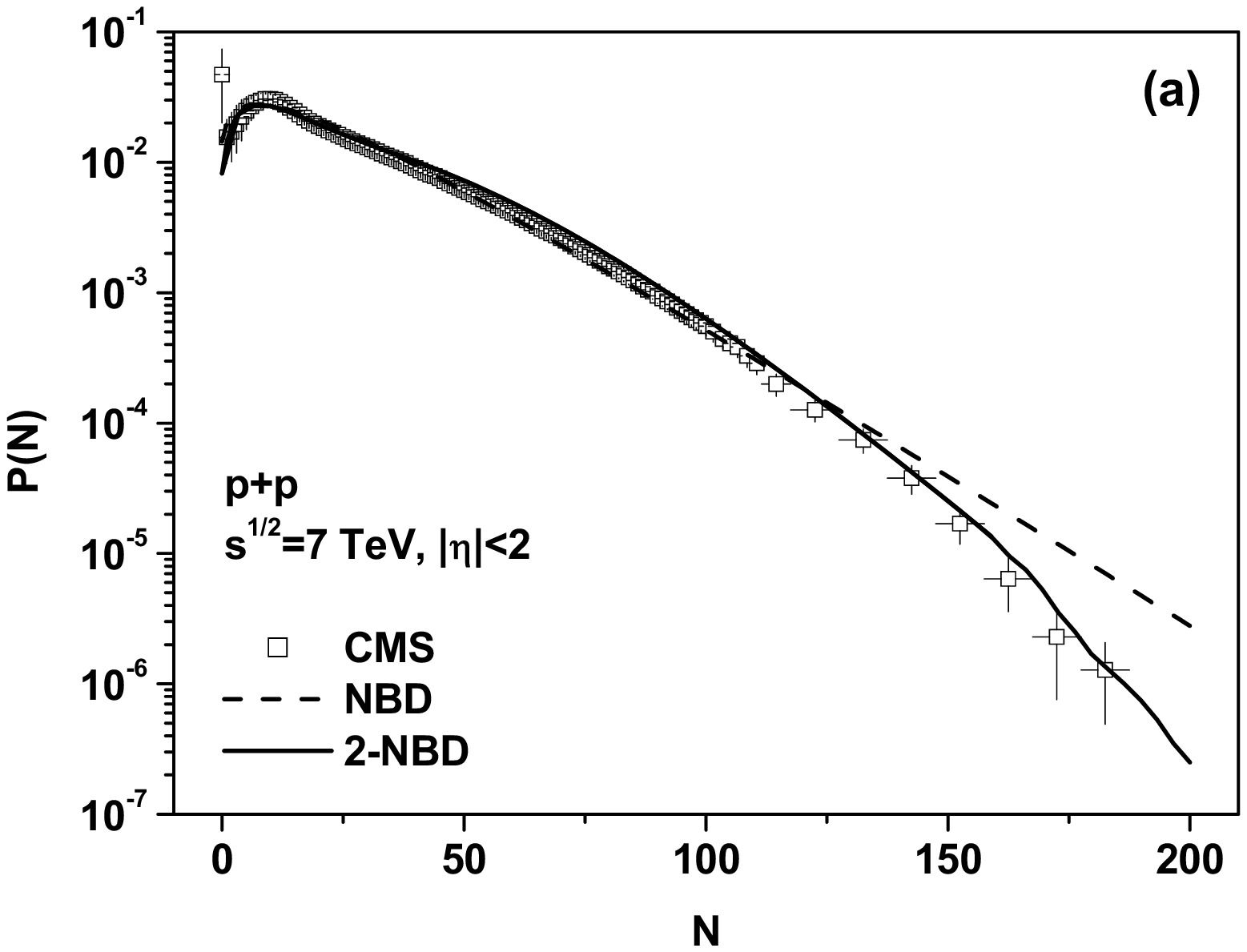}
\includegraphics[scale=0.25]{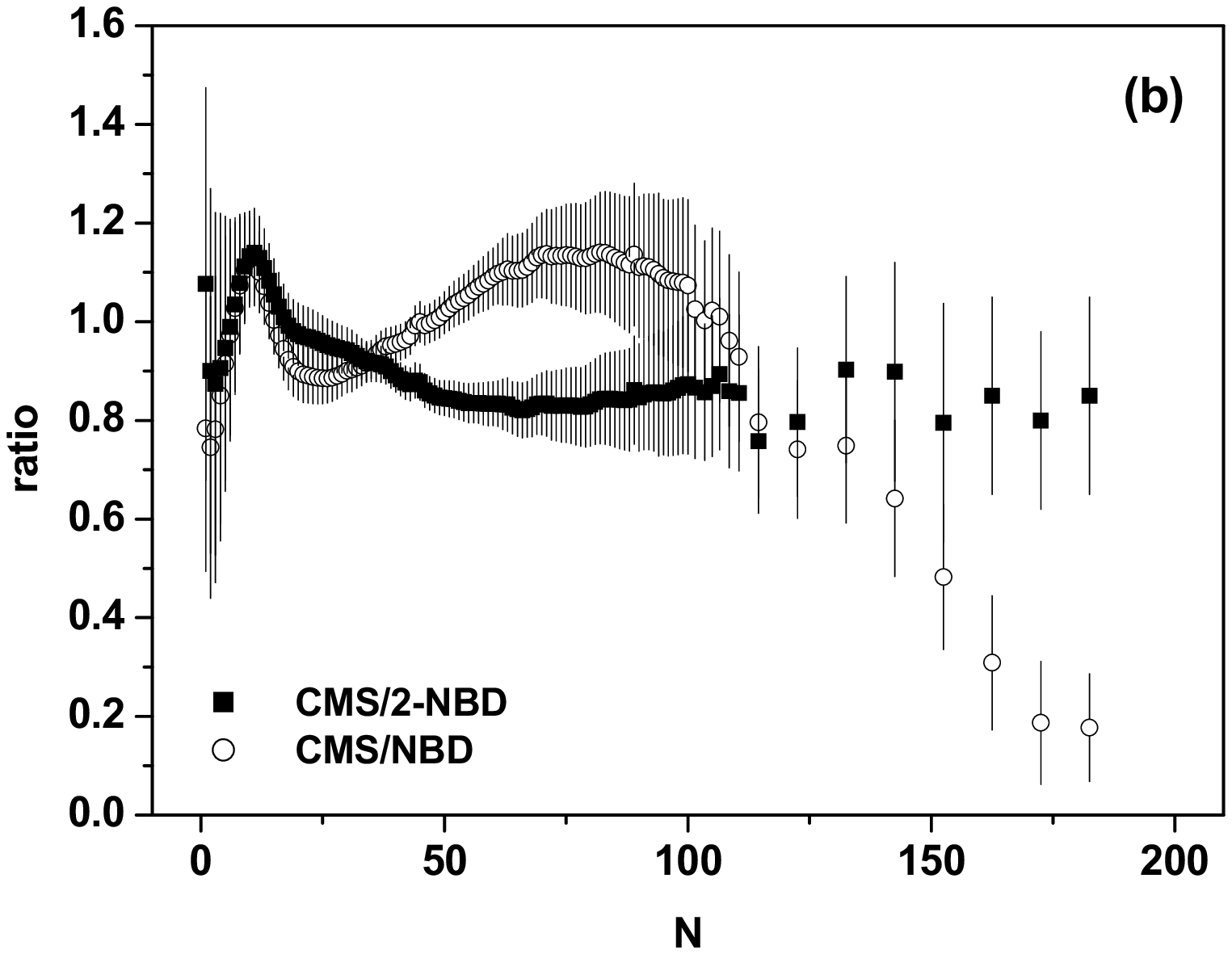}
\includegraphics[scale=0.25]{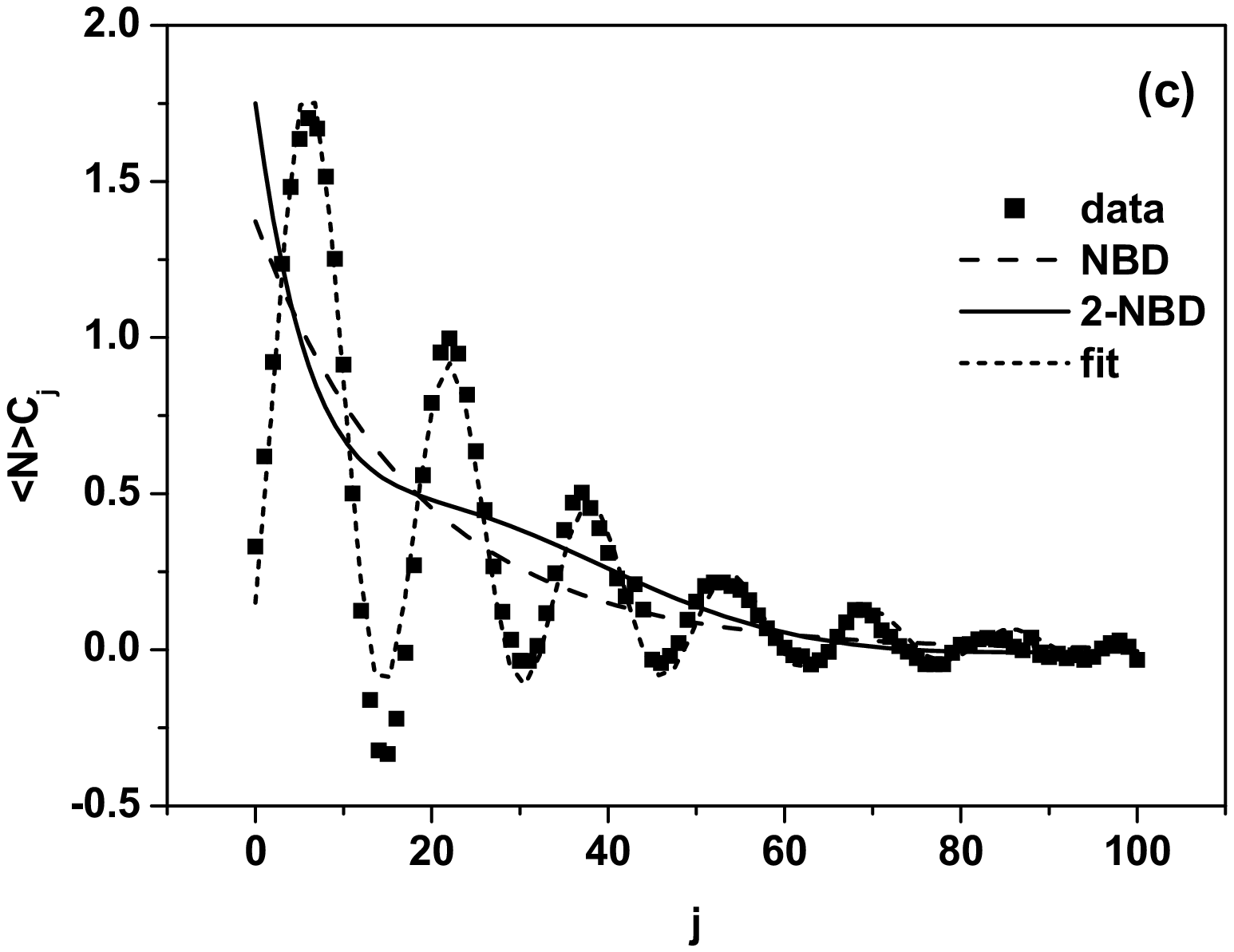}
\end{center}
\vspace{-5mm}
\caption{$(a)$ Charged hadron multiplicity distributions for the pseudorapidity range $|\eta| < 2$ at $\sqrt{s} = 7$ TeV, as given by the CMS experiment \cite{CMS-4} (points), compared with the NBD for parameters $\langle N\rangle = 25.5$ and $k = 1.45$ (dashed line) and with the $2$-component NBD (solid line) with parameters from \cite{NBD-PG} . $(b)$ Multiplicity dependence of the ratio $R = P_{CMS}(N)/P_{fit}(N)$ for the NBD (circles) and for the $2$-component NBD for the same data as in panel $(a)$ (squares). $(c)$ Coefficients $C_j$ emerging from the data and NBD fits presented in panel $(a)$. The data points are fitted by Eq. (\ref{CjFit}), see text for details.
}
\label{F3}
\end{figure}

Usually the first choice of distribution is the NBD. However, with growing energy and  number of produced secondaries it deviates more and more from the data for large $N$ and is replaced there either by combinations of two \cite{GU} , three \cite{Z} or multi-component NBDs \cite{DN} , or by some other form of $P(N)$ \cite{Kittel,DG,KNO-2} . To illustrate this problem,  in Fig. \ref{F3} $(a)$ data from \cite{CMS-4} are compared with a single NBD and $2$-NBD with parameters taken from \cite{NBD-PG} . The first impression is that improvement is quite substantial, but looking closer at the ratio $R = P_{CMS}(N)/P_{fit}(N)$ one recognizes that it concerns only the large $N$ tail, the small $n$ part remains essentially as bad as before. Following the same reasoning as in Section \ref{sec:Surprises} we take these observations seriously and assume that there is some additional information hidden in the $P(N)$. Therefore we concentrate now on the question of how to retrieve this information \cite{JPG} . For this purpose we shall use a more general form of recurrence relation given in Eq. (\ref{eq:dep}) and Table \ref{TablePN}, which is used in counting statistics when dealing with cascade stochastic processes \cite{CSP} . Its characteristic feature is that it connects all multiplicities by means of some coefficients $C_j$ which define the corresponding $P(N)$ in the following way:
\begin{equation}
(N + 1)P(N + 1) = \langle N\rangle \sum^{N}_{j=0} C_j P(N - j). \label{Cj}
\end{equation}
These coefficients contain the memory of particle $N+1$ about all $N-j$ previously produced particles. Eq. (\ref{Cj}) can be reversed and we thereby obtain a recurrence formula for the coefficients $C_j$ for some experimentally measured multiplicity distribution $P(N)$ \cite{JPG}  \footnote{A comment is needed at this point. In a recent review on the phenomenology of multiplicity distributions \cite{Alkin} our use of recurrence relation (\ref{Cj}) was criticized by attributing the observed fluctuation effect to the possible peculiarities of the experimental unfolding procedure used while preparing data \cite{Cowan} . However, to the best of our knowledge, such a statement was so far not substantiated by any known experimental analysis of the procedure used, therefore we assume that this is most likely a real effect connected with some dynamical features of the production mechanism. In fact in \cite{CSF,CSF1} the cascade stochastic processes leading to Eq. (\ref{Cj}) were successfully applied to multiparticle phenomenology.} :
\begin{equation}
\langle N\rangle C_j = (j+1)\left[ \frac{P(j+1)}{P(0)} \right] - \langle N\rangle \sum^{j-1}_{i=0}C_i \left[ \frac{P(j-i)}{P(0)} \right]. \label{rCj}
\end{equation}
As can be seen in Fig. \ref{F3} $(c)$ the coefficients $C_j$ obtained from the data presented in Fig. \ref{F3} $(a)$ show distinct oscillatory behaviour (with period roughly equal to $16$), gradually disappearing with $N$, which can be fitted by the following formula:
\begin{equation}
\langle N\rangle C_j = \left( a^2 + b^2\right)^{j/2} \sin\left[ c + j \arctan(b/a)\right] +d^j, \label{CjFit}
\end{equation}
with parameters: $a = 0.89$, $b = 0.37$, $c = 5.36$, $d = 0.95$. As seen in Fig. \ref{F3} $(c)$ such oscillations do not appear in the single NBD fit and there is only a small trace of them for the $2$-NBD fit presented there. Notice that for both models the ratio $R=data/fit$ presented in Fig. \ref{F3} $(b)$ signals that there are problems with the low multiplicity region. In Fig. \ref{Examples} one can see that this effect is also seen for different pseudorapidity windows (Fig. \ref{Examples} $(a)$) and in data from other experiments (see Fig. \ref{Examples} $(b)$ with ALICE \cite{ALICE-4} data, Fig. \ref{Examples} $(c)$ shows that in both cases the pattern is similar). Fig. \ref{Examples} $(d)$ shows the simplest (albeit unphysical) way to obtain oscillatory behaviour of an otherwise smooth distribution: it is enough to distort it at some point and this distortion propagates further. The fact that we do not observe any fluctuations for the single NBD is obvious once one realizes that for the NBD  dependence of the corresponding $C_j$ on the rank $j$ has the form \cite{JPG} :
\begin{equation}
C_j = \frac{k}{\langle N\rangle} p^{j+1} = \frac{k}{k + m} \exp( j \ln p). \label{CjNBD}
\end{equation}

\begin{figure}
\begin{center}
\resizebox{0.47\textwidth}{!}{ \includegraphics{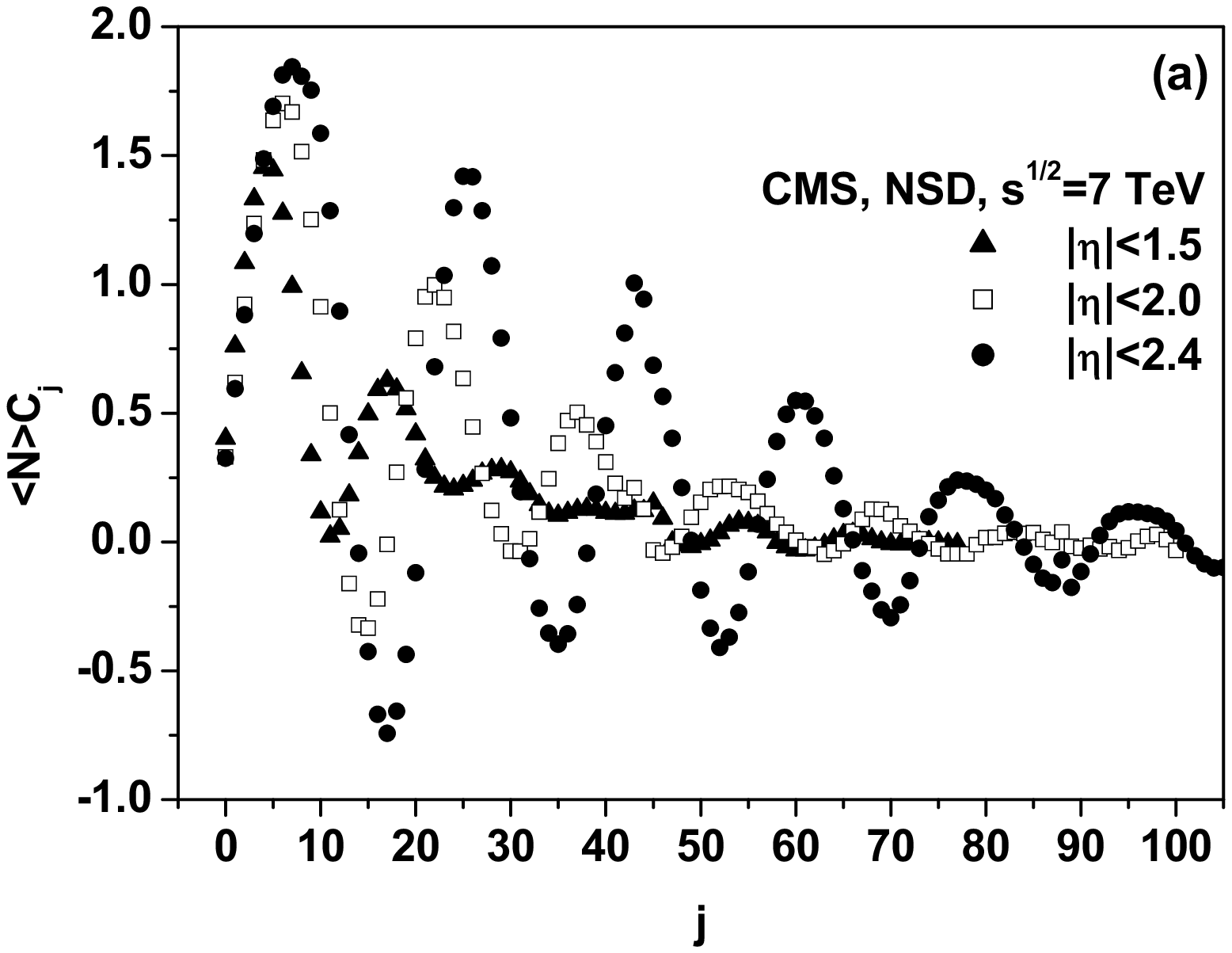}}
\resizebox{0.47\textwidth}{!}{ \includegraphics{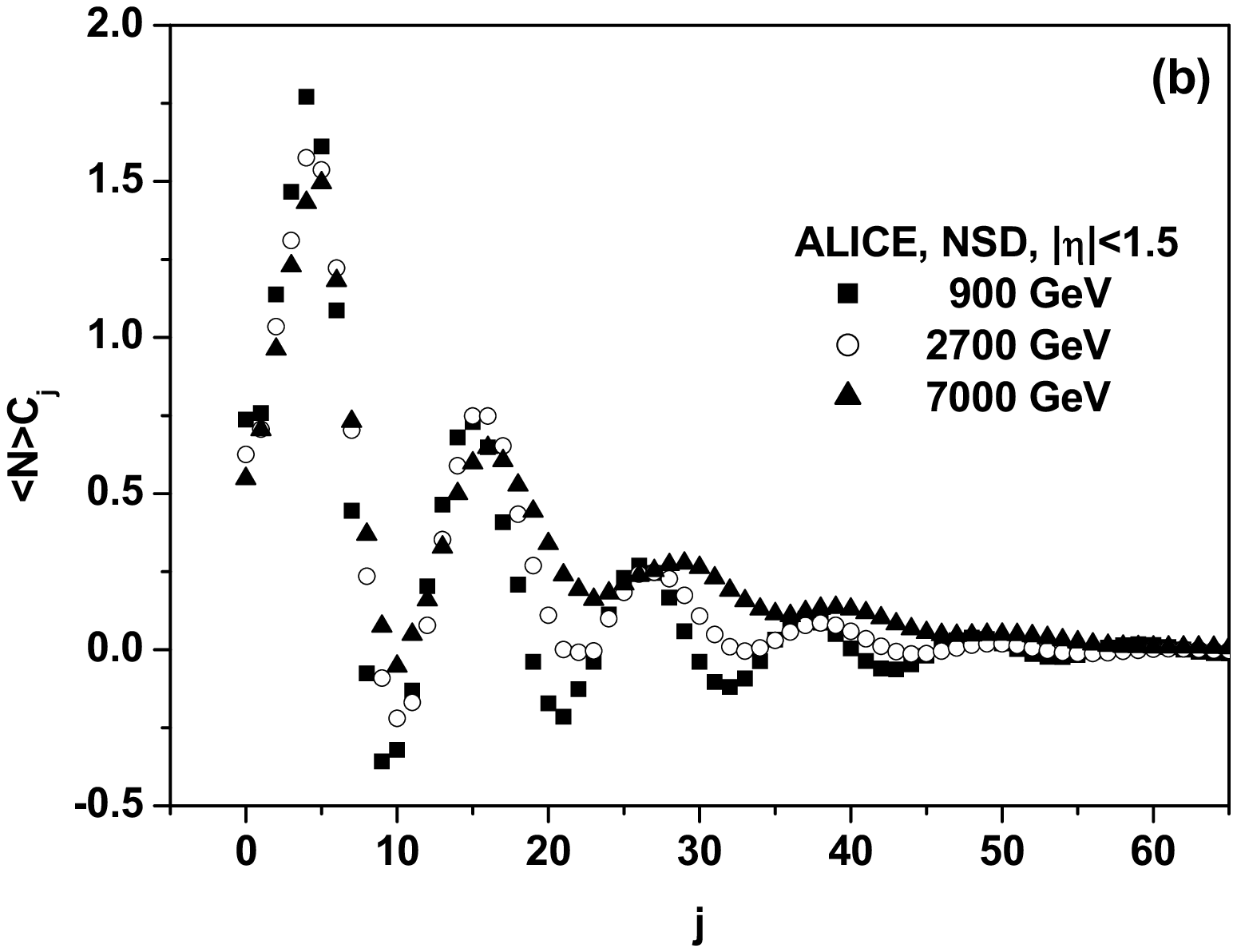} }
\vspace{-3mm}
\resizebox{0.47\textwidth}{!}{ \includegraphics{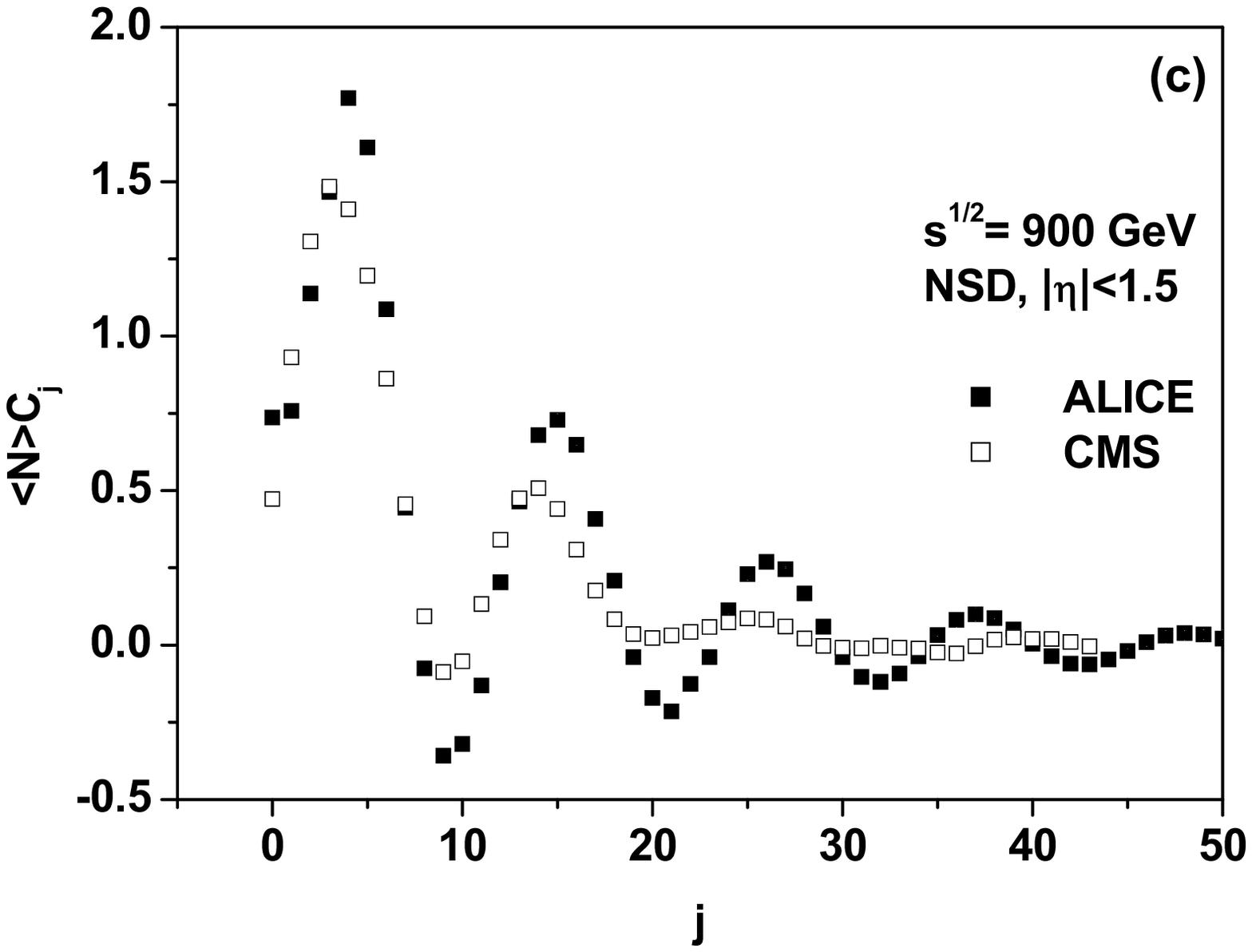} }
\resizebox{0.47\textwidth}{!}{ \includegraphics{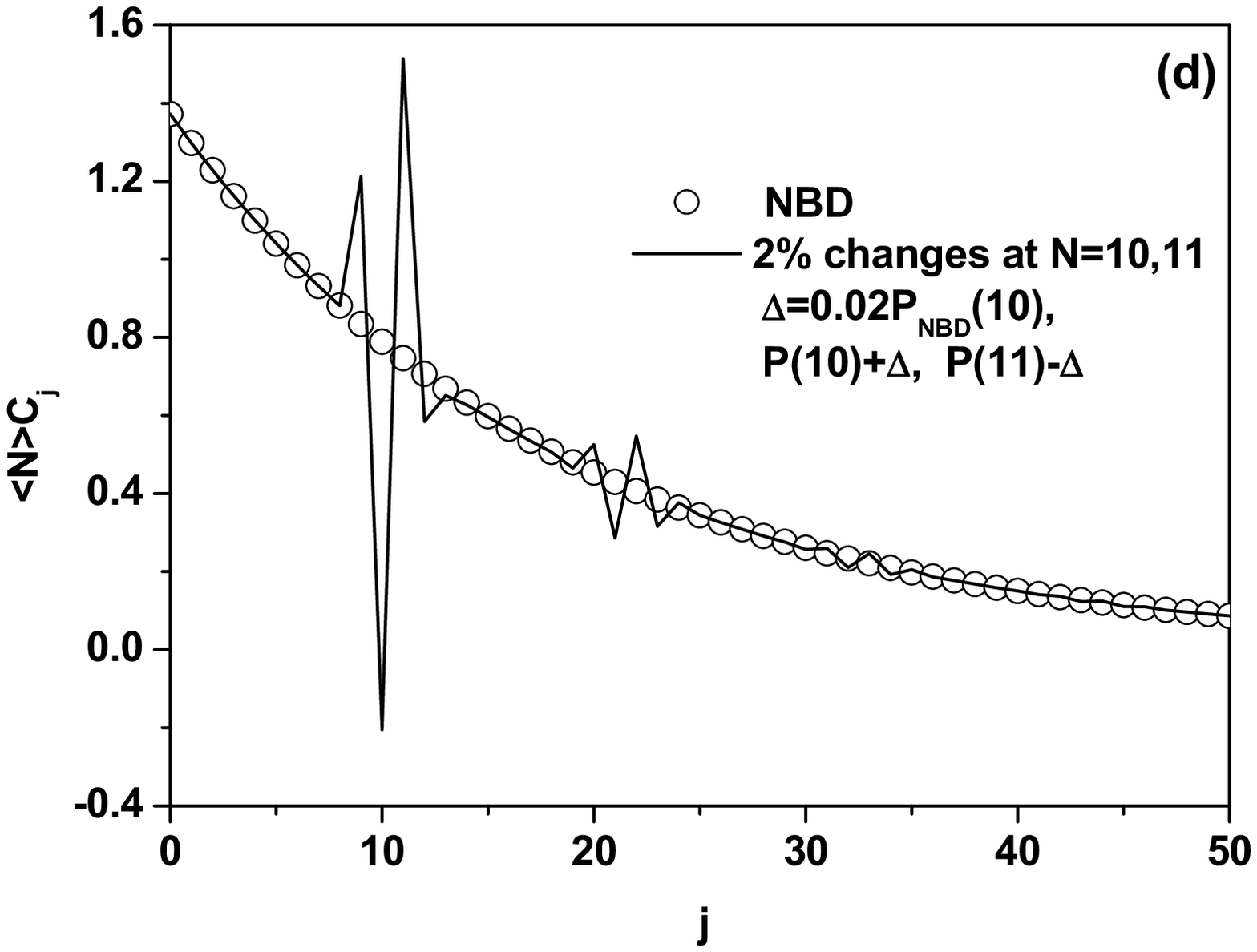} }
\vspace{-2mm}
\caption{$(a)$ Coefficients $C_j$ emerging from the CMS data at $\sqrt{s} = 7$ TeV for different pseudorapidity windows \cite{CMS-4} . $(b)$ Coefficients $C_j$ emerging from the ALICE data \cite{ALICE-4} . $(c)$ Comparison of coefficients $C_j$ emerging from the ALICE \cite{ALICE-4} and CMS \cite{CMS-4} data taken for $\sqrt{s} = 900$ GeV and for $|\eta| < 1.5$ pseudorapidity window. $(d)$ Illustration of how the oscillatory behavior of the coefficients $C_j$ emerges. See text for details.} \label{Examples}
\end{center}
\end{figure}

The coefficients $C_j$ are closely related to the so called {\it combinants} $C^{\star}_j$ introduced in \cite{KG} (see also \cite{Kittel,CombUse,Hegyi}) and defined in terms of the generating function $G(z)\sum^{\infty}_{N=0} P(N) z^N $ as
\begin{equation}
C^{\star}_j = \frac{1}{j!} \frac{d^j \ln G(z)}{d z^j}\bigg|_{z=0}\qquad {\rm or}\qquad \ln G(z) = \ln P(0) + \sum^{\infty}_{j=1} C^{\star}_j z^j. \label{CombDef}
\end{equation}
This means then that \cite{JPG}
\begin{equation}
C_j = \frac{j+1}{\langle N\rangle} C^{\star}_{j+1} \label{connection}
\end{equation}
and one can rewrite the recurrence relation, Eq. (\ref{Cj}),  in terms of the combinants $C_j^{\star}$:
\begin{equation}
(N + 1)P(N + 1) = \sum^{N}_{j=0} (j+1)C_j^{\star} P(N - j). \label{Cj_star}
\end{equation}
This allows us to express the coefficients $C_j$, which henceforth we shall call {\it modified combinants}, by the generating function $G(z)$ of $P(N)$:
\begin{equation}
\langle N\rangle C_j = \frac{1}{j!} \frac{ d^{j+1} \ln G(z)}{d z^{j+1}}\bigg|_{z=0}. \label{GF_Cj}
\end{equation}
This relation will be used in what follows when calculating the $C_j$ from distributions defined by some $G(z)$.

To continue our reasoning, note that, contrary to the NBD, the corresponding modified combinants for the BD oscillate rapidly,
\begin{equation}
C_j = (-1)^j \frac{K}{\langle N\rangle} \left( \frac{\langle N\rangle}{K-\langle N\rangle}\right)^{(j+1)} = \frac{(-1)^j}{1 - p}\left( \frac{p}{1 - p}\right)^{j}, \label{C_jBD}
\end{equation}
with a period equal to $2$. In Fig. \ref{F4} $(a)$ one can see that the amplitude of these oscillations depends on $p$, generally the $C_j$ increase with rank $j$ for $p > 0.5$ and decrease for $p < 0.5$. However, their general shape lacks the fading down feature of the $C_j$ observed experimentally. This indicates that the BD alone cannot explain data if used alone. We resort therefore to the idea of {\it compound distributions} (CD) \cite{Compound} \footnote{In fact, in an attempt to explain Bose-Einstein correlations we have used a combination of $k$ {\it elementary emitting cells (EEC)} producing particles according to a geometrical distribution \cite{BSWW} . For $k=const$  the resultant $P(N)$ was of NBD type, for $k$ distributed according to the BD it was a modified NBD. However, using this example we could not find a set of parameters providing both the observed $P(N)$ and oscillating $C_j$. Note that originally the NBD was seen as a compound Poisson distribution with the number of clusters given by a Poissonian distribution and the particle inside the clusters distributed according to a logarithmic distribution \cite{GVH} . }. From our point of view CD could, for example, describe a production process in which a number $M$ of some objects (clusters/fireballs/etc.) is produced following some distribution $f(M)$ (with generating function $F(z)$), which subsequently decay  independently into a number of secondaries, $n_{i = 1,\dots, M}$, always following some other (the same for all) distribution, $g(n)$ (with a generating function $G(z)$). The distribution $h( N ) = f\otimes  g$, where $ N =\sum_{i=0}^M n_i$, is a compound distribution of $f$ and $g$ for which
\begin{equation}
\langle N\rangle = \langle M\rangle \langle n\rangle\qquad {\rm and}\qquad Var(N) = \langle M\rangle Var(n) + Var(M) \langle n\rangle^2 \label{CD_moments}
\end{equation}
and its generating function $H(z)$ is equal to,
\begin{equation}
H(z) = F[G(z)]. \label{CD_GF}
\end{equation}
For the class of distributions of $M$ that satisfy our recursion relation Eq. (\ref{eq:dep}) the compound distribution $h=f\otimes  g$ is given by the so called Panjer's recursion relation \cite{Panjer} ,
\begin{equation}
Nh(N) = \sum^{N}_{j=1}[ \beta N + (\alpha - \beta)j ] g(j) h(N-j) = \sum^{N}_{j=1} C^{(P)}_j (N) h(N-j),
 \label{PP}
\end{equation}
with the initial value $h(0)=f(0)$. However, the coefficients $C^{(P)}_j$ occurring here depend on $N$, contrary to our recursion relation given by Eq. (\ref{Cj}) where the modified combinants, $C_j$, are independent of $N$. Moreover,  Eq. (\ref{Cj}) is not limited to the class of distributions satisfying Eq.(\ref{eq:dep}) but is valid for any distribution $P(N)$. For this reason the recursion relation Eq. (\ref{PP}) is not suitable for us.
\begin{figure}[h]
\begin{center}
\includegraphics[scale=0.35]{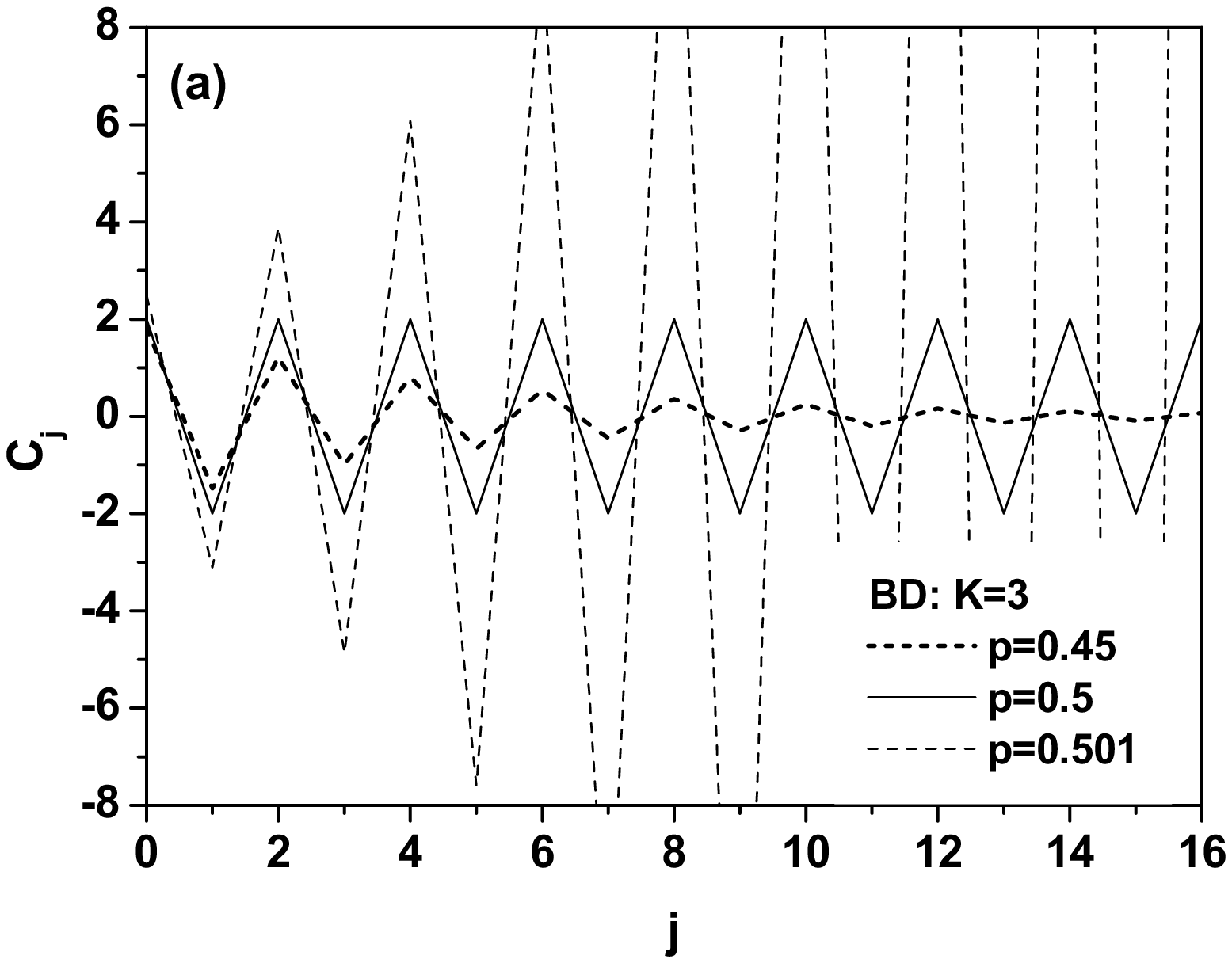}\hspace{5mm}
\includegraphics[scale=0.35]{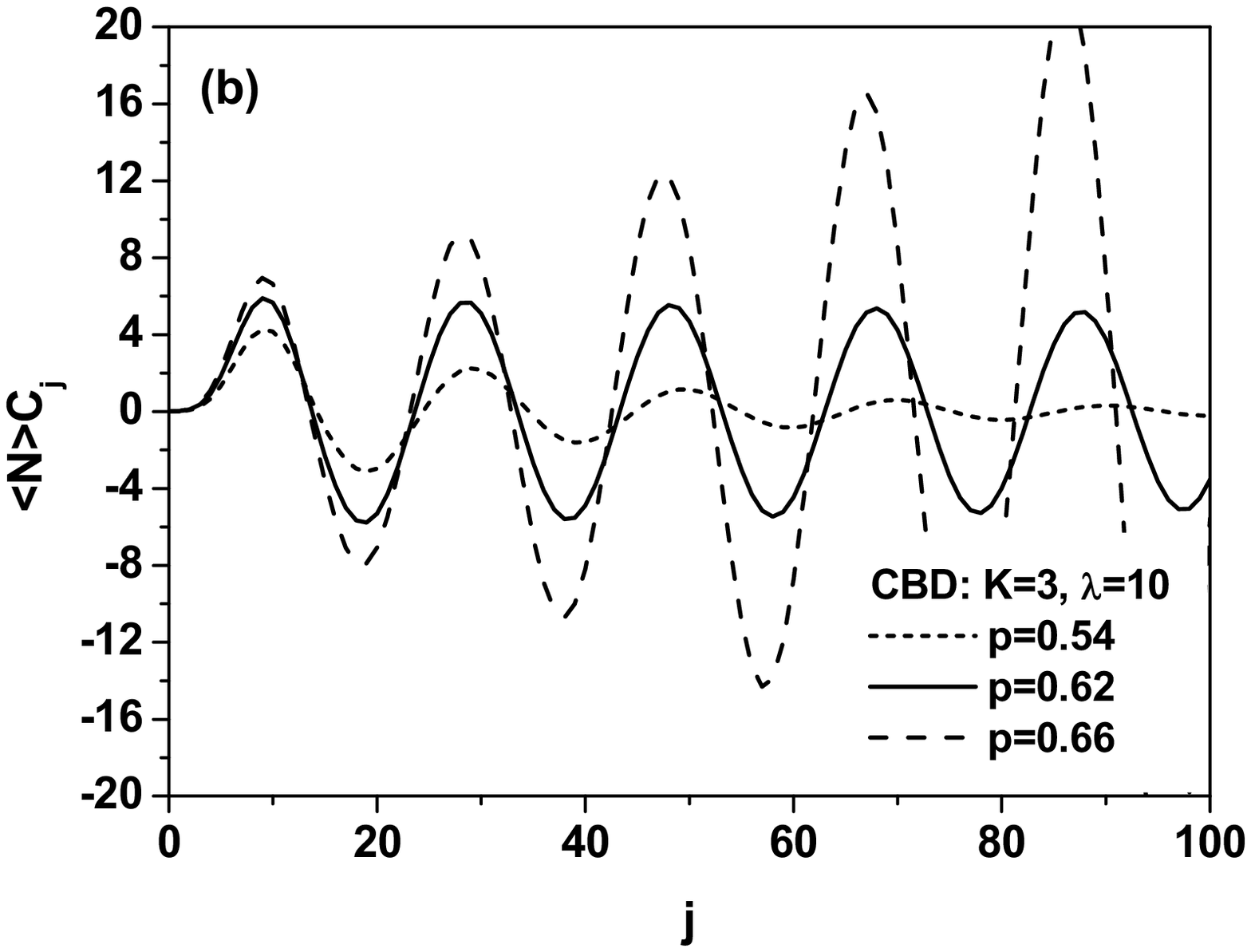}
\end{center}
\vspace{-4mm}
\caption{ $(a)$ Examples of $C_j$ for Binomial Distributions $(a)$ (from Eq. (\ref{C_jBD})). $(b)$ Compound Binomial Distributions from Eq. (\ref{C_BD_PD}).
}
\label{F4}
\end{figure}

To visualize the compound distribution in action we take $f$ as a Binomial Distribution with generating function $F(z) = (pz + 1 - p)^K$, and  $g$  a Poisson distribution with generating function $G(z) = \exp[ \lambda (z-1)]$. The generating function of the resulting Compound Binomial Distribution (CBD) is equal to:
\begin{equation}
H(z) = \left\{ p \exp[ \lambda (z-1)] + 1 -p \right\}^K \label{H_FG}
\end{equation}
and the corresponding modified combinants are:
\begin{eqnarray}
\langle N\rangle C_j &=& \frac{K \lambda^{j+2} \exp(-\lambda)}{j!} \sum^{j+2}_{i=1} \left[ \frac{p}{1 - p + p \exp(-\lambda)}\right]^i \frac{1}{i}\sum^{i}_{k=0} (-1)^{k+1}\binom{i}{k} k^{j+1} = \nonumber\\
&=& \frac{K \lambda^{j+2} \exp(-\lambda)}{j!} \sum^{j+2}_{i=1}\left[ \frac{p}{1 - p + p \exp(-\lambda)}\right]^i S(j+1,i) \label{C_BD_PD}
\end{eqnarray}
where
\begin{equation}
S(n,k) = \left\{ \begin{array}{c}
n\\
k
\end{array}\right\} = \frac{1}{k!} \sum_{i=0}^{k} (-1)^{k-i} \binom{k}{i}i^n \label{S}
\end{equation}
is the Stirling number of the second kind. Fig. \ref{F4} $(b)$ shows the above modified combinants for the CBD with $K=3$ and $\lambda =10$ calculated for three different values of $p$ in the BD: $p=0.54,~0.62,~0.66$. Note that in general the period of the oscillations is equal to $2\lambda$, i.e., in Fig. \ref{F4} $(b)$ where $\lambda = 10$ it is equal to $20$. The multiplicity distribution in this case is
\begin{eqnarray}
P(0) &=& \left( 1 - p + p e^{-\lambda}\right)^K,\\   \label{P(N)}
P(N)\! &=&\! \frac{1}{N!} \frac{d^N H(z)}{dz^N}\bigg|_{z=0}\!\! =\! \frac{1}{N!}\sum^K_{i=1}i!\binom{K}{i}\left( \lambda p e^{-\lambda}\right)^i \left( 1 - p + p e^{-\lambda}\right)^{K-i} S(N,i) \nonumber
\end{eqnarray}
(the proper normalization comes from the fact that $H(1)=1$). This shows that the choice of a BD as the basis of the used CD is crucial to obtain oscillatory $C_j$ (for example, a compound distribution formed from a NBD and some other NBD provides smooth $C_j$).

Unfortunately, such a single component CBD (depending on three parameters: $p$, $K$ and $\lambda$, $P(N) = h(N;p,K,\lambda)$, does not describe the experimental $P(N)$. We return therefore to the idea of using a multicomponent version of the CBD, for example a $3$-component CBD defined as (with $w_i$ being weights):
\begin{equation}
P(N) = \sum_{i=1,2,3} w_i h\left(N; p_i, K_i, \lambda_i\right);\qquad \qquad \sum_{1=1,2,3} w_i = 1.  \label{3CBD}
\end{equation}
The results of using Eq. (\ref{3CBD}) (with parameters: $\omega_1 = 0.34$, $\omega_2 = 0.4$, $\omega_3 = 0.26$; $p_1 = 0.22$, $p_2 = 0.22$, $p_3 = 0.12$; $K_1 = 10$, $K_2 = 12$, $K_3 = 30$ and $\lambda_1 = 4$, $\lambda_2 = 9$, $\lambda_3 = 14$) are presented in Fig. \ref{FigCBD-1}. As one can see this time the fit to $P(N)$ is quite good and the modified combinants $C_j$ follow an oscillatory pattern as far as the period of the oscillations is concerned, albeit their amplitudes still decay too slowly.
\begin{figure}[h]
\begin{center}
\includegraphics[scale=0.35]{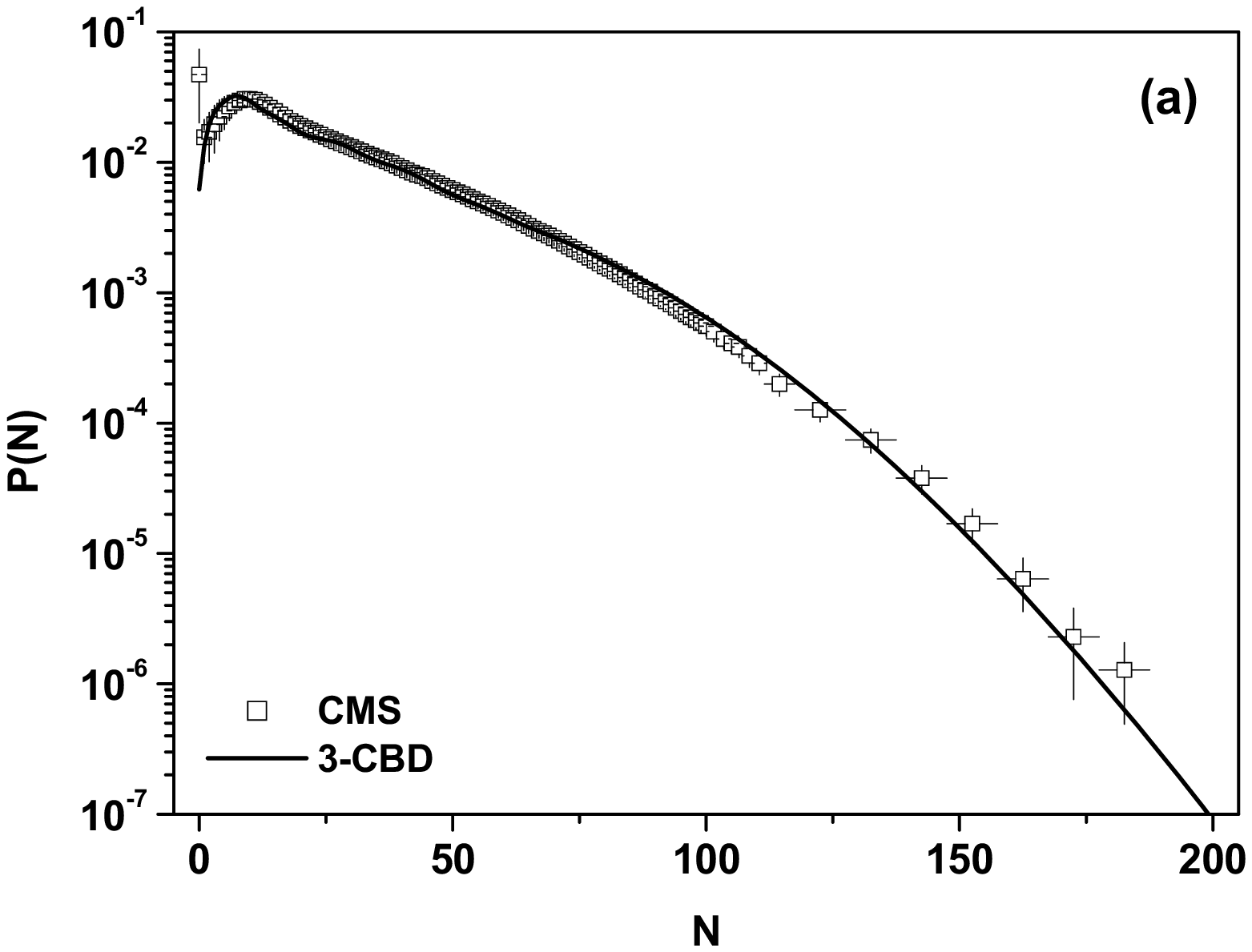}
\includegraphics[scale=0.35]{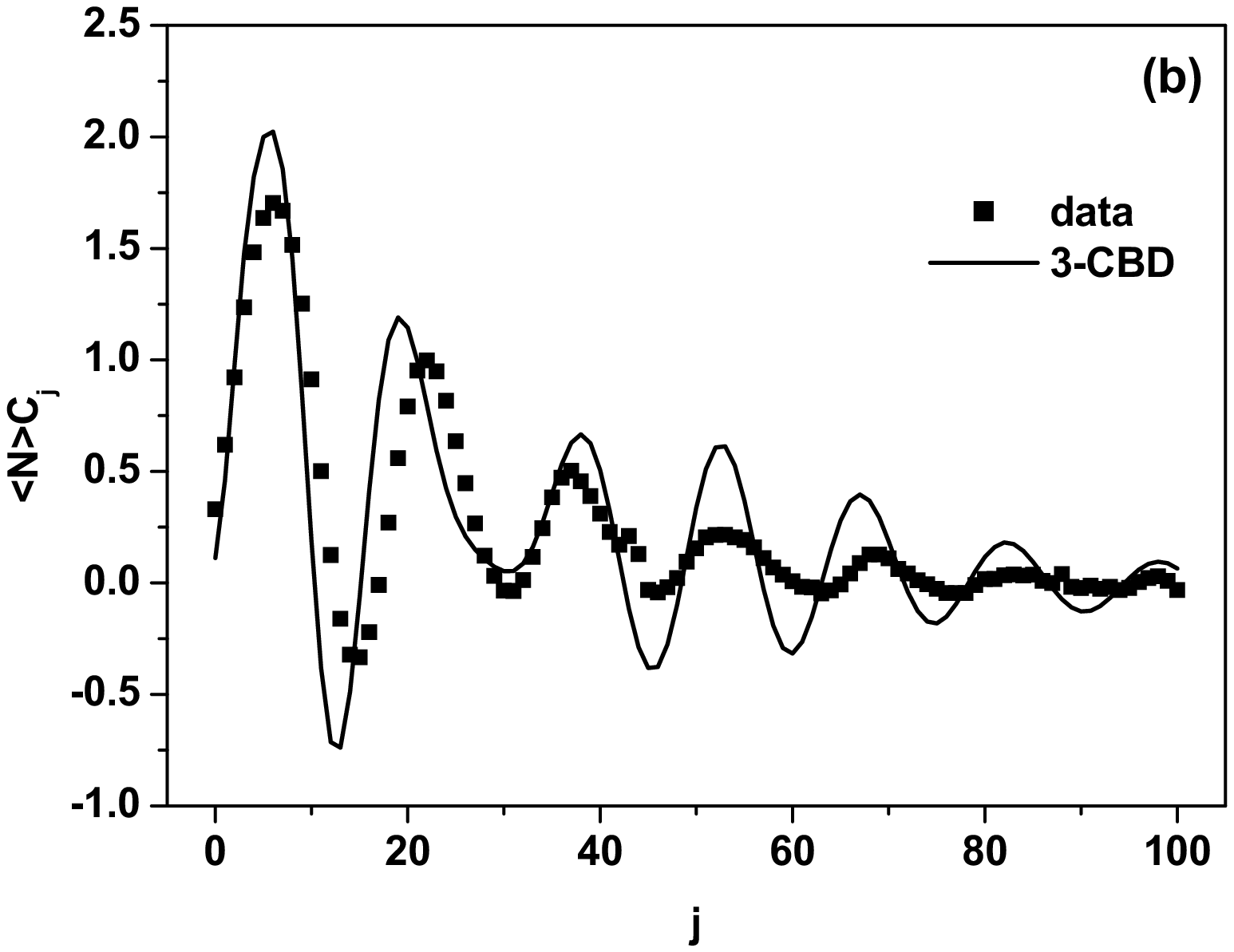}
\end{center}
\vspace{-5mm}
\caption{$(a)$  Charged hadron multiplicity distributions for $|\eta| < 2$ at $\sqrt{s} = 7$ TeV, as given by the CMS experiment \cite{CMS-4} (points), compared with a $3$-component CBD, Eq. (\ref{3CBD}). $(b)$ Coefficients $C_j$ emerging from the CMS data used in panel $(a)$ compared with the corresponding $C_j$ obtained from the  $3$-component compound binomial distribution ($3$-CBD).}
\label{FigCBD-1}
\end{figure}

For comparison we present a $2$-component version of a CBD compounded of a BD and a NBD whose generating function is
\begin{equation}
H(z) = \left[ p\left( \frac{1 - p'}{1 - p's}\right)^k + 1 - p\right]^K,\qquad {\rm where}\qquad p' = \frac{m}{m + k}. \label{2-CBD}
\end{equation}
In this case we have
\begin{equation}
P(N) = \sum_{i=1,2} w_i h\left(N; p_i, K_i, \lambda_i\right);\qquad \qquad \sum_{1=1,2} w_i = 1.  \label{2CBD}
\end{equation}
The results of using Eq. (\ref{2CBD}) (with parameters:  $K_1 = K_2 = 3$, $p_1 = p_2 = 0.7$, $k_1 = 4$, $k_2 = 2.2$, $m_1 = 6$, $m_2 = 18.5$ and $w_1 = w_2 = 0.5$) look even better than before. This means that using multicomponent compound distributions based on BD (responsible for oscillations of $C_j$) and some other distribution providing damping of oscillations for large $N$, one could probably describe the data. However, this immediately rises the problem of the interpretation of such an approach for which we do not find at the moment a solution.
\begin{figure}[t]
\begin{center}
\includegraphics[scale=0.35]{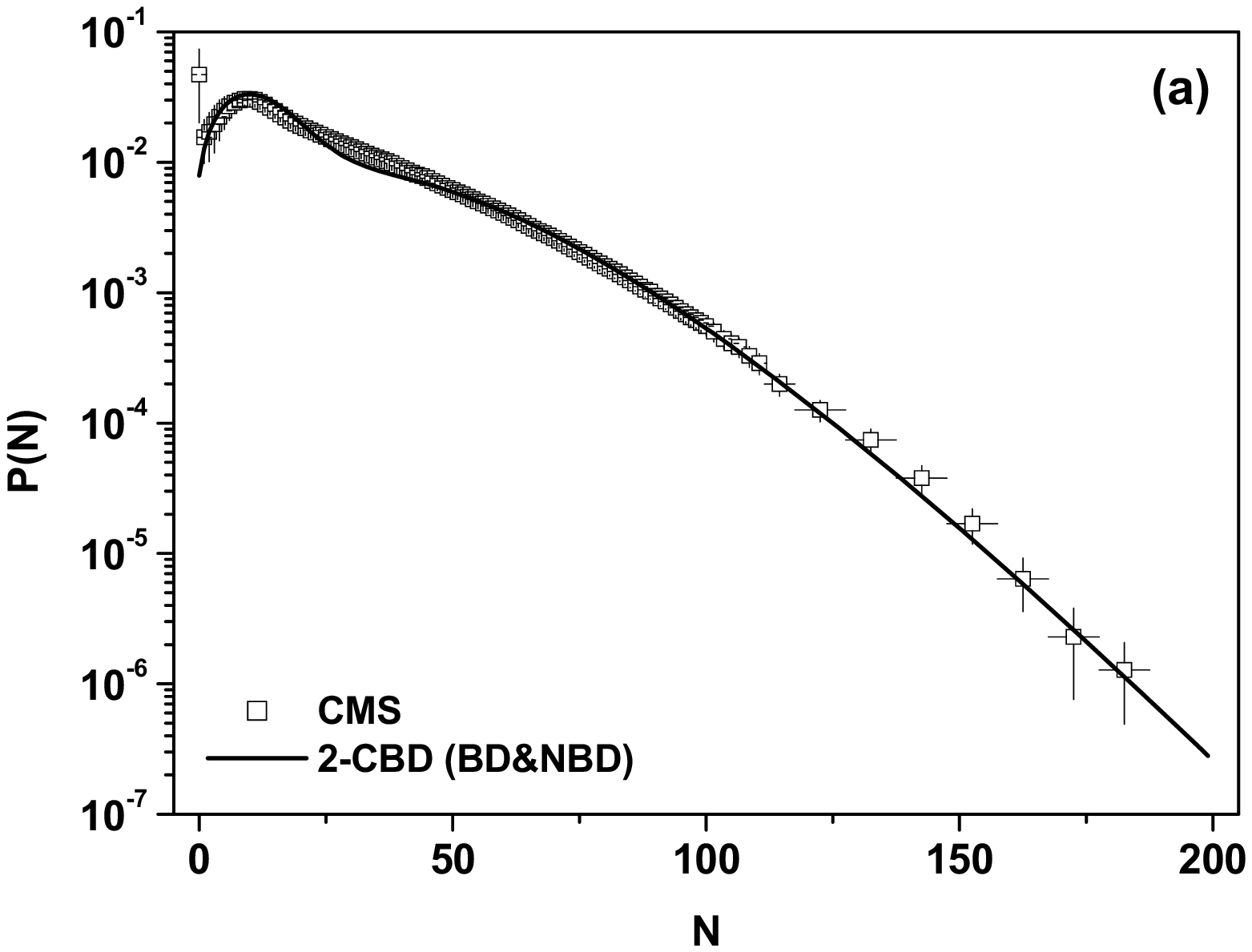}
\includegraphics[scale=0.35]{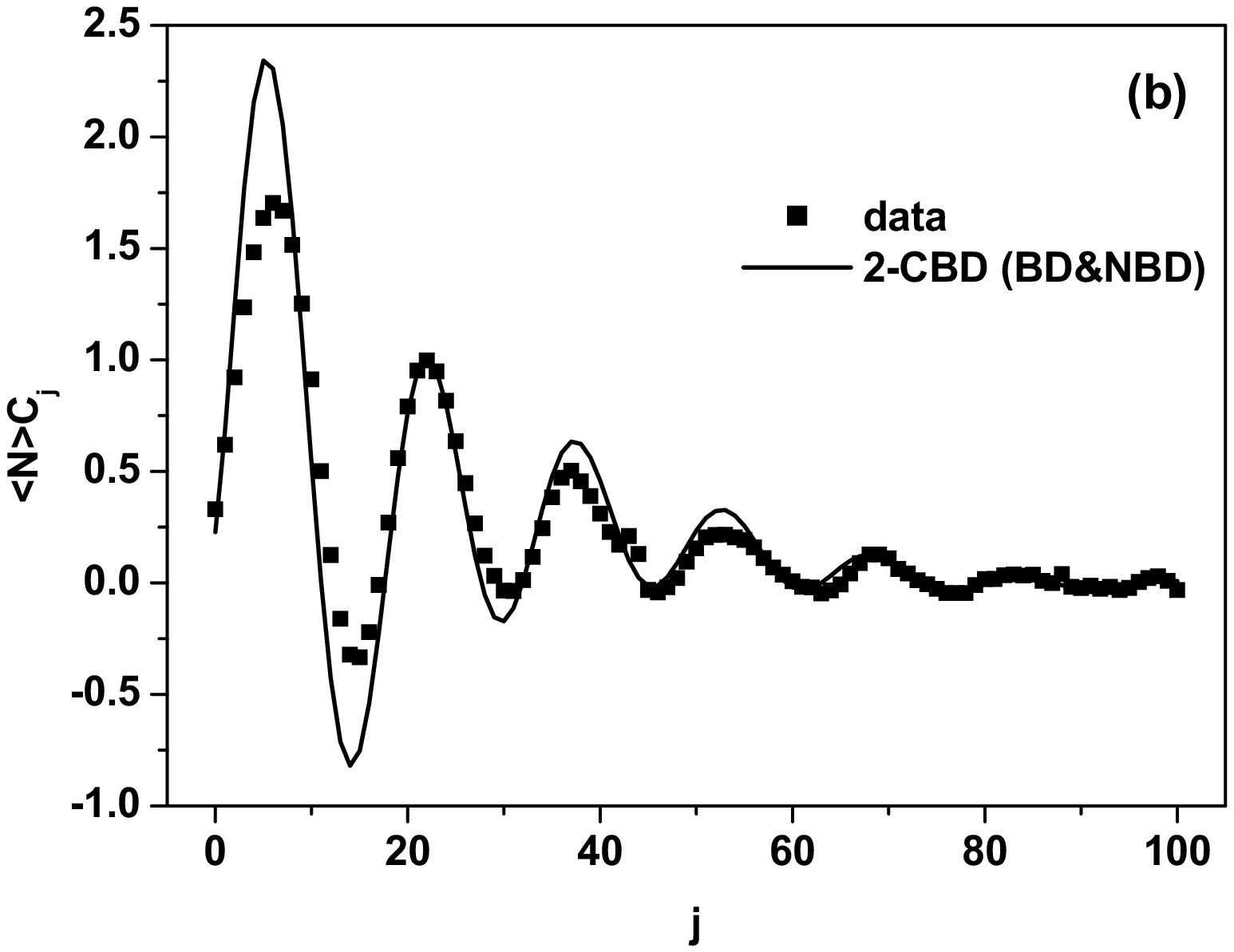}
\end{center}
\vspace{-5mm}
\caption{$(a)$  Charged hadron multiplicity distributions for $|\eta| < 2$ at $\sqrt{s} = 7$ TeV, as given by the CMS experiment \cite{CMS-4} (points), compared with a $3$-component CBD, Eq. (\ref{2CBD}). $(b)$ Coefficients $C_j$ emerging from the CMS data used in panel $(a)$ compared with the corresponding $C_j$ obtained from the  $2$-component compound binomial distribution ($2$-CBD). }
\label{FigCBD-2}
\end{figure}
\begin{figure}[h]
\begin{center}
\includegraphics[scale=0.35]{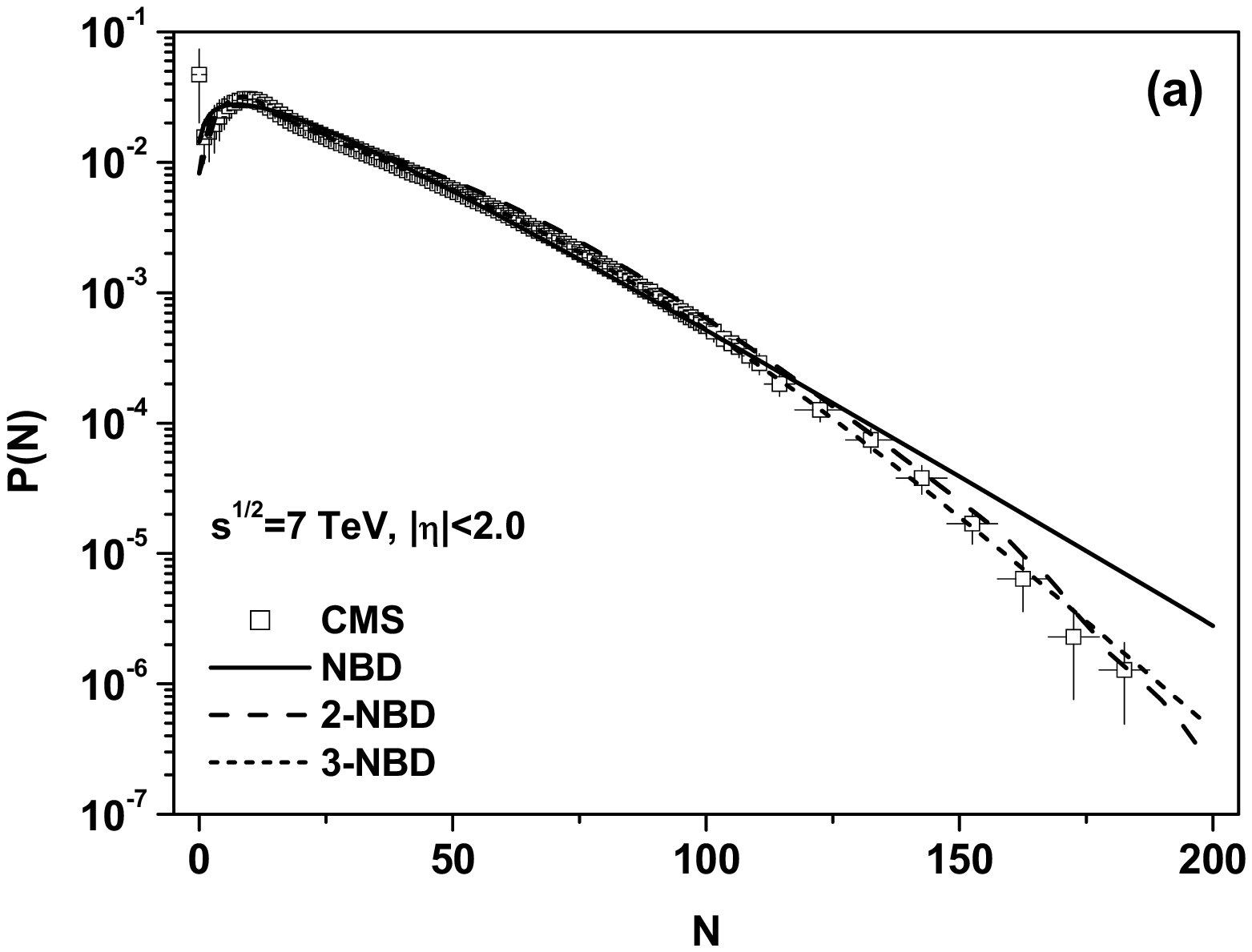}
\includegraphics[scale=0.35]{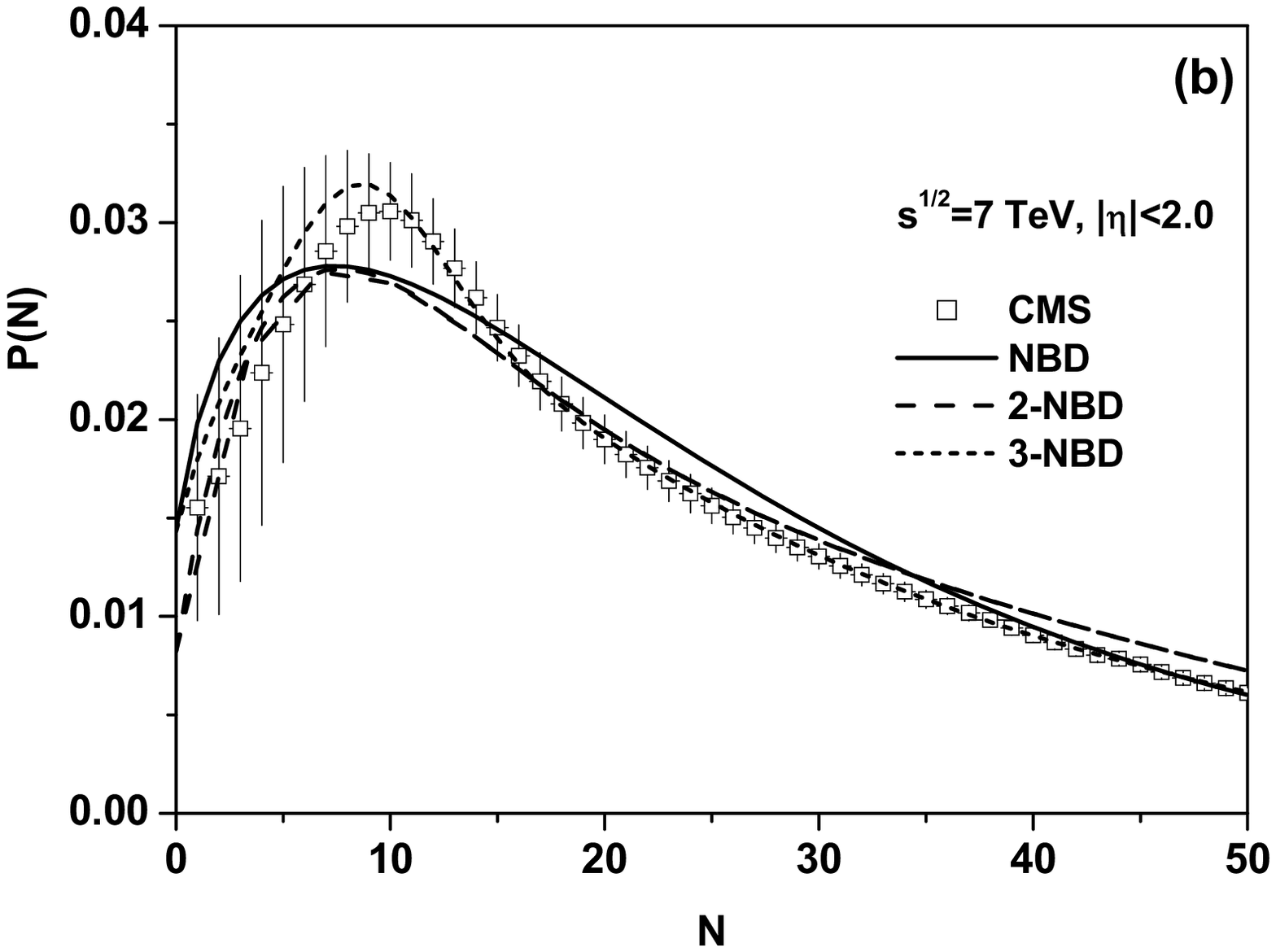}
\vspace{-5mm}
\includegraphics[scale=0.35]{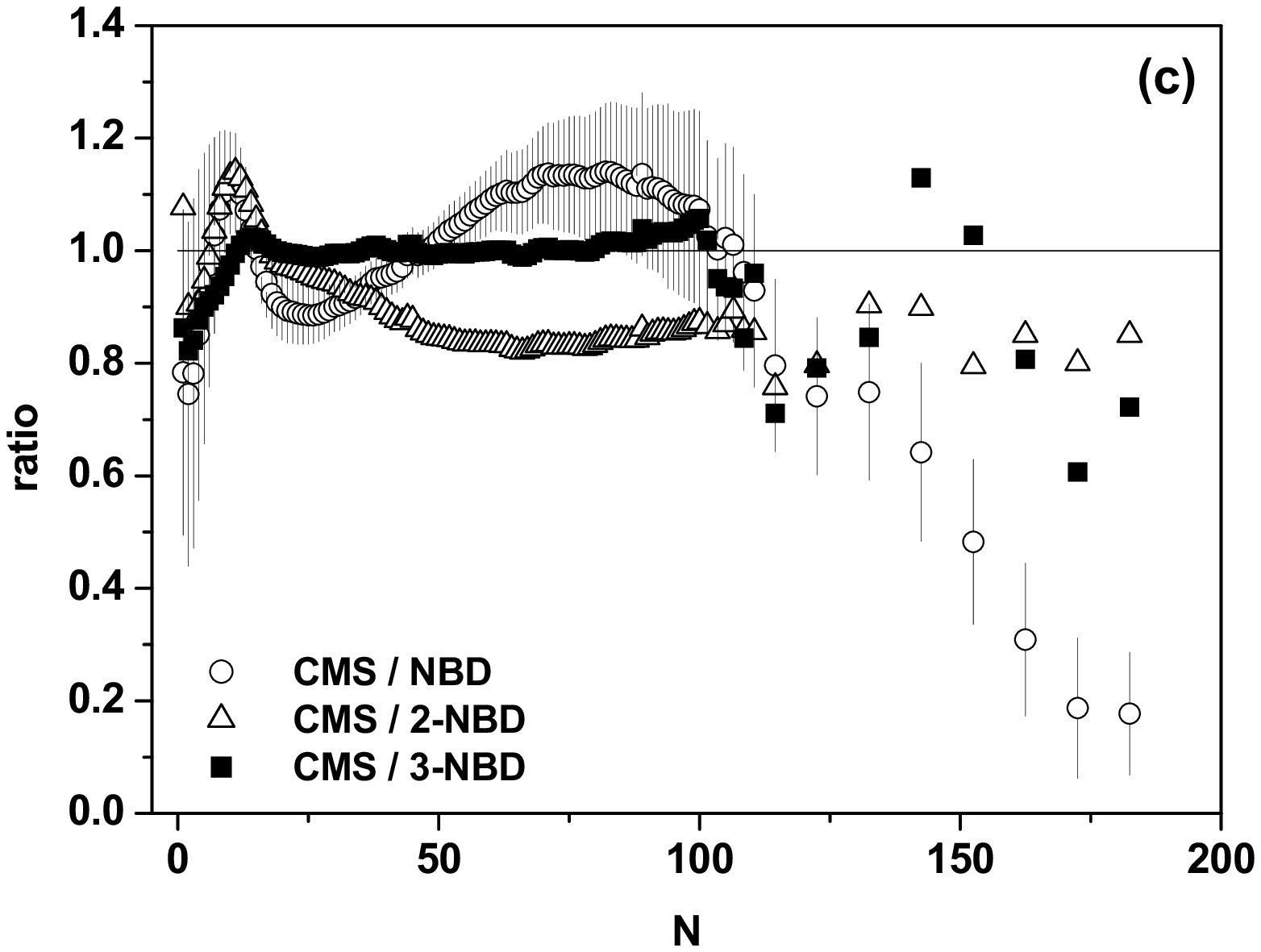}
\includegraphics[scale=0.35]{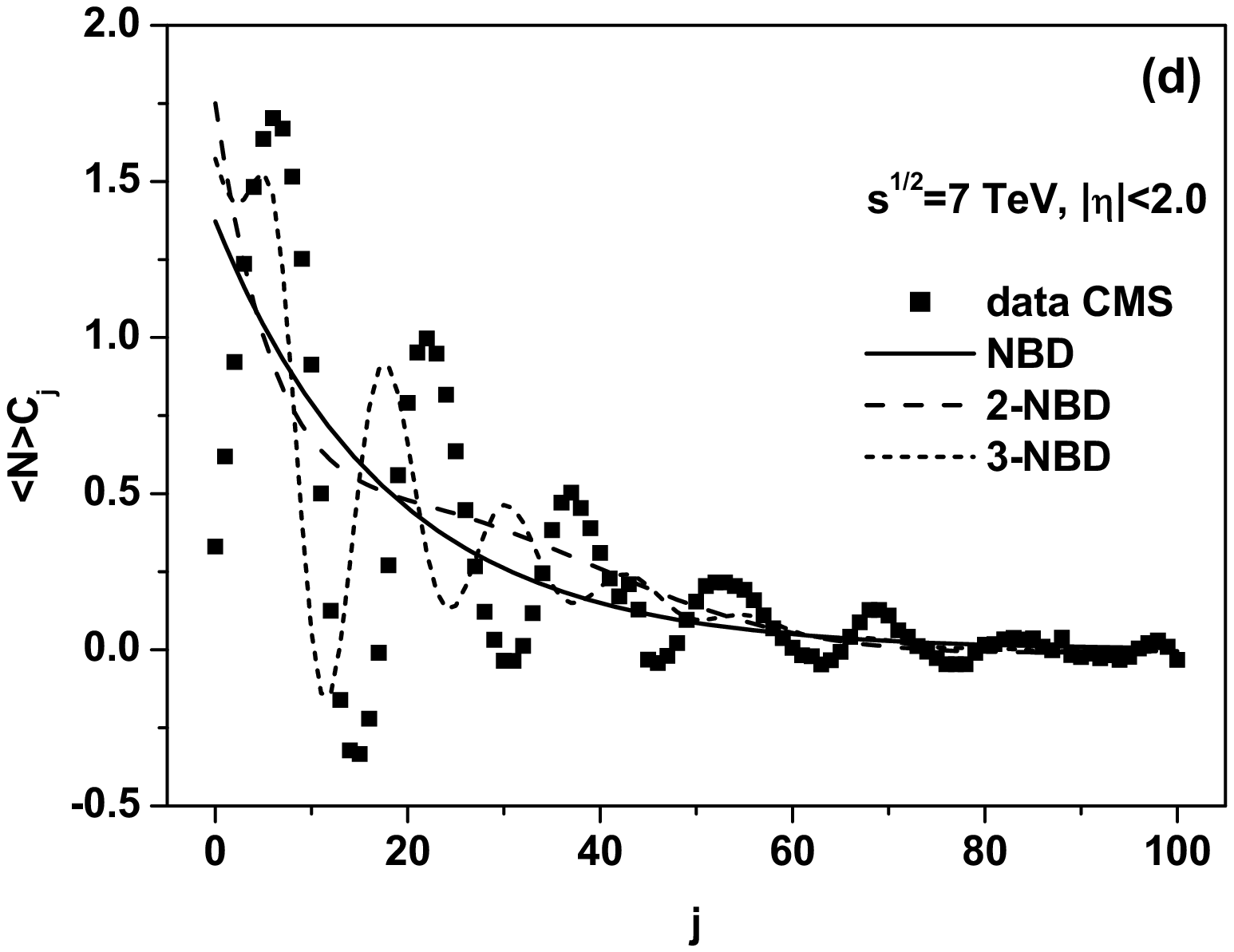}
\end{center}
\vspace{-3mm}
\caption{Results using $3$-component $P(N)$ composed from $3$ NBD as proposed in \cite{Z} (with parameters the same as in \cite{Z}).}
\label{Zbor}
\end{figure}
\begin{figure}[h]
\begin{center}
\includegraphics[scale=0.35]{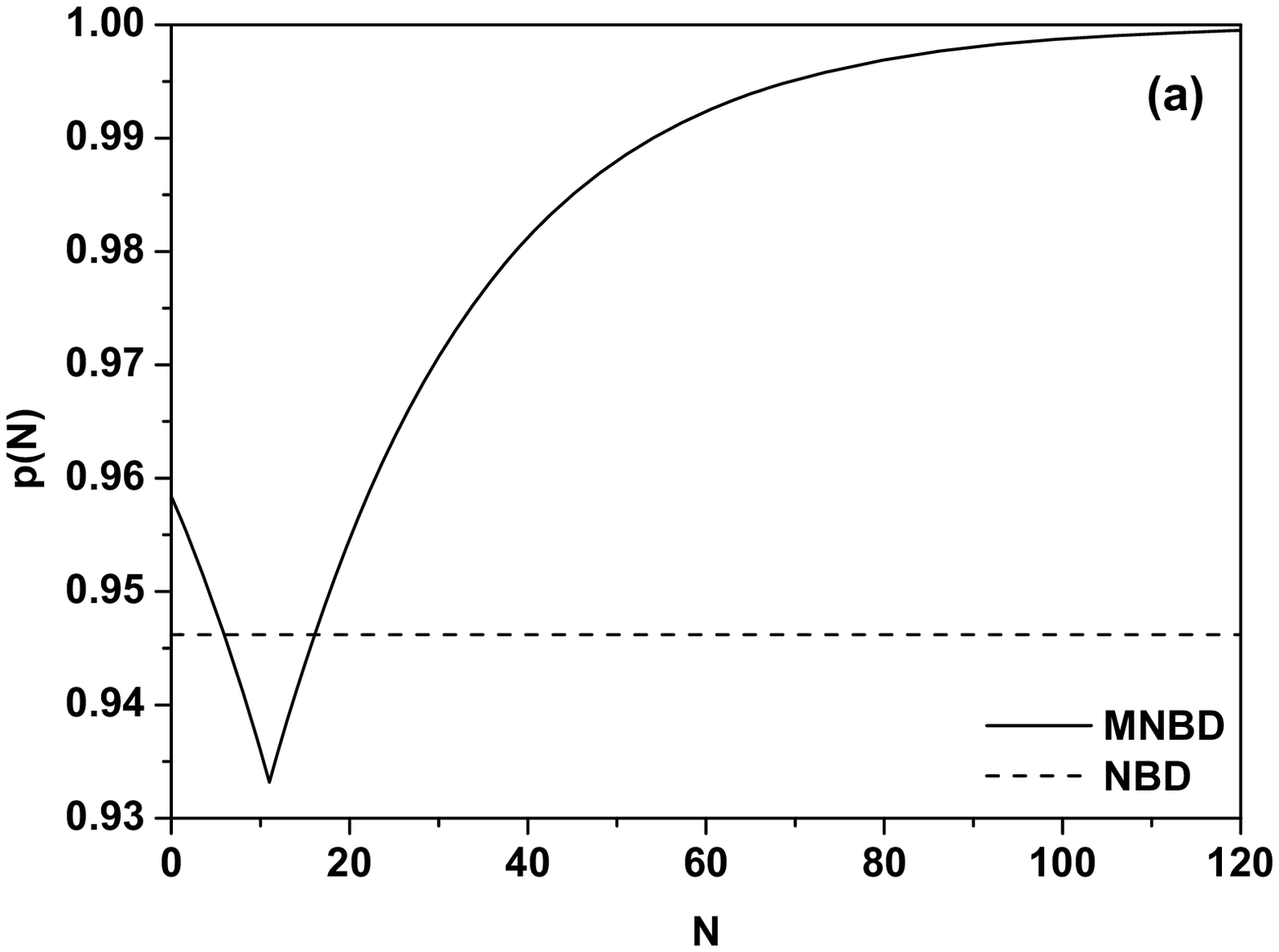}
\includegraphics[scale=0.35]{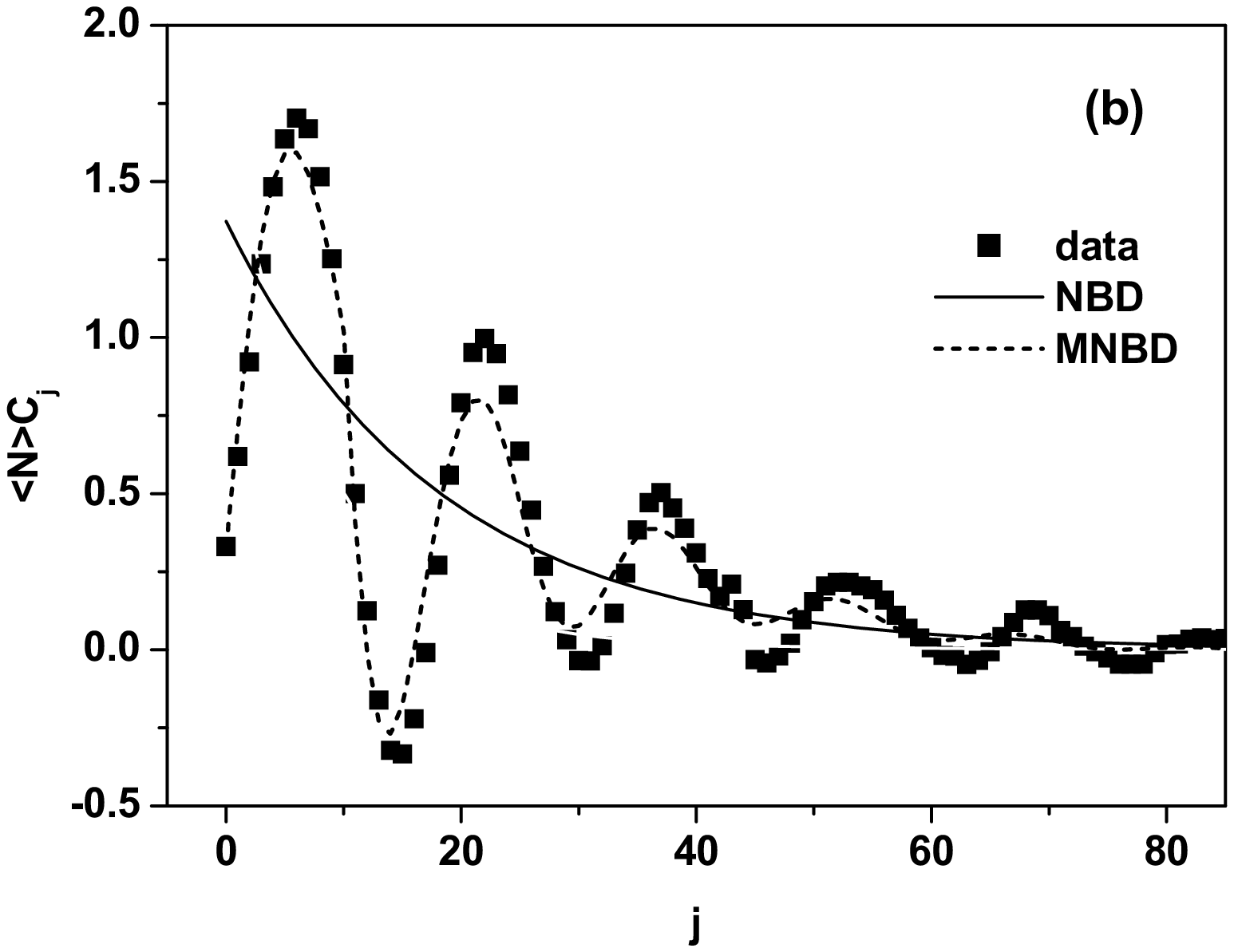}
\end{center}
\vspace{-5mm}
\caption{$(a)$  The modified probability of particle emission in the MNBD proposed in \cite{JPG} corresponding to the parameters in Eq. (\ref{Non}) equal to:  $c = 20.252$, $a_1 = 0.044$, $a_2 = 1.04\cdot 10^{-9}$ and $b = 11$. $(b)$  Coefficients $C_j$ emerging from the MNBD fit to the CMS data \cite{CMS-4} taken for $\sqrt{s}=7$ TeV and pseudorapidity window $|\eta|<2$ compared with the $C_j$ obtained from the single NBD and from the $2$-component NBD ($2$-NBD) fits to the CMS data with parameters from \cite{NBD-PG}.}
\label{FigJPG}
\end{figure}

These examples show that most probably single distributions (simple or compound) are not able to describe oscillations of $C_j$ and $P(N)$ at the same time. Therefore, we end by returning to the multi-NBD attempts mentioned before, this time to the $3$-NBD fit proposed in \cite{Z} , apparently the most successful so far. Its characteristic feature is the claim that in the data there is a place for a third component aiming to describe the low $N$ events (see \cite{Z}  for details). Fig. \ref{Zbor} shows our results based on using the parameters from \cite{Z} . As one can see, albeit $P(N)$ seems to be now very nicely reproduced, the ratio $R = data/fit$ signals that there are still some problems with the low $N$ part, and the oscillations, which are now present, still do not reproduce the data as well as those in Figs. \ref{FigCBD-1} and \ref{FigCBD-2}.

The problem of the origin of the oscillations  of $C_j$ remains therefore still open. We therefore end by returning to our paper \cite{JPG} where we have fitted both the $P(N)$ and the oscillations of the  modified combinants $C_j$  {\it par force} by suitably modifying the parameter $m$ in the probability of particle emission, $p = m/(m+k)$, in the NBD form of $P(N)$ used (see Table \ref{TablePN}, such an approach was used because changes in the parameter $k$ would result in changing the form of the original NBD). The best result was obtained for
\begin{equation}
m = c\exp\left[ a_1 | N - b| + a_2( N - b)^4\right], \label{Non}
\end{equation}
(with new free parameters: $c$, $a_1$, $a_2$ and $b$, see \cite{JPG} for details). This corresponds to a rather complicated, nonlinear and non monotonic form of $g(N)$ in the recurrence relation  Eq. (\ref{eq:dep}) and to $p = p(N)$ in the form shown in Fig. \ref{FigJPG} $(a)$. The proposed form of $p(N)$ is valid for $N < N_{max}$, with some maximum cut-off due to the normalization of $P(N)$. Note that the proposed modification is located in the region of small multiplicities $N$. After it $p(N)$ grows steadily. This is the price to keep only one term in $P(N)$ and to obtain the desired flat ratio of $R = data/fit$, resembling that in Fig. \ref{Zbor} $(c)$ but without the dip clearly visible there for very small $N$. So far such a spout-like form of the modification used in \cite{JPG} is just the simplest possible choice of parametrization that brings agreement with data and we cannot offer for the time being any plausible interpretation of such behavior. However, as one can see in Fig. \ref{FigJPG} $(b)$, this time all the coefficients $C_j$ calculated from this MNND fit very well those obtained from the data.

\section{Summary}
\label{sec:Summary}

Our review is focused on presenting and discussing some intriguing aspects of multiparticle production processes. In particular we concentrate on examples of the surprising efficiency of the nonextensive statistical approach (in particular on its scaling and self-similarity properties) and on the possible oscillatory behaviour apparently hidden in the multiplicity distributions.  Some summarizing remarks are in order here.

The connection between the Tsallis distribution in energy and the NBD form of the resultant $P(N)$ derived in Section \ref{sec:TtoP} can be regarded as a consequence of using a gamma distribution for clusters formed before fragmentation. Whereas the former arises from the fluctuations of temperature in a Boltzmann-Gibbs distribution, the latter arises from the fluctuations of mean multiplicity in a Poissonian distribution. The common feature is that in both cases fluctuations are caused by a gamma distribution which is stable under the size distribution, i.e., exhibits self-similarity and scaling behavior (actually, the NBD is also a  self-similar distribution \cite{CaT-1,CaT-2}). This indicates once more that self-similarity encountered in the processes under  consideration is the physical ground of the observed similarities discusses here.

Concerning the conjecture that jets, being a part of all produced particles, are approximately similar to inelastic collisions, as discussed in Section \ref{sec:SelfSim}, one must remember that in reality we do not observe the whole process of hadronization (analyzing all its subprocesses) to be able really to speak of self-similarity in multiparticle production processes. Our conjecture is based only on the information that we have on one such subprocess, i.e., on the production in jets. The observed similarities between them and multiparticle production in inelastic collisions in total is therefore the basis of our claim that we are dealing with a process which shows the same statistical properties at many scales, this is our self-similarity.

The power law and quasi power-law distributions discussed Section \ref{sec:oscillations} are ubiquitous in many different, apparently very disparate, branches of science (like, for example, earthquakes, escape probabilities in chaotic maps close to crisis, biased diffusion of tracers in random systems, kinetic and dynamic processes in random quenched and fractal media, diffusion limited aggregates, growth models, or stock markets near financial crashes, to name only a few \cite{WW-CSF81}). In most cases, they are decorated with log-periodic oscillations of different kinds \cite{Sornette,WWln,WW-CSF81} . This means therefore that oscillations of certain variables constitute a {\it universal phenomenon} which should occur in a large class of stochastic processes, independently of the microscopic details, including transverse momentum distributions at LHC energies. Therefore,  either the exponent $n$ of the power-law distributions becomes complex, or the scale parameter $T$ exhibits some specific log-periodic oscillations. Thus this means that either the system and/or the underlying physical mechanisms have characteristic {\it scale invariance behavior} (resulting in a complex power index $n$) or we observe a {\it sound wave in hadronic matter} (resulting in temperature oscillations)  which has a self similar solution (in log-periodic form). In the former case the discrete scale invariance and its associated complex exponents $n$ can appear spontaneously, without a pre-existing hierarchical structure \cite{Sornette} . In the latter case the corresponding wave equation has {\it self-similar solutions of the second kind} connected with the so called {\it intermediate asymptotic} (observed in phenomena which do not depend on the initial conditions because sufficient time has already passed, although the system considered is still out of equilibrium) \cite{BZ1,BZ2,B1,B2} . This suggests that both in $p+p$ and $Pb+Pb$ collisions one is dealing with an inhomogeneous medium with the density and the velocity of sound both depending on the position, and this can have some interesting experimental consequences. Note that the idea of oscillating $T$  was discussed recently in \cite{Hindusi} (although in a different context).

Concerning  the presence of oscillations in counting statistics in the multiplicity distributions $P(N)$, as discussed in Section \ref{sec:Combinants}, it is also a well established phenomenon. The known examples include oscillations of the higher-order cumulants of the probability distributions describing the transport through  a double quantum dot, oscillations in quantum optics (in the photon distribution function in slightly squeezed states) (see \cite{PNAS} for more information and references). In elementary particle physics oscillations of the so called $H_q$ moments, which represent ratios of the cumulants to factorial moments, also have a long history \cite{Kittel,DG,KNO-2} . However, our expectation that the oscillations of the modified combinants, $C_j$, could also be observed (and successfully measured) in multiparticle production processes is new. To see them one must first deduce $C_j$ from the experimental data on the multiplicity distribution $P(N)$ using the recurrence relation ({\ref{Cj}). Note that, contrary to the $H_q$ moments, the $C_j$ are independent of the multiplicity distribution $P(N)$ for $N > j$. In the case when the $C_j$ show oscillatory behavior,  they can be used to search for some underlying dynamical mechanism which could be responsible for it. The present situation is such that the measured $P(N)$ are most frequently described by NBD, the modified combinants of which do not oscillate. However, with increasing collision energy and increasing multiplicity of produced secondaries the NBD are not able to describe data properly. We propose therefore to use the modified combinants  $C_j$ obtained from the measured $P(N)$ together with the fitted multiplicity distributions $P(N)$ to allow for a more detailed quantitative description of the complex structure of the multiparticle production process. We argue that the observed strong oscillations of the coefficients $C_j$ of the $pp$ data at LHC energies indicate the compound character of the measured distributions $P(N)$ with a central role played by the BD which provides the oscillatory character of the $C_j$. This must be supplemented by some other distribution in such a way that the compound distribution fits both the observed $P(N)$ and $C_j$ deduced from it. However, at the moment we are not able to get reasonable fits to both $P(N)$ and $C_j$.  In fact, in \cite{Comb-BS} it was shown that combinants can serve as a useful tool to distinguish between two types of hadronizing sources: thermally equilibrated sources without restriction for multiplicity of produced particles - for them  combinants do not oscillate (and $P(N)$ is essentially a NBD) and sources with restriction for the number of emitted secondaries - for them combinants oscillate (and $P(N)$ follows essentially a BD). However, so far there have been no attempts to join both possibilities.  Therefore, these oscillations still await their physical justification, i.e., identification of some physical process (or combination of processes) which would result in such a phenomenon.

We close by noting that both phenomena discussed here describe, in fact, different dynamical aspects of the multiparticle production process at high energies. The quasi power-like distributions and the related log-periodic oscillations are related to events with rather small multiplicities of secondaries produced with large and very large momenta; they are called {\it hard collisions} and essentially probe the collision dynamics towards the edge of the phase space. The multiparticle distributions collect instead all produced particles, the majority of which come from the so called {\it soft collisions} concentrated in the middle of the phase space. In this sense, both the phenomena discussed provide us with complementary new information on these processes and, because of this, they should be considered, as much as possible, jointly. Because of some similarities observed between  hadronic, nuclear and $e^{+}e^{-}$ collisions \cite{Kittel,Sarkisyan,Bzdak} (see also Chapter $20$ of \cite{PDG2016}), one might expect that the phenomena discussed above will also appear in these reactions. However, this is a separate problem, too extensive and not yet much discussed, to be presented here.

\section*{Acknowledgments}

This research  was supported in part (GW) by the National Science Center (NCN) under contract 2016/22/M/ST2/00176.  We would like to thank Dr Nicholas Keeley for reading the manuscript.


\begin{thebibliography}{000} 

\bibitem{CM-1} C. Michael and L. Vanryckeghem, {\it J. Phys.  G\/} {\bf 3}, L151 (1977).

\bibitem{CM-2} C. Michael, {\it Prog. Part. Nucl. Phys.} {\bf 2}, 1 (1979).

\bibitem{H} R. Hagedorn, {\it Riv. Nuovo Cimento} {\bf 6}, 1 (1983).

\bibitem{UA1} UA1 Collab. (G. Arnison, {\it et al}.), {\it Phys. Lett. B \/} {\bf 118}, 167 (1982).

\bibitem{Tsallis-1} C. Tsallis, {\it J. Stat. Phys.} {\bf 52},479 (1998).

\bibitem{Tsallis-2} C. Tsallis, {\it Introduction to Nonextensive Statistical Mechanics}, (Springer, New York, 2009).

\bibitem{Contemporary} C. Tsallis, {\it Contemporary Physics}  {\bf 55}, 179 (2014).

\bibitem{UA1had} UA1 Collaboration. (C. Albajar,{\it et al}.),  {\it Nucl. Phys.  B\/} {\bf 335}, 261 (1990).

\bibitem{CMS-1} CMS Collaboration. (V. Khachatryan, {\it et al}.), {\it JHEP} {\bf 02}, 041 (2010).

\bibitem{CMS-2} CMS Collaboration. (V. Khachatryan, {\it et al}.), {\it Phys. Rev. Lett.} {\bf 105}, 022002 (2010).

\bibitem{CMS-3} CMS Collaboration. (V. Khachatryan, {\it et al}.), {\it JHEP} {\bf 08}, 086 (2011).

\bibitem{ATLAS-1} ATLAS Collaboration (G. Aad, {\it et al}.), {\it New J. Phys.} {\bf 13}, 053033 (2011).

\bibitem{ATLAS-2} ATLAS Collaboration (G. Aad, {\it et al}.), {\it Phys. Rev. D\/} {\bf 84}, 054001 (2011).

\bibitem{ATLAS-3} ATLAS Collaboration (G. Aad, {\it et al}.), {\it Eur. Phys. J. C\/} {\bf 71}, 1795 (2011) 1795.

\bibitem{ALICE-1} ALICE Collaboration. (K. Aamodt, {\it et al}.), {\it Phys. Lett. B\/} {\bf 693}, 53 (2010).

\bibitem{ALICE-2} ALICE Collaboration. (K. Aamodt, {\it et al}.), {\it Eur. Phys. J. C\/} {\bf 71}, 2662 (2013).

\bibitem{ALICE-3} ALICE Collaboration. (B. Abelev, {\it et al}.), {\it Phys. Lett. B\/} {\bf 720}, 52 (2013).

\bibitem{PHENIX-1} PHENIX Collaboration. (A. Adare, {\it et al}.), {\it Phys. Rev. D\/} {\bf 83}, 052004 (2011).

\bibitem{PHENIX-2} PHENIX Collaboration. (A. Adare, {\it et al}.), {\it Phys. Rev. C\/} {\bf 83}, 064903 (2011).

\bibitem{STAR} STAR Collaboratotion. (J. Adams, {\it et al}.),  {\it Phys. Rev. D\/} {\bf 74}, 032006 (2006).

\bibitem{RWW-1} M. Rybczy\'nski, Z. W\l odarczyk and G. Wilk, {\it Nucl. Phys. B\/ (Proc. Suppl.)} {\bf 97}, 81 (2001).

\bibitem{RWW-2} F. S. Navarra, O. V. Utyuzh, G. Wilk and Z. W\l odarczyk, {\it Phys. Rev. D\/}  {\bf 67}, 114002 (2003).

\bibitem{RWW-3} G. Wilk and Z. W\l odarczyk, {\it J. Phys. G\/} {\bf 38}, 065101 (2011).

\bibitem{RWW-4} M. Rybczy\'nski, Z. W\l odarczyk and G. Wilk, {\it J. Phys. G\/} {\bf 39}, 095004 (2012).

\bibitem{RWW-5} M. Rybczy\'nski and Z. W\l odarczyk, {\it Eur. Phys. J.  C\/} {\bf 74}, 2785 (2014).

\bibitem{WWrev-1} G. Wilk and Z. W\l odarczyk, {\it Eur. Phys. J. A\/} {\bf 40}, 299 (2009).

\bibitem{WWrev-2} G. Wilk and Z. W\l odarczyk,  {\it Eur. Phys. J. A\/} {\bf 48}, 161 (2012).

\bibitem{RWW-6} G. Wilk and Z. W\l odarczyk, {\it Cent. Eur. J. Phys.} {\bf 10}, 568 (2012).

\bibitem{Wibig-1} T. Wibig, {\it J. Phys. G\/} {\bf 37}, 115009 (2010).

\bibitem{Wibig-2} T. Wibig, {\it Eur. Phys. J. C\/} {\bf 74}, 2966 (2014).

\bibitem{Betall-1} K. \"Urm\"osy, G. G. Barnaf\"oldi and T. S. Bir\'o, {\it Phys. Lett. B\/} {\bf 701}, 111 (2012).

\bibitem{Betall-2} K. \"Urm\"osy, G. G. Barnaf\"oldi and T. S. Bir\'o, {\it Phys. Lett. B\/} {\bf 718}, 125 (2012).

\bibitem{Betall-3} T. S. Bir\'o, G. G. Barnaf\"oldi and P. V\'an, {\it Eur. Phys. J. A\/} {\bf 49}, 110 (2013).

\bibitem{Betall-4} T. S. Bir\'o, G. G. Barnaf\"oldi and P. V\'an, {\it Physica A\/} {\bf 417}, 215 (2015).

\bibitem{JCl-1} J. Cleymans and D. Worku, {\it J. Phys. G\/} {\bf 39}, 025006 (2012).

\bibitem{JCl-2} J. Cleymans and D. Worku, {\it Eur. Phys. J. A\/} {\bf 48}, 160 (2012).

\bibitem{JCl-3} M. D. Azmi and J. Cleymans, {\it J. Phys. G\/} {\bf 41}, 065001 (2014).

\bibitem{AD-1} A. Deppman, {\it Physica A\/} {\bf 391}, 6380 (2012).

\bibitem{AD-2} A. Deppman,  {\it J. Phys. G\/} {\bf 41}, 055108 (2014).

\bibitem{AD-3} I. Sena and A. Deppman, {\it Eur. Phys. J. A\/} {\bf 49}, 17 (2013).

\bibitem{AD-4} L. Marques, E. Andrade-II and A. Deppman, {\it Phys. Rev. D\/} {\bf 87}, 114022 (2013).

\bibitem{Others-1} P. K. Khandai, P. Sett, P. Shukla and V. Singh, {\it Int. J. Mod. Phys. A\/} {\bf 28}, 1350066 (2013).

\bibitem{Others-2} P. K. Khandai, P. Sett, P. Shukla and V. Singh, {\it J. Phys. G\/} {\bf 41}, 025105 (2014).

\bibitem{Others-3} B.-C. Li, Y.-Z. Wang and F.-H. Liu, {\it Phys. Lett. B\/} {\bf 725}, 352 (2013).

\bibitem{WalRaf} D. B. Walton and J. Rafelski, {\it Phys. Rev. Lett.} {\bf 84}, 31 (2000).

\bibitem{BMNSW} M. Biyajima, T. Mizoguchi, N. Nakajima, N. Suzuki and  Wilk, {\it  Eur. Phys. J. C\/} {\bf 48}, 597 (2006).

\bibitem{SS-BC} C. Beck and E. G. D. Cohen, {\it Physica A\/} {\bf 322}, 267 (2003).

\bibitem{SS-S} F. Sattin, {\it Eur. Phys. J. B\/} {\bf 49}, 219 (2006).

\bibitem{WW-APPB46} G. Wilk and Z. W\l odarczyk, {\it Acta Phys. Pol.} {\bf 46}, 1103 (2015).

\bibitem{WW-Entropy14} G. Wilk and Z. W\l odarczyk, {\it Entropy } {\bf 17}, 384 (2015).

\bibitem{WW-Entropy17} G. Wilk and Z. W\l odarczyk, {\it Entropy } {\bf 19}, 670 (2017).

\bibitem{WW-CSF81} G. Wilk and Z. W\l odarczyk, {\it Chaos Sol. Fract.} {\bf 81}, 487 (2015).

\bibitem{Just-1} O. J. E. Maroney, {\it Phys. Rev. E\/} {\bf 80}, 061141 (2009).

\bibitem{Just-2} T. S. Bir\'o, K. \"Urm\"ossy, Z. Schram, {\bf J. Phys. G\/} {\bf 37}, 094027 (2010).

\bibitem{Just-3} T. S. Bir\'o, P. V\'an, {\it Phys. Rev. E\/}  {\bf 83}, 061147 (2011).

\bibitem{Bbook} T. S. Bir\'o, {\it Is there a Temperature? Conceptual Challenges at High Energy, Acceleration and
                Complexity}, (Springer,  New York Dordrecht Heidelberg London, 2011).

\bibitem{JRGW-1} J. Ro\.zynek and G. Wilk, {\it Eur. Phys. J. A\/} {\bf 52}, 12 (2016).

\bibitem{JRGW-2} J. Ro\.zynek and G. Wilk, {\it Eur. Phys. J. A\/} {\bf 52}, 294 (2016).

\bibitem{AD-DM} E. Megias, D.P. Menezes and A. Deppman, {\it Physica A\/} {\bf 421} ,15 (2015).

\bibitem{F} H. Feshbach, {\it Physics Today} {\bf 40}, 9 (1987).

\bibitem{Hill-book} T. L. Hill, {\it  Thermodynamics of Small Systems (Parts 1 and 2)}, (New York: Benjamin, 1964).

\bibitem{TG} T. Giebultowicz, {\it Nature} {\bf 408} 299 (2000).

\bibitem{AKR-CSP-SA} A. K. Rajagopal, C. S. Pande and S. Abe, {\it Nanothermodynamics — a generic approach to
material properties at nanoscale}; Invited Presentation at the Indo-USWorkshop on
Nanoscale Materials: from Science to Technology (Puri, India, April 2004) [available at https://arxiv.org/abs/cond-mat/0403738].

\bibitem{HT} H. Touchette,  Temperature Fluctuations and Mixtures of Equilibrium States in the Canonical Ensemble, in    {\it Nonextensive Entropy—Interdisciplinary Applications}, ed.~ M. Gell-Mann and C.~ Tsallis (Oxford University Press, 2004),  p.~159.

\bibitem{G-MCP} V. García-Morales, J. Cervera and J. Pellicer, {\it Phys. Lett. A\/} {\bf 336}, 82 (2005).

\bibitem{Kodama-1} T. Kodama, H.-T. Elze, C.E. Augiar and T. Koide, {\it Europhys. Lett.} {\bf 70}, 439 (2005).

\bibitem{Kodama-2} T. Kodama, {\it J. Phys. G\/} {\bf 31}, S1051 (2005).

\bibitem{WWcov} G. Wilk and Z. W\l odarczyk, {\it Physica A\/} {\bf 390}, 3566 (2011).

\bibitem{IndPart} W. Feller, {\it An introduction to probability theory and its
                  applications}, Vol. II, (John Wiley and Sons Inc., New York 1966).

\bibitem{1overE-1} M. Sibata, {\it Phys. Rev. D\/} {\bf 24}, 1847 (1981).

\bibitem{1overE-2} C. Y. Wong, {\it Phys. Rev. D\/} {\bf 30}, 972 (1984).

\bibitem{Reif} F. Reif, {\it Statistical Physics}, (McGraw-Hill Book Company,  New York, 1967).

\bibitem{Feynm} R. P. Feynman, {\it Statistical Mechanics. A set of Lectures}, (W.A. Benjamin Inc. Menlo Park,
                California, 1972).

\bibitem{CR} G. Wilk and Z. W\l odarczyk, {\it Phys. Rev. D\/} {\bf 50}, 2318  (1994).

\bibitem{CRLF} G. Wilk and Z. W\l odarczyk, {\it Nuclear Physics B\/ (Proc. Suppl.)} {\bf 75A}, 191 (1999).

\bibitem{SS-WW-1} G. Wilk and Z. W\l odarczyk, {\it Phys. Rev. Lett.} {\bf 84}, 2770  (2000).

\bibitem{SS-WW-2} G. Wilk and Z. W\l odarczyk, {\it Chaos Solitons Fractals} {\bf 13}, 581 (2002).

\bibitem{WWTout} G. Wilk and Z. W\l odarczyk, {\it Phys. Rev.  C\/} {\bf 79}, 054903 (2009).

\bibitem{WWTin} G. Wilk and Z. W\l odarczyk, {\it Cent. Eur. J. Phys.} {\bf 8} 726 (2010).

\bibitem{BJ1}  T. S. Bir\'o and A. Jakov\'ac, {\it Phys. Rev. Lett.} {\bf 94}, 132302 (2005).

\bibitem{BJ2} T. S. Bir\'o, G. Purcel and K. \"Urm\"osy, {\it Eur. Phys. J. A\/} {\bf 40}, 325 (2009).

\bibitem{SS-B} C. Beck, {\it Phys. Rev. Lett.} {\bf 87}, 180601 (2001).

\bibitem{SS-BS} A. G. Bashkirov and A. D. Sukhanov, {\it J. Exp. Theor. Phys.} {\bf 95}, 440 (2002).

\bibitem{PLA} G. Wilk and Z. W\l odarczyk, {\it Phys. Lett.  A\/} {\bf 379}, 2941 (2015).

\bibitem{WWnets-1} G. Wilk and Z. W\l odarczyk, {\it Acta Phys. Pol. B\/} {\bf 35}, 871 (2004).

\bibitem{WWnets-2} G. Wilk and Z. W\l odarczyk, {\it Acta Phys. Pol. B\/} {\bf 36}, (2005).

\bibitem{nets-1} C. Tsallis, {\it Eur. Phys. J. ST\/} {\bf 161}, 175 (2008).

\bibitem{nets-2} D. J. B. Soares, C. Tsallis, A. M. Mariz and L. R.  da Silva, {\it Europhys. Lett.} {\bf 70}, 70 (2008).

\bibitem{WWnets-3} G. Wilk and Z. W\l odarczyk, {\it Acta Phys. Pol. B\/} {\bf 35}, 2141 (2004).

\bibitem{MGMG} M. Ga\'zdzicki and M.I. Gorenstein, {\it Phys. Lett. B\/} {\bf 517}, 250 (2001).

\bibitem{WWAIP-2013} G. Wilk and Z. W\l odarczyk, {\it AIP Conf. Proc.} {\bf 1558}, 893 (2013).

\bibitem{AT} C. Anteneodo and C. Tsallis, {\it J. Math. Phys.} {\bf 44}, 5194 (2003).

\bibitem{GR} S.M. Papalexiou, and D. Koutsoyiannis, Entropy maximization, p-moments and power-type distributions in nature,
             in {\it European Geosciences Union General Assembly 2011, Geophysical Research Abstracts}, Vol. {\bf 13}, Vienna, EGU2011-6884, doi:10.13140/RG.2.2.16999.24484, European Geosciences Union, 2011. Available              at http://itia.ntua.gr/1127.

\bibitem{TfromS} E. Rufeil Fiori and A. Plastino, {\it Physica A\/} {\bf 392}, 1742 (2013).

\bibitem{R} A. Rostovtsev, On a geometric mean and power-law statistical distributions, arXiv:cond-mat/0507414 (2005).

\bibitem{WWjets} G. Wilk and Z. W\l odarczyk, {\it Phys. Lett. B\/} {\bf 727}, 163 (2013).

\bibitem{NBDder-1} P. Carruthers and C. S. Shih, {\it Int. J. Mod. Phys. A\/} {\bf 2}, 1447 (1986).

\bibitem{NBDder-2} C. Vignat and A. Plastino, {\it Phys. Lett. A\/} {\bf 360}, 415 (2007).

\bibitem{NBDder-3} R. A. Fisher, {\it Ann. Eugenics} {\bf 11}, 182 (1941).

\bibitem{KNO-1} Z. Koba, H. B. Nielsen and P. Olesen, {\it Nucl. Phys. B\/} {\bf 40}, 319 (1972).

\bibitem{KNO-2} J. F. Fiete Grosse-Oetringhaus and K. Reygers, {\it J.  Phys. G\/}  {\bf 37}, 083001 (2010).

\bibitem{Q} G. Wilk and Z. W\l odarczyk, {\it Physica A\/} {\bf 376}, 279 (2007).

\bibitem{WWCT} C-Y. Wong, G. Wilk, L. J. L. Cirto and C. Tsallis, {\it Phys. Rev. D\/} {\bf 91}, 114027 (2015).

\bibitem{WW12} C. Y. Wong and G. Wilk, {\it Acta Phys. Pol. B\/} {\bf 43}, 2047 (2012).

\bibitem{WW13} C. Y. Wong and G. Wilk, {\it Phys. Rev. D\/}  {\bf 87}, 114007 (2013).

\bibitem{MW-limited} M. Rybczy\'nski and Z. W\l odarczyk, {\it Eur. Phys. J. A\/} {\bf 51}, 80 (2015).

\bibitem{CMS-5}  CMS Collaboration. (V. Khachatryan, {\it et al}.), {\it Phys. Rev. Lett.} {\bf 107},132001 (2011).

\bibitem{CDF} CDF Collaboration. (T. Aaltonen, {\it et al.}), {\it  Phys. Rev. D\/} {\bf 78}, 052006 (2008) [{\it Erratum:
              Phys. Rev. D\/} {\bf 79}, 119902 (2009)].

\bibitem{D0}   D0 Collaboration, (B. Abbott, {\it et al.}), {\it  Phys. Rev. D\/} {\bf 64}, 032003 (2001).

\bibitem{UA2} UA2 Collaboration, (P. Bagnaia, {\it et al.}), {\it  Phys. Lett. B\/} {\bf 138}, 430 (1984).

\bibitem{AKW} A. Wr\'oblewski, {\it Acta Phys. Polon. B\/} {\bf 4}, 857 (1973).

\bibitem{G-G} C. Geich-Gimbel, {\it Int. J. Mod. Phys. A\/} {\bf 4}, 1527 (1989).

\bibitem{Satz} H. Satz, {\it Int. J. Mod. Phys. E\/} {\bf 21}, 1230006 (2012).

\bibitem{Kas-1} D. Sornette and R. Cont, {\it J. Phys.  I France} {\bf 7}, 431 (1997).

\bibitem{Kas-2} C. Anteneodo and C. Tsallis, {\it J. Math. Phys.}  {\bf 44}, 5914  (2003).

\bibitem{SsH-1} R. Hagedorn and R. Ranft, {\it Suppl. Nuovo\ Cim.} {\bf 6}, 169 (1968).

\bibitem{SsH-2} R. Hagedorn, {\it Nucl. Phys. B\/} {\bf 24}, 93 (1970).

\bibitem{SsQCD-1} J. D. Bjorken, {\it Phys. Rev. D\/} {\bf 45}, 4077 (1992) 4077.

\bibitem{SsQCD-2} G. Gustafson and A. Nilsson, {\it Nucl. Phys. B\/} {\bf 355}, 106 (1991).

\bibitem{SsQCD-3} W. Ochs, {\it Acta Phys. Polon. B\/} {\bf 22}, 203 (1991).

\bibitem{TF-1} A. Deppman, {\it Phys. Rev. D\/} {\bf 93}, 93 (2016).

\bibitem{TF-2} A. Deppman, {\it Universe} {\bf 3}, 62 (2017).

\bibitem{TF-3} A. Deppman, E. Meg\'ias, D. P. Menezes and T. Frederico, Fractal structure and non extensive statistics,
               arXiv:1801.01160v1   [cond-mat.stat-mech].

\bibitem{NA49-1} NA49 Collab. (C. Alt {\it et al}.), {\it Phys. Rev. C\/} {\bf 77}, 034906 (2008).

\bibitem{NA49-2} NA49 Collab. (S. V. Afanasiev {\it et al}.), {\it Phys. Rev. C\/} {\bf 66}, 054902 (2002).

\bibitem{RHIC} B. De, S. Bhattacharyya, G. Sau and S.K. Biswas, {\it Int. J. Mod. Phys. E\/} {\bf 16}, 1687 (2007).

\bibitem{BBB}  T. S. Bir\'o, G. G. Barnaf\"oldi, G. B\'ir\'o and K. M. Shen, {\it J. Physics: Conf. Series} {\bf 779},
               012081 (2017).

\bibitem{SBW}  K. M. Shen, T. S. Bir\'o and E.-K. Wang, {\it Physica A\/} {\bf 492}, 2353 (2018).

\bibitem{JL} J. Lindhard, 'Complementarity' between energy and temperature, in  {\it The Lesson of Quantum Theory},
             eds. by J.de Boer, E. Dal and O. Ulfbeck (North-Holland, Amsterdam, 1986).

\bibitem{BH} N. Bohr, in {\it Collected Works}, ed. J. Kalekar (North-Holland, Amsterdam, 1985), Vol. 6, pp.
             316-330 and 376-377.

\bibitem{UL} J. Uffink and J. van Lith, {\it Found. Phys.} {\bf 29}, 655 (1999).

\bibitem{L} B. H. Lavenda, {\it Found. Phys. Lett.} {\bf 13}, 487 (2000).

\bibitem{Q1} K. Chandra Kar, {\it Phys. Rev.} {\bf 21}, 672 (1923).

\bibitem{Gor} V. V. Begun, M. I. Gorenstein, M. Hauer, V. P. Konchakovski and O. S. Zozulya,
              {\it Phys. Rev. C\/} {\bf 74}, 044803 (2006).

\bibitem{WWW} G. Wilk, Z. W\l odarczyk and W. Wolak, {\it Acta Phys. Pol. B\/} {\bf 42}, 1277 (2011).

\bibitem{NA49-3} NA49 Collab. (C. Alt {\it et al}.), {\it Phys. Rev. C\/} {\bf 77}, 024903 (2008).

\bibitem{PN} A. K. Dash and B. M. Mohanty, {\it J. Phys. G\/} {\bf 37}, 025102 (2010).

\bibitem{BiroRev} G. Bír\'o, G. G. Barnaf\"oldi, T. S. Bir\'o, K. \"Urm\"ossy and \'A. Tak\'acs,
                  {\it Entropy} {\bf 19}, 88 (2017).

\bibitem{WWln} G. Wilk and Z. W\l odarczyk, {\it Physica  A \/} {\bf 413}, 53 (2014).

\bibitem{RWWlnA} M. Rybczy\'nski, G. Wilk and Z. W\l odarczyk, {\it Eur. Phys. J. Web Conf.} {\bf 90}, 1002 (2015).

\bibitem{SoftHard1} G. G. Barnaf\"oldi, K. \"Urm\"ossy and G.~Bir\'o, {\it J. Phys. Conf. Ser.} {\bf 612}, 012048 (2015).

\bibitem{SoftHard2} K. \"Urm\"ossy, G. G. Barnaf\"oldi, Sz. Harangoz\'o, T. S. Bir\'o and  Z. Xu, {\it  J. Phys. Conf.
                    Ser.} {\bf 805}, 012010 (2017).

\bibitem{Sornette} D. Sornette, {\it Phys. Rep.} {\bf 239}, 239 (1998).

\bibitem{Lang} N. G. van Kampen, {\it Stochastic Processes in Physics and Chemistry}, (Elsevier, Amsterdam, 2007), pp. 193-231.

\bibitem{WWwaves} G. Wilk and Z. W\l odarczyk, {\it Physica A\/} {\bf 486}, 579 (2017).

\bibitem{HydroWaves} D. A. Foga\c ca, L. G. Ferreira Filho and F. S. Navarra, {\it Nucl. Phys. A\/} {\bf 819}, 150 (2009).

\bibitem{Landau} L. D.Landau and E. M. Lifshitz, {\it Fluid mechanics}, (Pergamon Press, Oxford, 1987).

\bibitem{BZ1} G. I. Barenblatt and Ya. B. Zeldovich, {\it Russian Math. Surv.} {\bf 26}, 45 (1971).

\bibitem{BZ2} G. I. Barenblatt and Ya. B. Zeldovich, {\it Ann. Rev. Fluid Mech.} {\bf 4}, 285 (1972).

\bibitem{B1} G. I. Barenblatt,  {\it Scaling, self-similarity, and intermediate asymptotics}, (Cambridge University Press,
             1996).

\bibitem{B2} G. I. Barenblatt, {\it Scaling}, (Cambridge University Press, 2013).

\bibitem{BLASTMODEL} M. Lv, Y. G. Ma, G. Q. Zhang, J. H. Chen and D. Q. Fanga, {\it Phys. Lett. B\/} {\bf 733}, 105 (2014).

\bibitem{NA61} K. Grebieszkow, {\it PoS DIS2014}, 018 (2014), [arXiv:1407.3690 [hep-ex]].

\bibitem{GGS} M. Ga\'zdzicki, M. Gorenstein and P. Seyboth, {\it Acta Phys. Polon. B\/} {\bf 42}, 307 (2011).

\bibitem{Hill} T. L. Hill, {\it Statistical Mechanics}, (McGraw-Hill, New York, 1956) (also reprinted  (Dover, New York,
               1987)).

\bibitem{Balescu} R. Balescu, {\it Equilibrium and Nonequilibrium Statistical Mechanics}, (New York: Wiley Interscience,
                  1975).

\bibitem{RecentNA61}  NA61/SHINE Collaboration. (A. Aduszkiewicz {\it et al}.) {\it Eur. Phys. J.  C \/} {\bf 76}, 635
                     (2016).

\bibitem{FS} F. H. Stillinger, {\it J. Chem. Phys.} {\bf 109}, 3983 (1998).

\bibitem{RW} M. Rybczy\'nski and Z. W\l odarczyk, {\it J. Phys. Conf. Ser.} {\bf 5}, 238 (2005).

\bibitem{Kittel} W. Kittel and E. A. De Wolf, {\it  Soft Multihadron Dynamics}, (World Scientific, Singapore, 2005).

\bibitem{JPG} G. Wilk and Z. W\l odarczyk, {\it J. Phys. G\/} {\bf 44}, 015002 (2017).

\bibitem{CMS-4} CMS Collaboration. (V. Khachatryan, {\it et al}.), {\it JHEP} {\bf 01}, 079 (2011).

\bibitem{GU} A. Giovannini and R. Ugoccioni, {\it Phys. Rev. D\/} {\bf 68}, 034009 (2003).

\bibitem{Z} I. J. Zborovsky, {\it J. Phys. G\/} {\bf 40}, 055005 (2013).

\bibitem{DN} I. M. Dremin and V. A. Nechitailo, {\it Phys. Rev. D\/} {\bf 70}, 034005 (2004).

\bibitem{DG} I. M. Dremin and J. W. Gary, {\it Phys. Rep.} {\bf 349}, 301 (2001).

\bibitem{NBD-PG} P. Ghosh, {\it Phys. Rev. D\/} {\bf 85}, 054017 (2012).

\bibitem{CSP} B. E. A. Saleh and M. K. Teich, {\it Proc. IEEE} {\bf 70}, 229 (1982).

\bibitem{ALICE-4} ALICE Collaboration. (J. Adam, {\it et al}.), {\it Eur. Phys. J. C\/} {\bf 77}, 33 (2017).

\bibitem{Alkin} A. Alkin, {\it Ukr. J. Phys.} {\bf 62}, 743 (2017) [arXiv:1710.01979].

\bibitem{Cowan} C. Cowan, {\it Statistical Data Analysis}, (Clarendon Press, 1988).

\bibitem{CSF} V. D. Rusov, T. N. Zelentsova, S. I. Kosenko, M. M. Ovsyanko and I. V. Sharf,
              {\it Phys. Lett. B\/} {\bf 504}, 213 (2001).

\bibitem{CSF1} V. D. Rusov and I. V. Sharf, {\it Nucl. Phys. A\/} {\bf 764} 460 (2006).

\bibitem{KG}  S. K. Kauffmann and M. Gyulassy, {\it J. Phys. A\/} {\bf 11}, 1715 (1978).

\bibitem{CombUse} A. B. Balantekin and J.E. Seger, {\it Phys. Lett. B\/} {\bf 266}, 231 (1991).

\bibitem{Hegyi} S. Hegyi, {\it Phys. Lett. B\/} {\bf 463}, 126 (1999).

\bibitem{BSWW} M. Biyajima, N. Suzuki, G. Wilk and Z. W\l odarczyk, {\it Phys. Lett. B\/} {\bf 386}, 297 (1996).

\bibitem{GVH} A. Giovannini and L. Van Hove, {\it Z. Phys. C\/} {\bf 30}, 391 (1986).

\bibitem{Compound} B. Sundt and R. Vernic, {\it Recursions for Convolutions and Compound Distributions with Insurance
                   Applications}, (Springer-Verlag Berlin Heidelberg, 2009).

\bibitem{Panjer} H. H. Panjer, {\it ASTIN Bull.} {\bf 12}, 22 (1981).

\bibitem{CaT-1} G. Calucci and D. Treleani, {\it Phys. Rev. D\/} {\bf 57}, 602 (1998).

\bibitem{CaT-2} M. A. Braun and C. Pajares, {\it Phys. Lett. B\/} {\bf 444}, 435 (1998) 435.

\bibitem{Hindusi} T. Bhattacharyya, P. Garg, R. Sahoo and P. Samantray, {\it Eur. Phys. J. A\/} {\bf 52}, 283 (2016).

\bibitem{PNAS} C. Flindt, C. Fricke, F. Hohls, T. Novotny, K. Netocny, T. Brandes and R. J. Haug, {\it Proc. Natl. Acad.
               Sci. U. S. A.} {\bf 106}, 10116 (2009).

\bibitem{Comb-BS} A. B. Balantekin and J .E. Seger, {\it Phys. Lett. B\/} {\bf 266} 231 (1991).

\bibitem{Sarkisyan} E. K. Sarkisyan, A. N. Mishra, R. Sahoo, S. Alexander and A. S. Sakharov, {\it Phys. Rev. D\/}
                    {\bf 93}, 054046 (2016).

\bibitem{Bzdak} A. Bzdak, {\it Phys. Rev. D\/}, {\bf 96}, 036007 (2017).

\bibitem{PDG2016} Particle Data Group. ( C. Patrignani, {\it et~al}.), {\it Chinese Phys. C\/} {\bf 40}, 100001 (2016).




\end{thebibliography}
\end{document}